\documentclass[a4paper,fleqn,usenatbib]{mnras}

\usepackage{Times}

\usepackage[T1]{fontenc}
\usepackage{ae,aecompl}

\usepackage{graphicx}	
\usepackage{amsmath}	
\usepackage{amssymb}	

\title[The origin of scatter in SFR(M$_{\rm star}$)]{The origin of scatter in the star formation rate - stellar mass relation}

\author[J. Matthee et al.]{Jorryt Matthee$^{1,2}$\thanks{E-mail: mattheej@phys.ethz.ch}\thanks{Zwicky Fellow} \& Joop Schaye$^{1}$\\
$^{1}$ Leiden Observatory, Leiden University, P.O.\ Box 9513, NL-2300 RA Leiden, The Netherlands\\
$^{2}$ Department of Physics, ETH Z\"urich,Wolfgang-Pauli-Strasse 27, 8093 Z\"urich, Switzerland\\}

\pubyear{2018}

\begin{document}
\label{firstpage}
\pagerange{\pageref{firstpage}--\pageref{lastpage}}
\maketitle

\begin{abstract}
Observations have revealed that the star formation rate (SFR) and stellar mass (M$_{\rm star}$) of star-forming galaxies follow a tight relation known as the galaxy main sequence. However, what physical information is encoded in this relation is under debate. Here, we use the EAGLE cosmological hydrodynamical simulation to study the mass dependence, evolution and origin of scatter in the SFR-M$_{\rm star}$ relation. At $z=0$, we find that the scatter decreases slightly with stellar mass from 0.35 dex at M$_{\rm star} \approx 10^9$ M$_{\odot}$ to 0.30 dex at M$_{\rm star} \gtrsim 10^{10.5}$ M$_{\odot}$. The scatter decreases from $z=0$ to $z=5$ by 0.05 dex at M$_{\rm star} \gtrsim 10^{10}$ M$_{\odot}$ and by 0.15 dex for lower masses. We show that the scatter at $z=0.1$ originates from a combination of fluctuations on short time-scales (ranging from 0.2-2 Gyr) that are presumably associated with self-regulation from cooling, star formation and outflows, but is dominated by long time-scale ($\sim 10$ Gyr) variations related to differences in halo formation times. Shorter time-scale fluctuations are relatively more important for lower-mass galaxies. At high masses, differences in black hole formation efficiency cause additional scatter, but also diminish the scatter caused by different halo formation times. While individual galaxies cross the main sequence multiple times during their evolution, they fluctuate around tracks associated with their halo properties, i.e. galaxies above/below the main sequence at $z = 0.1$ tend to have been above/below the main sequence for $\gg1$ Gyr.
\end{abstract}




\section{Introduction}
Observations of large samples of star-forming galaxies have shown that their star formation rate (SFR) and stellar mass (M$_{\rm star}$) are closely related \citep[e.g.][]{Brinchmann2004,Elbaz2011,Rodighiero2014,Speagle2014,Schreiber2015}. This relation is sometimes called the `main sequence of galaxies' \citep{Noeske2007,Whitaker2012}. The existence of a main sequence (with a scatter of only $\approx 0.3$ dex) indicates that the majority of galaxies (of fixed stellar mass) have SFRs that lie within a factor of two and it may be explained as an effect of the self-regulating nature of star-formation due to the interplay between gas accretion, star-formation and feedback driven outflows \citep[e.g.][]{Schaye2010,Dave2011,Haas2013,Lilly2013,Tacchella2016,RP2016}. 

However, what information can be derived from the existence of the main sequence is still under debate \citep[e.g.][]{Kelson2014,Abramson2015}. Is the main sequence some ``attractor-solution", where galaxies fluctuate rapidly around the median relation because of stochasticity in star-formation events on short time-scales \citep[e.g.][]{Peng2010,Behroozi2013}, meaning that the star formation histories of galaxies with the same stellar mass are similar? Or does the main sequence simply show a ``population average" at a certain age of the Universe \citep[e.g.][]{Gladders2013,Abramson2016}, and do galaxies at fixed stellar mass have very diverse star formation histories (SFHs) on longer time-scales? Or is it a combination of both? Which effects are most important at different mass-scales and time-scales?

Crucial information may be encoded in the scatter in the SFR-M$_{\rm star}$ relation, and its dependence on stellar mass and cosmic time. What makes the growth rates of galaxies different, what are the important time-scales? Are the differences between galaxies at different positions along the main sequence systematic or stochastic? How important is the influence of the environment inside the halo (satellite galaxies) and at large radii (large-scale overdensity)? What is the role of the dark matter accretion history? These questions are the subject of this paper.

In order to investigate the physical origin of scatter in the SFR-M$_{\rm star}$ relation, we study galaxies in the cosmological hydrodynamical EAGLE simulation \citep{Schaye2014,Crain2015,McAlpine2015}. A benefit of using simulations is that we can trace galaxies during their evolution and also compare the evolution of the stellar mass with the evolution of the dark matter halo mass (particularly in matched dark matter only simulations) and the large-scale environment. EAGLE has been designed to reproduce the $z\approx0$ galaxy stellar mass function and galaxy sizes, but simultaneously reproduces many other galaxy scaling relations \citep[e.g.][]{Crain2015,Lagos2015,Schaye2014,Bahe2016,Segers2016}. Importantly for this work, this includes the growth of the stellar mass density, the evolution of the star formation rate function \citep{Katsianis2017} and the  the SFR-M$_{\rm star}$ relation \citep{Furlong2015}. A limitation from EAGLE is that short time-scale fluctuations in SFRs (particularly below 100 Myr) are challenging to measure due to the mass resolution and the time resolution of the simulation output. Likewise, frequently used observational SFR indicators also measure SFRs averaged over (at least) the past 100 Myr, such as the UV+IR that is used in e.g. \cite{Chang2015}.

This paper is structured as follows. We first explain the methods in \S $\ref{sec:methods}$, including the simulation set-up and the way scatter is measured. We then characterise the scatter in the SFR-M$_{\rm star}$ relation in \S $\ref{sec:mass_scatter}$. This includes the dependence on mass, comparison with observations in the local Universe and the evolution of the scatter. We explore the relative importance of fluctuations in the SFRs on long and short time-scales in \S $\ref{burstiness}$. Then, in \S $\ref{sec:cosmological_origin}$, we investigate the relation between the scatter and the growth histories of galaxies and their dark matter haloes. We study the relation with supermassive black hole mass as a tracer of the impact of AGN feedback in \S $\ref{sec:BH}$. Our results are discussed in \S $\ref{sec:discussion}$, including potential observational tests, and we summarise the paper in \S $\ref{sec:summary}$.

\section{Methods} \label{sec:methods}
\subsection{The EAGLE simulations}
We study the properties of simulated galaxies from the EAGLE cosmological hydrodynamical simulation \citep{Schaye2014,Crain2015}\footnote{Galaxy catalogues and merger-trees are available through \cite{McAlpine2015}.}. EAGLE is simulated with the smoothed particle hydrodynamic $N$-body code {\sc Gadget 3} \citep{Springel2005Gadget}, with extensive modifications in the hydrodynamic solver and time-stepping \citep{Durier2012,Hopkins2013,SchallerSPH}. We use the (100 cMpc)$^3$ reference model, which includes $2 \times 1504^3$ particles with masses $9.7\times10^6$ M$_{\odot}$ (dark matter) and $1.8\times10^6$ M$_{\odot}$ (initial baryonic). As several important physical processes are unresolved at these mass scales, the following sub-grid models are included: radiative cooling of gas \citep{Wiersma2009cooling}, the formation of star particles \citep{SchayeVecchia2008}, chemical enrichment by stellar mass loss \citep{Wiersma2009enrich}, feedback from star formation  \citep{VecchiaSchaye2012}, and the growth of black holes and the feedback associated with AGN activity \citep{Springel2005BH,BoothSchaye2009,Rosas2015,Schaye2014}.

Galaxies have been identified as subhalos in Friends-of-Friends (FoF) halos \citep[e.g.][]{Einasto1984} using the {\sc subfind} \citep{Springel2001,Dolag2008} algorithm. The subhalo that is at the minimum gravitational potential in a FoF halo is defined to be the central galaxy. The edges of subhalos are identified through saddle points in the density distribution and only gravitationally bound particles are members of a subhalo. We measure the evolution of halo masses by linking galaxies between redshift snap/snipshots (snipshots are snapshots stored at finer time steps but containing fewer variables) with their most-massive progenitors using the merger-trees described in \cite{Qu2016}. Following \cite{Schaye2014}, we measure the stellar mass and SFR of a galaxy by summing over the particles within a 30 pkpc radius from the minimum of the gravitational potential, mimicking typical apertures used in observations. Similarly to other analyses, the {\it instantaneous} SFR in EAGLE is measured directly from the dense gas mass \citep{SchayeVecchia2008}. The density threshold for star formation increases with decreasing metallicity following the fits of \cite{Schaye2004} for the critical density required for the formation of a cold, molecular gas phase. In order to benefit from a higher time resolution in SFR histories, SFRs in \S $\ref{sec:variety}$ and \S $\ref{sec:relative_timescales}$ are computed using the difference in initial (i.e. zero-age main sequence) stellar masses of subhalos between two snipshots. The caveats related to this measurement are discussed in detail in \S $\ref{sec:variety}$.

We compare the properties of the reference simulation (including baryon physics) with a matched dark matter only version of EAGLE (DMO). This version was run with the same initial conditions and resolution as the reference model and allows to distinguish between causation and correlation because DMO properties cannot be affected by baryonic effects. Halos are matched between the simulations by finding the DMO halo that includes the 50 most-bound dark matter particles in the reference model (see \citealt{Schaller2014} for details). Halos are successfully matched if at least 25 of these particles are members of a single FoF group. More than $99$\% of the haloes that are included in our analysis are matched successfully. We have verified that all our results remain very similar if only properties from the reference simulation are used.

\begin{figure}
\includegraphics[width=8.5cm]{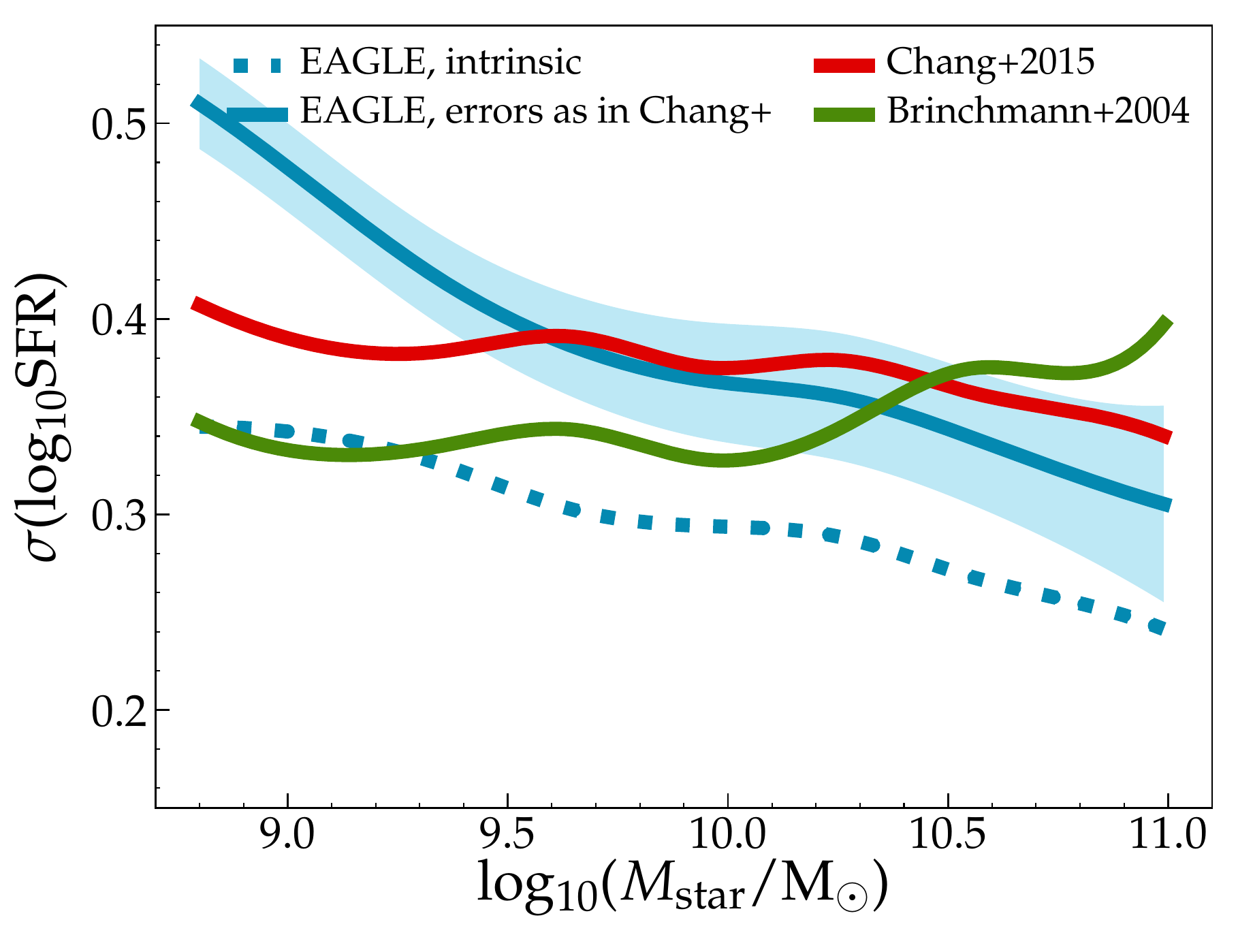}
\caption{\small{The mass dependence of the scatter in the SFR at fixed stellar mass for star-forming galaxies at $z=0$ in EAGLE (blue) and observed at $0.02<z<0.04$ from SDSS (red; \citealt{Chang2015}). For a proper comparison, we apply the \citet{Chang2015} measurement uncertainties to EAGLE. The shaded region indicates the additional uncertainty in EAGLE due to cosmic variance. The dashed blue line shows the intrinsic scatter in EAGLE. The green line shows the scatter measured in the SDSS sample when using H$\alpha$ and D$_n$4000 SFRs (\citealt{Brinchmann2004}). The predicted scatter in SFR at fixed stellar mass decreases slowly with stellar mass, in agreement with the observations.}} 
\label{fig:scatter_SFR_SDSS}
\end{figure}

\subsection{Measuring scatter}
As the goal of this paper is to study the origin of scatter in the SFR-M$_{\rm star}$ relation, we remove quenched galaxies from our analysis. While a significant fraction of quenched galaxies (in particularly quenched satellites) are devoid of star-forming gas (such that these galaxies formally have SFR=0 M$_{\odot}$ yr$^{-1}$), others still have a reduced, but non-zero SFR. In order to select star-forming galaxies that define the main relation between SFR and M$_{\rm star}$, we select galaxies with SFR/M$_{\rm star}\equiv$  sSFR $>10^{-11}$ yr$^{-1}$ at $z=0$. This selection threshold evolves with redshift to account for the overall change in specific SFRs as detailed in \S $\ref{sec:evolution}$.

We measure the scatter from the residuals with respect to the relation determined using the non-parametric local polynomial regression (LPR) method \citep{LPRBOOK}. In this method, for each data point X = [log$_{10}$ M$_{\rm star}$, log$_{10}$ SFR] a fitted value is obtained using a local linear fit with the least squares method, for which only the nearest half in (log$_{10}$ M$_{\rm star}$, log$_{10}$ SFR) distance of the other data points are used and points are weighted by their distance in (log$_{10}$ M$_{\rm star}$, log$_{10}$  SFR) space to the data point X, see \cite{Matthee2017} for full details. The residual is the vertical offset from the fitted value at the respective data point. The benefit of this method is that no assumption on the functional form of the relation is required. However, in practice the results are similar to a simple log$_{10}$ SFR $\propto$ log$_{10}$ M$_{\rm star}$ relation. Therefore, the scatter in the residuals is very similar to the scatter in the specific SFR.


\section{The amount of scatter} \label{sec:mass_scatter}
Before studying the physical origin of the scatter in the SFR-M$_{\rm star}$ relation, we measure the magnitude of the scatter as a function of stellar mass in both the EAGLE simulation and a sample of star-forming galaxies from the Sloan Digital Sky Survey (SDSS, e.g. \citealt{Padmanabhan2008,Alam2015}). For SDSS, we use the measurements from \cite{Chang2015}, who combined the SDSS optical photometry with WISE infrared photometry and simultaneously and self-consistently modelled the dust attenuation and re-emission. As shown in Figure 5 of \cite{SchallerSPH}, there is very good agreement between the normalisation and slope of the main sequence in \cite{Chang2015} and in the EAGLE simulation. As recommended by \cite{Chang2015}, we only include galaxies that have FLAG = 1 (meaning they have reliable aperture corrections). As in EAGLE, we select star-forming galaxies with sSFR $>10^{-11}$ yr$^{-1}$. In order to compare the measurements from \cite{Chang2015} with older SDSS measurements (DR7), we also perform our analysis on the same galaxy sample with SFRs based on a combination of H$\alpha$ and D$_n$4000 measurements (the latter being mostly important at M$_{\rm star}>10^{10}$ M$_{\odot}$) from \cite{Brinchmann2004}. These are based on the flux measured with fibre-spectroscopy and data-driven corrections to take emission at larger radii into account based on measured broad-band colours.

\subsection{Mass dependence in EAGLE and SDSS} \label{sec:eagle_sdss}
For EAGLE, we split the sample of galaxies at $z=0$ in stellar mass bins of 0.3 dex from log$_{10}$(M$_{\rm star}$/M$_{\odot}) = 8.8$ to $11.2$ and measure the 1$\sigma$ standard deviation, $\sigma$(log$_{10}$ SFR), from the residual values in each bin. The lower limit of the mass-range that we study is based on the resolution limit of EAGLE (such that galaxies are resolved with $>500$ star particles; but note that analyses that do include lower-mass galaxies find that the EAGLE simulation is in agreement with the faint-end of the stellar mass and SFR functions; e.g. \citealt{Furlong2015,Katsianis2017}). The upper limits are set by the simulation volume (10$^6$ cMpc$^3$), such that each bin includes at least 50 galaxies. Next, we shift bin edges by $\pm0.1$ dex, repeat the measurement, and interpolate the measured standard deviations in the bins to show how the 1$\sigma$ scatter varies with stellar mass. Choosing a smaller bin-width increases the uncertainty, while a larger bin-width smooths the trends.

For a proper comparison with the observations, we take the additional scatter due to measurement errors into account as follows. We assume that the uncertainties in the masses and SFRs in EAGLE depend on both mass and SFR similarly as in the observational sample from \cite{Chang2015}. For each EAGLE galaxy, we select the 100 galaxies in the Chang et al. sample (within the redshift range $0.02<z<0.04$ as described below) that are closest in log$_{10}$ M$_{\rm star}$, log$_{10}$ SFR - space, and compute the median uncertainty in the log$_{10}$ M$_{\rm star}$ and log$_{10}$ SFR of those galaxies. This results in uncertainties of $0.11^{+0.01}_{-0.01}$ dex and $0.21^{+0.26}_{-0.09}$ dex, in mass and SFR respectively, with the highest uncertainties found in galaxies with low mass and low SFR. Then, we perturb the mass and SFR of each galaxy in EAGLE 1000 times within the typical observational error (assuming the errors in log$_{10}$ SFR are gaussian), re-select galaxies in each bin, and re-measure the scatter. This results in 1000 realisations of the mass-dependence of the scatter, from which we obtain the median and the 1$\sigma$ confidence intervals. We account for cosmic variance due to the limited simulation volume by measuring the scatter in jackknife realisations of the galaxies in eight sub-boxes of (50 cMpc)$^3$. The intrinsic scatter in EAGLE is $\approx0.30$ dex, while the mimicked `observed' scatter decreases from $\approx0.5$ dex at M$_{\rm star}\approx10^9$ M$_{\odot}$ to $\approx0.3$ dex at M$_{\rm star}\approx10^{11}$ M$_{\odot}$.

Several observational biases complicate the analysis of the SDSS sample. First, due to their limited sensitivity, observations suffer from incompleteness, particularly at low stellar masses. As the sensitivity limit is in observed flux, it is related to a complex combination of SFR, mass, dust attenuation and stellar age. It is therefore likely that incompleteness biases the measured scatter \citep[e.g.][]{Reddy2012}. Second, observations are not performed within an infinitesimally small redshift slice. As the typical SFR of galaxies (at all masses) increases with redshift \citep[e.g.][]{MadauDickinson2014}, this means that sources with a higher redshift will have a higher SFR at fixed stellar mass (increasing the observed scatter). Due to the shape of the galaxy luminosity function, the redshift-comoving volume relation and the redshift dependence of mass-incompleteness, the sample is dominated by slightly higher-redshift galaxies at the high-mass end, and by slightly lower-redshift galaxies at the low-mass end \citep[see also][]{Gunawardhana2013,Boogaard2018}. Such biases can be overcome by restricting the sample to a narrow redshift slice, at the lowest redshift possible. 

As motivated in Appendix $\ref{appendix_SDSS}$, we select the subset of SDSS galaxies with redshifts between $z=0.02$ and $z=0.04$,  which is complete at stellar masses $\gtrsim10^{8.8}$ M$_{\odot}$. Combined with the survey footprint, this corresponds to a comoving volume of $\approx4\times10^6$ Mpc$^3$. As a result, the SDSS sample consists of 33,769 galaxies. Using the same method as used for EAGLE, we find an observed scatter that decreases from $\approx 0.40$ dex at M$_{\rm star} =10^9$ M$_{\odot}$ to $\approx 0.36$ dex at M$_{\rm star} >3\times10^{10}$ M$_{\odot}$. This is similar to the results from \cite{Chang2015}, who measure a scatter of 0.39 dex. 

As shown in Fig. $\ref{fig:scatter_SFR_SDSS}$, we find that the scatter in the SFR-M$_{\rm star}$ relation decreases slightly with increasing stellar mass, both in EAGLE and in the SDSS analysis by \cite{Chang2015}. The slope of the observed trend is consistent with the predictions when we add observational errors to the EAGLE measurements at M$_{\rm star}>10^{9.5}$ M$_{\odot}$, but shallower for lower masses. This highlights the importance of accurate understanding of the mass and SFR dependence of the uncertainties when inferring the intrinsic scatter from observations. 

Remarkably, the scatter is smaller when SFR measurements from \cite{Brinchmann2004} are used, particularly at M$_{\rm star}<10^{10}$. This is counter-intuitive as SFR measurements at these masses come mostly from H$\alpha$ emission-line luminosities, which are sensitive to shorter time-scales than SFRs measured by \cite{Chang2015}. One potential reason is that \cite{Chang2015} under-estimate the SFRs in galaxies with relatively low SFRs (see \citealt{Salim2016} for a detailed explanation). Indeed, we find that the distribution of sSFRs at fixed mass is skewed towards asymmetric values more in the \cite{Chang2015} measurements than in the measurements from \cite{Brinchmann2004}. The scatter in SFRs from \cite{Brinchmann2004} increases towards high masses, with a rapid increase coinciding with the mass where SFRs are mostly measured using D$_n$4000 measurements. Therefore, this additional scatter can be explained by a larger uncertainty in the SFR, potentially enhanced by more uncertain fibre-corrections at the high-mass end.

We have tested the robustness of these results by changing the stellar mass binning, splitting the SDSS sample in four sub-regions on the sky, changing the upper redshift limit in SDSS (ranging from $z=0.03-0.05$), measuring the scatter in log$_{10}$ sSFR instead of using the residuals, but find that the results are unchanged. Finally, we note that the intrinsic scatter in the SFR-M$_{\rm star}$ relation in EAGLE is slightly larger than the scatter in the Illustris simulation measured by \cite{Sparre2015}, who finds a mass-independent scatter of $\approx 0.27$ dex, and in semi-analytical models, such as the 0.2 dex in e.g. \cite{Dutton2010,Mitra2017}.

\begin{figure}
\includegraphics[width=8.5cm]{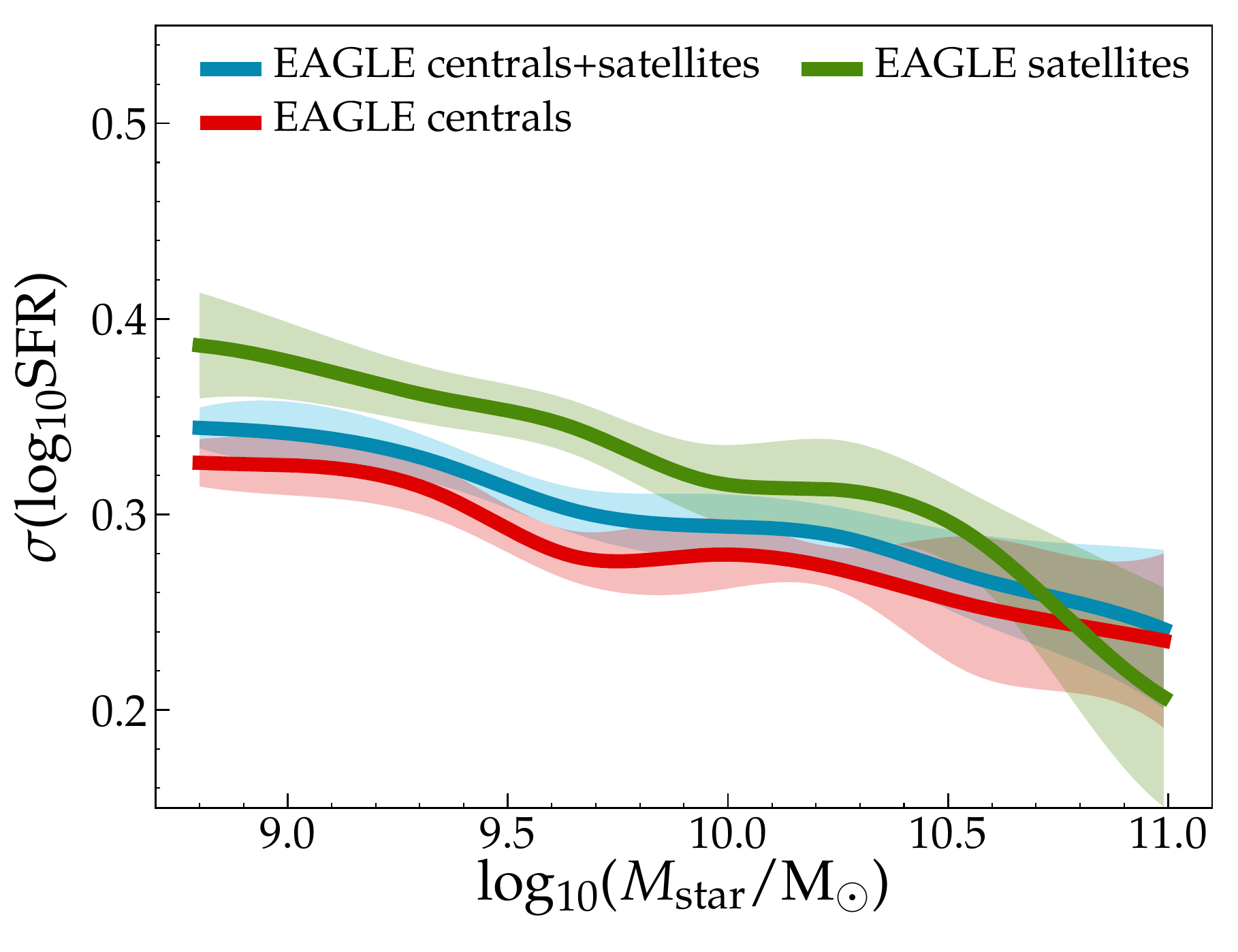}
\caption{\small{Scatter in SFR at fixed stellar mass of star-forming galaxies galaxies in EAGLE at $z=0$. The blue line shows the analysis including all galaxies, while the red (green) line includes only central (satellite) galaxies. Star-forming satellites add only $\approx0.04$ dex of scatter to the SFR-M$_{\rm star}$ relation. This means that when becoming a satellite the SFR is either only mildly affected or so significantly that the satellite is removed from the star-forming galaxy sample.}} 
\label{fig:scatter_SFR_satellites}
\end{figure}

\subsection{Scatter due to star-forming satellites} \label{sec:satellites}
Galaxies that become satellites in a larger halo may experience several physical processes that influence their SFR, such as ram pressure stripping of gas, tidally induced star formation, harassment and/or strangulation \citep[e.g.][]{vandenBosch2008,Peng2012,Bluck2016}. Thus, satellite interactions may both (temporally) increase or decrease their SFRs. We investigate the contribution of satellites to the scatter in the SFR-M$_{\rm star}$ relation by removing satellite galaxies from the EAGLE analysis. We show in Fig. $\ref{fig:scatter_SFR_satellites}$ that the scatter decreases by only $\approx 0.04$ dex up to masses of M$_{\rm star} \approx 10^{10.5}$ M$_{\odot}$ (at higher masses there are only a handful of star-forming satellite galaxies due to the limited simulation volume). Recall that this scatter is measured for star-forming galaxies only. In EAGLE, the fraction of satellites that are quenched (i.e. sSFR $< 10^{-11}$ M$_{\odot}$ yr$^{-1}$) decreases from $\approx50$ \% at M$_{\rm star}= 10^9$ M$_{\odot}$ to $\approx35$ \% at M$_{\rm star} =10^{10}$ M$_{\odot}$ and increases to $\approx 50$ \% at M$_{\rm star} = 10^{11}$ M$_{\odot}$. Therefore, a significant fraction of satellites has not been included in this analysis. Our results indicate that satellite-specific processes are either weak, or strong and rapid (such that the satellites are not selected as star-forming galaxies anymore). We note that the distribution of sSFRs for satellite galaxies is relatively symmetric at sSFR $> 10^{-11}$ yr$^{-1}$, indicating that satellite-specific processes may also enhance SFR and that the larger scatter compared to central galaxies does not originate from a distribution that is skewed to lower values.\footnote{It is remarkable that, at the highest stellar masses, $\sigma(\Delta$log$_{10}$SFR) is smaller in both subsets of centrals and satellites than in the combined sample of centrals and satellites. The reason is that the median of the distribution of $\Delta$log$_{10}$ SFR of satellites at these masses peaks slightly below zero, while the distribution for central galaxies is centred around zero. The combination of the two distributions results in a larger scatter than each distribution individually.  }

\begin{figure}
\includegraphics[width=8.6cm]{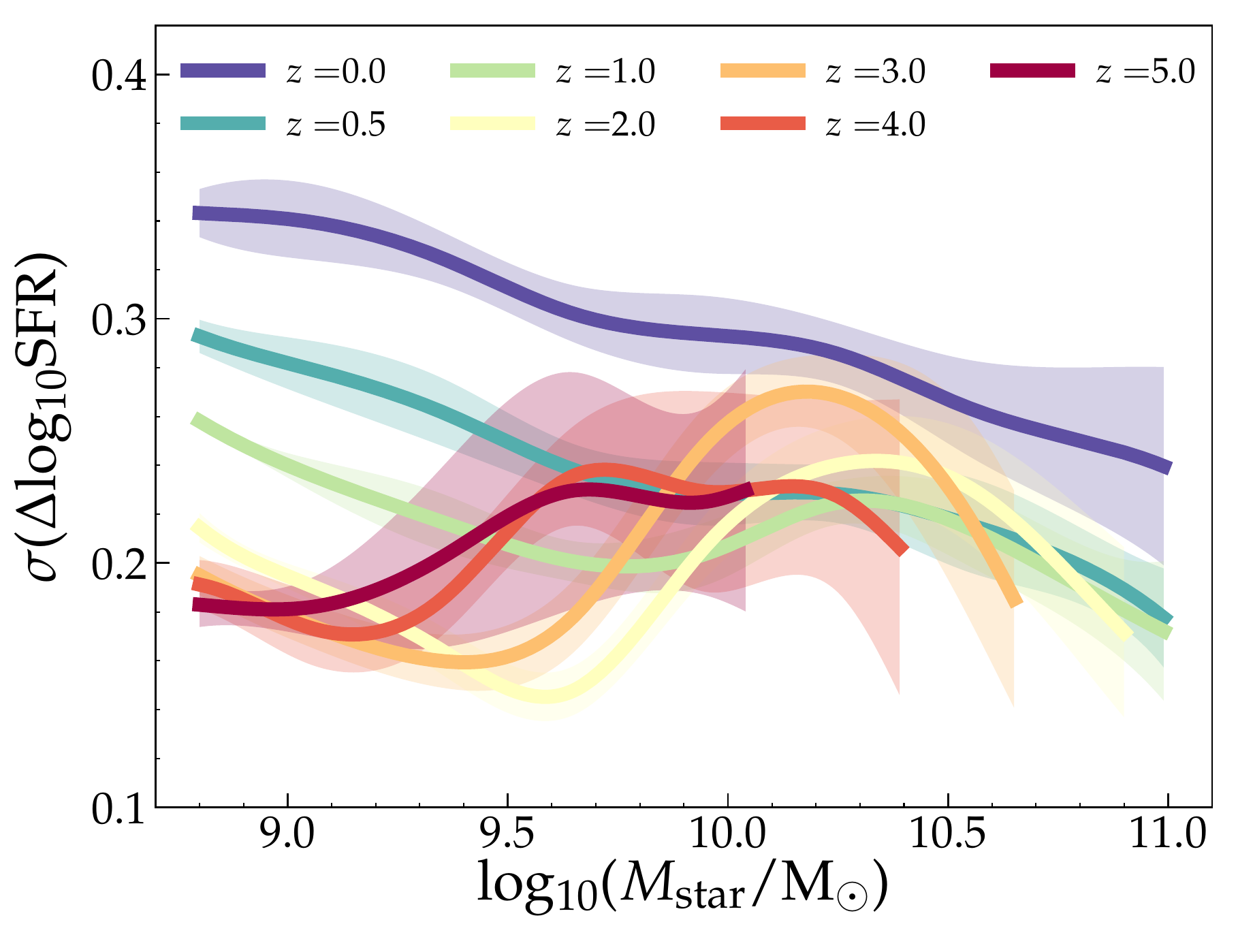}\\ 
\caption{\small{Evolution of the mass dependence of scatter in the SFR-M$_{\rm star}$ relation. Star-forming galaxies are selected with an evolving sSFR threshold as described in the text and we only show mass-ranges where there are $>50$ galaxies per bin of 0.3 dex width. The scatter decreases with increasing redshift, particularly at lower stellar masses. At high masses the decrease is less prominent and within the estimated uncertainty due to cosmic variance.}}
\label{fig:scatter_SFR_evolution}
\end{figure}

\subsection{Evolution} \label{sec:evolution}
We measure the evolution of the magnitude and mass dependence of the scatter in the SFR-M$_{\rm star}$ relation in EAGLE in redshift slices with $z=0-5$. As the typical SFR of galaxies increases at $z>0$ \citep[e.g.][]{Whitaker2012,Sobral2013,MadauDickinson2014},  we now change the threshold sSFR value that determines whether a galaxy is star-forming with redshift. At each redshift slice, we compute the median sSFR of all galaxies with M$_{\rm star}>10^{9}$ M$_{\odot}$ and set the threshold to log$_{10}$(sSFR$_{\rm median}$)-0.65. At $z=0.0$, this threshold corresponds to  sSFR $> 10^{-11}$ yr$^{-1}$, and it increases to sSFR $> 10^{-10.4,-10.0,-9.6,-9.4,-9.1,9.0}$ yr$^{-1}$ at $z=0.5, 1, 2, 3, 4, 5$, respectively. We include both centrals and satellites. 

As shown in Fig. $\ref{fig:scatter_SFR_evolution}$, we find that the scatter decreases from $z=0$ to $z=5$ by 0.05 dex at M$_{\rm star} \gtrsim 10^{10}$ M$_{\odot}$. This evolution occurs mostly between $z=0$ and $z=1$, although we note that the amount of evolution is of the same order as the uncertainties associated with cosmic variance. These results are roughly consistent with the observations from e.g. \cite{Whitaker2012}, \cite{Shivaei2015}, \cite{Schreiber2015}, which typically are mass-complete (especially at $z>2$) at M$_{\rm star} > 10^{10}$ M$_{\odot}$ and who find a relatively constant scatter. On the other hand,  \cite{Kurczynski2016,Santini2017} report observational indications that the scatter increases with cosmic time, while \cite{Pearson2018} finds that the scatter decreases strongly at $z>2$. 

At lower stellar masses, the scatter in the SFR-M$_{\rm star}$ relation in EAGLE decreases more strongly with increasing redshift, for example, it decreases from $\approx 0.35$ dex at $z=0$ to $\approx 0.25 (0.20)$ dex at $z\approx 1(>2)$ for  M$_{\rm star} \approx10^9$ M$_{\odot}$. These results differ from those of \cite{Mitra2017}, whose analytic model predicts that the scatter at these low masses increases with redshift under the assumption that fluctuations in the SFR follow fluctuations in halo accretion rates. The difference with \cite{Mitra2017} is mostly at low redshift, where they predict a scatter of $\approx0.20$ dex at M$_{\rm star} = 10^{9-9.5}$ M$_{\odot}$ at $0.5<z<1.0$, significantly lower than EAGLE. At high-redshift the results are similar.
We note that the scatter in the linear SFR at fixed mass, as opposed to the scatter in log$_{10}$ SFR considered so far, increases towards higher redshift at low masses. 

As we discuss and show in more detail in \S $\ref{sec:BH}$, the increasing scatter at M$_{\rm star} \approx10^{9.8}$ M$_{\odot}$ at high redshift is due to supermassive black hole growth impacting the SFHs, even after applying a relatively high sSFR threshold to remove quenched galaxies. The stellar mass above which the scatter rapidly increases (most clearly seen at $z>1$; and similar to observational indications in \citealt{Guo2013}) decreases slightly with increasing redshift. This indicates that the stellar mass above which AGN feedback becomes important is slightly lower at higher redshift.

\begin{figure}
\includegraphics[width=8.4cm]{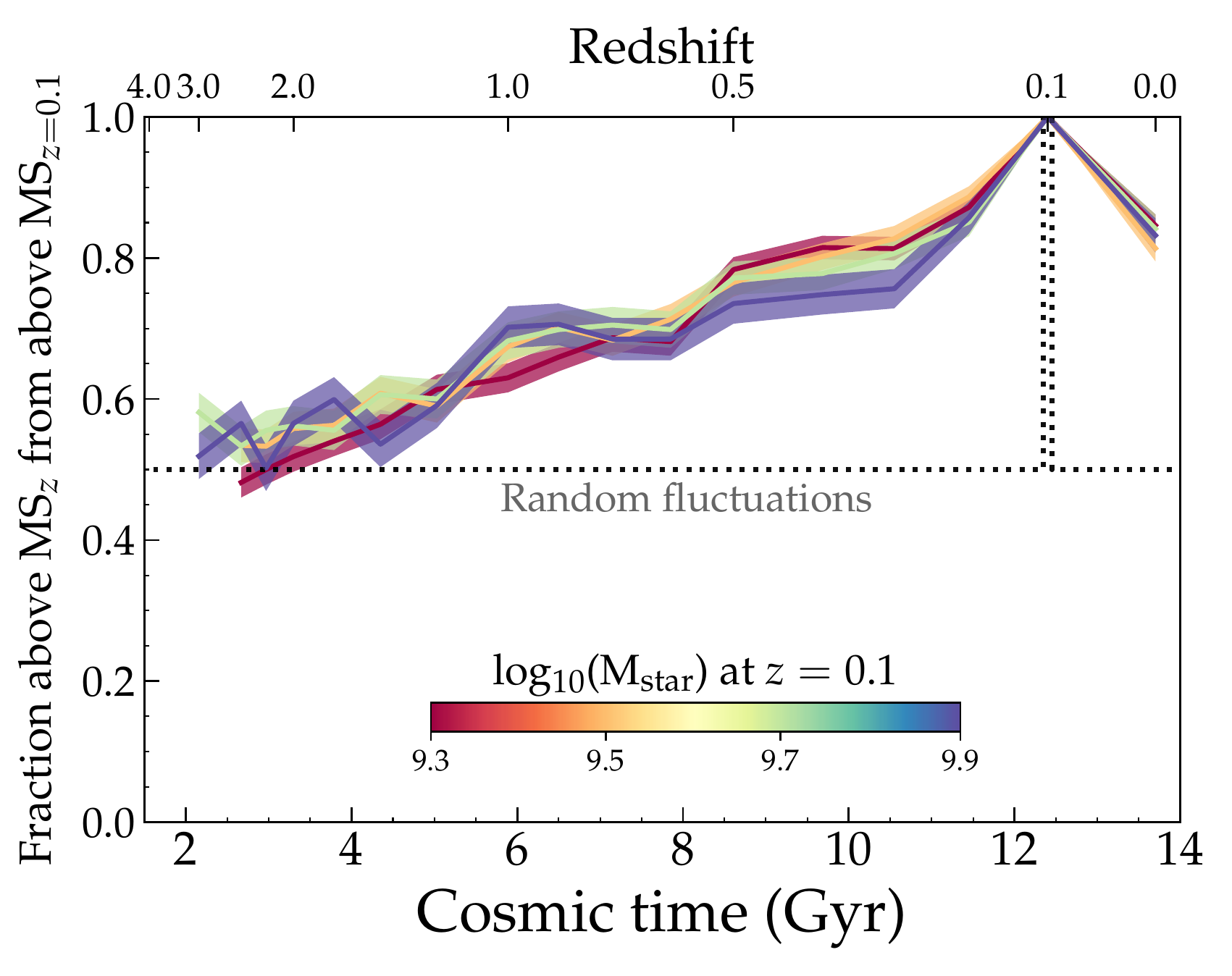} 
\includegraphics[width=8.4cm]{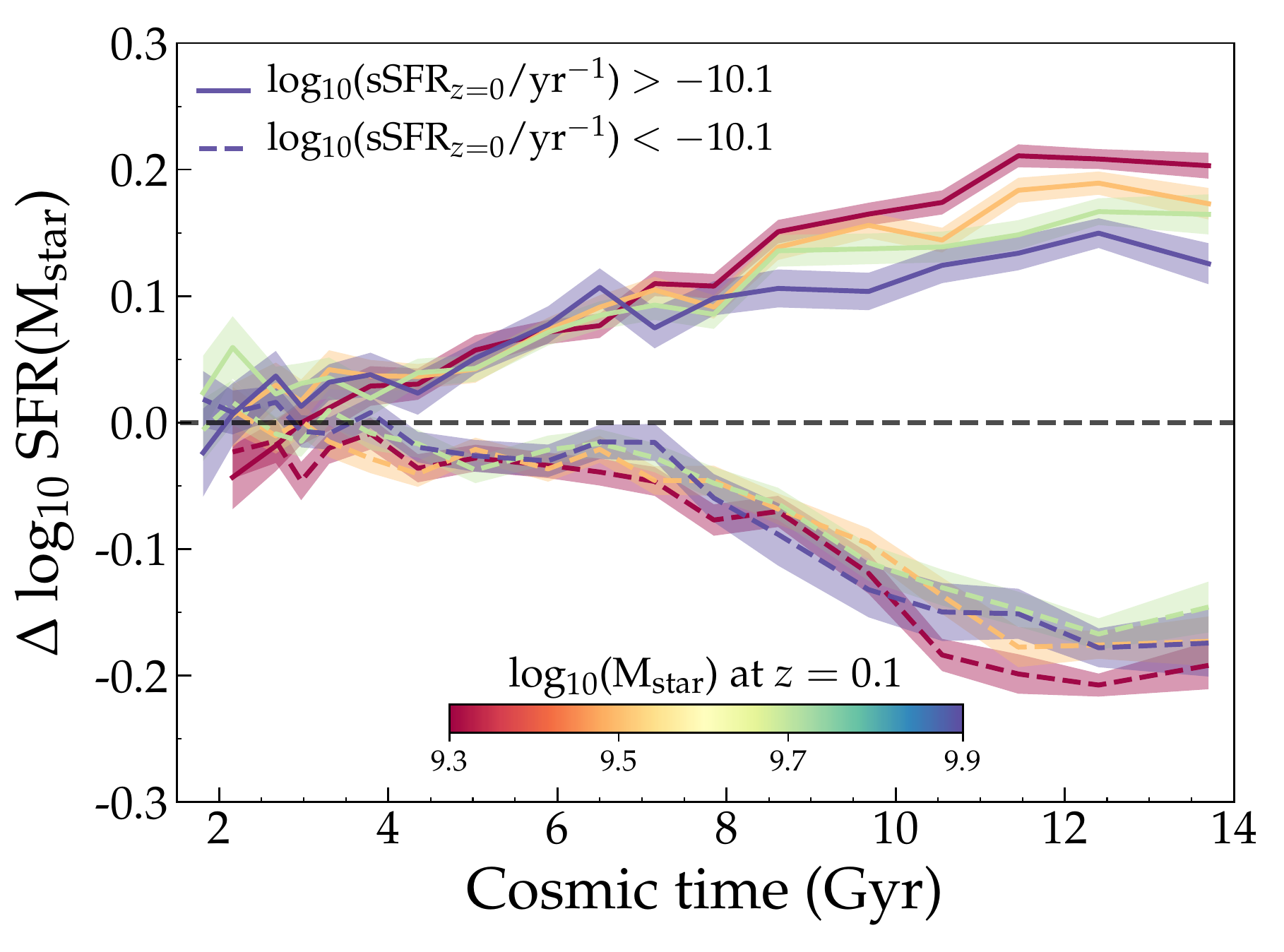} 

\caption{\small{{\it Top}: The fraction of galaxies that are above the main sequence at a specific snapshot among the sample of central, star-forming galaxies that are above the main sequence at $z=0.1$, in bins of $z=0.1$ stellar mass. By definition, this fraction is 1.0 at $z=0.1$. The shaded region shows the binomial uncertainty on the fraction. For each mass bin, we only show the results if the stellar mass of the median main progenitor in each bin is resolved with at least 100 star particles (M$_{\rm star}>10^8$ M$_{\odot}$). The horizontal dotted line indicates the expectation if SFRs have no `memory' (i.e. the SFR of a galaxy at a certain redshift is drawn randomly from a normal distribution around the main sequence at the specific redshift). It is clear that there is a correlation between the SFRs of galaxies between different points in cosmic time. {\it Bottom:} Median difference to the SFR-M$_{\rm star}$ relation at each redshift for central star-forming galaxies binned by stellar mass and split into `above' (solid) and `below' (dashed) the main sequence at $z=0.1$. The shaded region shows the uncertainty on the median. Relatively independently of stellar mass, the {\it median} galaxy that is above/below the main sequence has been above/below the main sequence for the prior $\approx 10$ Gyr. Note that the medians wash out short time-scale fluctuations that are also present (see Fig. $\ref{fig:path_individual}$).  }}
\label{fig:memory}
\end{figure}

\section{How long galaxies remain above/below the main sequence} \label{burstiness}
In this section, we investigate on what time-scales central, star-forming galaxies change their position relative to the evolving SFR-M$_{\rm star}$ relation. In particular, we measure the SFRs and stellar masses of the progenitors of galaxies that are star-forming centrals at $z=0$ at multiple redshifts up to $z=4$ and compare how the locations of galaxies in the SFR-M$_{\rm star}$ plane change relative to the main sequence at each specific redshift slice. We first focus on the {\it typical} (median) fluctuations depending on galaxies' present-day stellar mass and sSFR and then investigate the role of more stochastic fluctuations in individual galaxies.

\subsection{Median fluctuations}
The first results are shown in Fig. $\ref{fig:memory}$. In the top panel, we select the sub-set of galaxies that are above the main sequence at $z=0.1$ (in bins of stellar mass), and measure the fraction of these galaxies that are also above the main sequence at other points in cosmic time. If the SFRs of galaxies had `no memory' (i.e. if at each point in cosmic time their SFR were drawn randomly from the distribution of sSFRs around the main sequence at its specific mass at that time), one would expect this fraction to be 50 \% at $z\neq 0.1$. However, we clearly find that the majority of galaxies that lie above the main sequence at $z=0.1$ are also above the main sequence at earlier times (up to $z\approx2$, or $\approx10$ Gyr before $z=0.1$). This means that {\it in the median}, a galaxy's SFR {\it remembers} its past SFR. 

We note that the SFRs of individual galaxies do fluctuate between `above' and `below' the main sequence on shorter time-scales (see below), but that their population average displays a clear degree of long time-scale coherence depending on their location relative to the SFR-M$_{\rm star}$ relation (see also \citealt{Abramson2016}). That long time-scale fluctuations are present is also illustrated in the bottom panel of Fig. $\ref{fig:memory}$, which shows that the {\it median} galaxy above the main sequence has been above the main sequence for a long time, $\approx10$ Gyr. Similarly, the median galaxy below the main sequence (but with a sSFR$>10^{-11}$ yr$^{-1}$ at $z=0.1$) has tended to be below the main sequence for most of the history of the universe.

These results show that `oscillations' of galaxies through the `main sequence' contain an oscillation-mode with a period similar to the Hubble time. This oscillation time-scale is longer than for galaxies in the hydrodynamical simulations presented in \cite{Tacchella2016}, who find a typical time-scale of $\approx0.4\times t_{\rm Hubble}$, where $ t_{\rm Hubble} \approx$ 14 Gyr at $z=0$. We investigate these oscillations in more detail in \S $\ref{sec:relative_timescales}$.

\begin{figure}
\includegraphics[width=8.6cm]{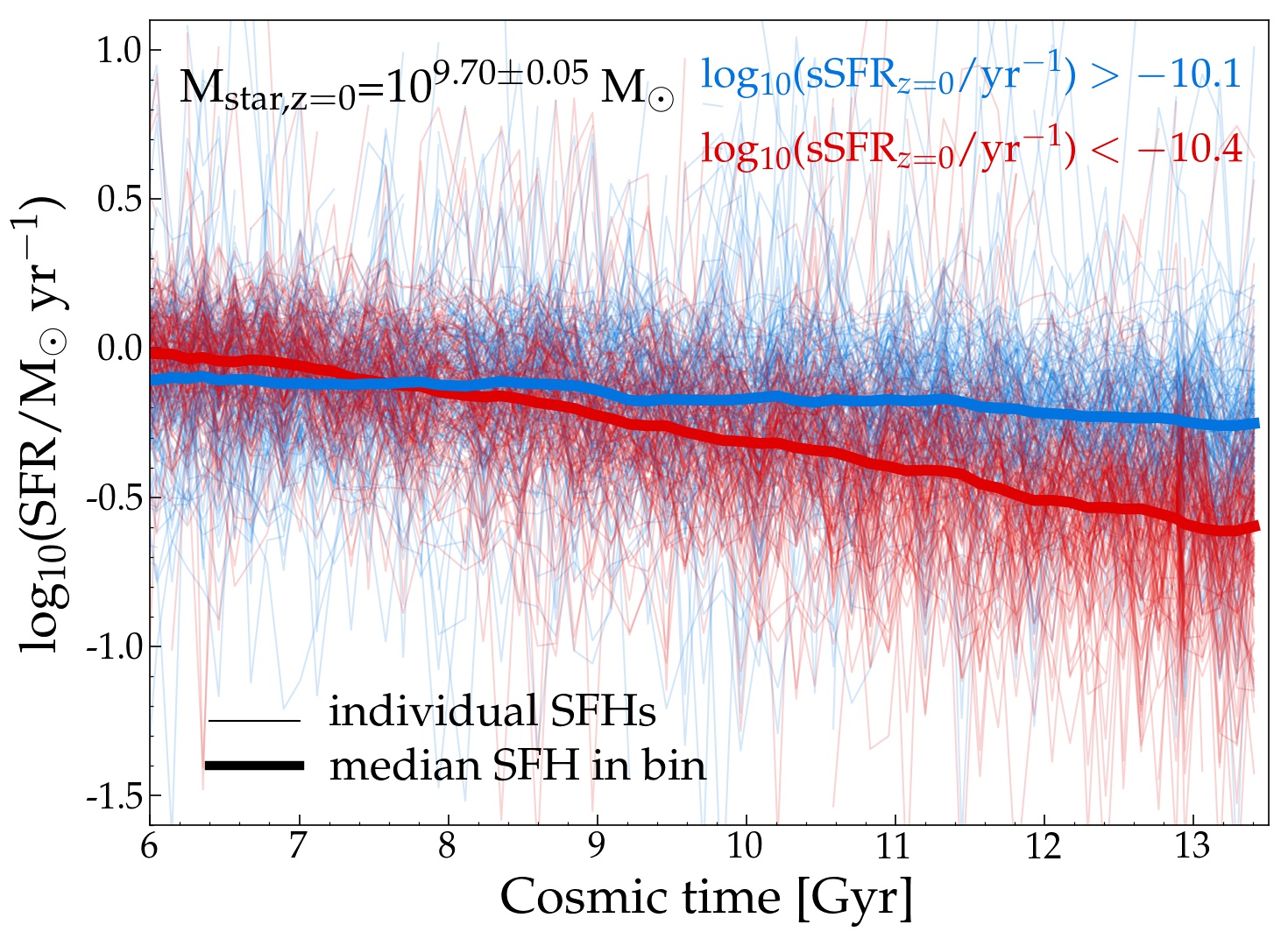} 
\caption{\small{Star formation histories of galaxies with M$_{\rm star}=10^{9.70\pm0.05}$ M$_{\odot}$ at $z=0$, split by their current SFR. Thin lines show paths of individual galaxies, while thick lines show the median in each bin of present-day sSFR. This figure highlights that median SFHs (like those shown in Fig. $\ref{fig:memory}$) smooth out short time-scale fluctuations of typically $\approx0.3$ dex. The median paths do recover the long time-scale variations that depend on the current SFR.}} \label{fig:path_individual}
\end{figure}

\subsection{The variety in SFHs of individual galaxies} \label{sec:variety}
A major limitation of our median\footnote{We note that we find qualitatively similar results when using {\it mean} stacking.} stacking method (e.g. Fig. $\ref{fig:memory}$) is that it may wash out short-time scale fluctuations in SFRs, such as those observed in high-resolution zoom-simulations \citep[e.g.][]{Hopkins2014,Muratov2015} and observations \citep[e.g.][]{Guo2016}. In particular, if the fluctuations for different galaxies in the same bin are not in phase, they will on average cancel (for an example of a similar effect occuring in stacked radial SF-profiles, see \citealt{Orr2017}). This would for example be the case if galaxies were to fluctuate around relatively parallel tracks in SFR-M$_{\rm star}$ space. 

Indeed, we find that the SFRs of individual galaxies fluctuate significantly on short time-scales, with variations in log$_{10}$SFR of $\approx0.2-0.3$ dex over 100 Myr, as illustrated in Fig. $\ref{fig:path_individual}$. In this figure, we show the SFRs of individual galaxies with M$_{\rm star, z=0} = 10^{9.70\pm0.05}$ M$_{\odot}$ measured in the simulation {\it snipshots} (which have a spacing of $\sim 100$ Myr from cosmic time 5.4-13.5 Gyr), coloured by their present-day sSFR. It is clear that the median SFH in bins of sSFR are insensitive to fluctuations on short time-scales, but that systematic differences on long time-scales can be identified with median SFHs.

\begin{figure*}
\begin{tabular}{ccc}
\includegraphics[width=6cm]{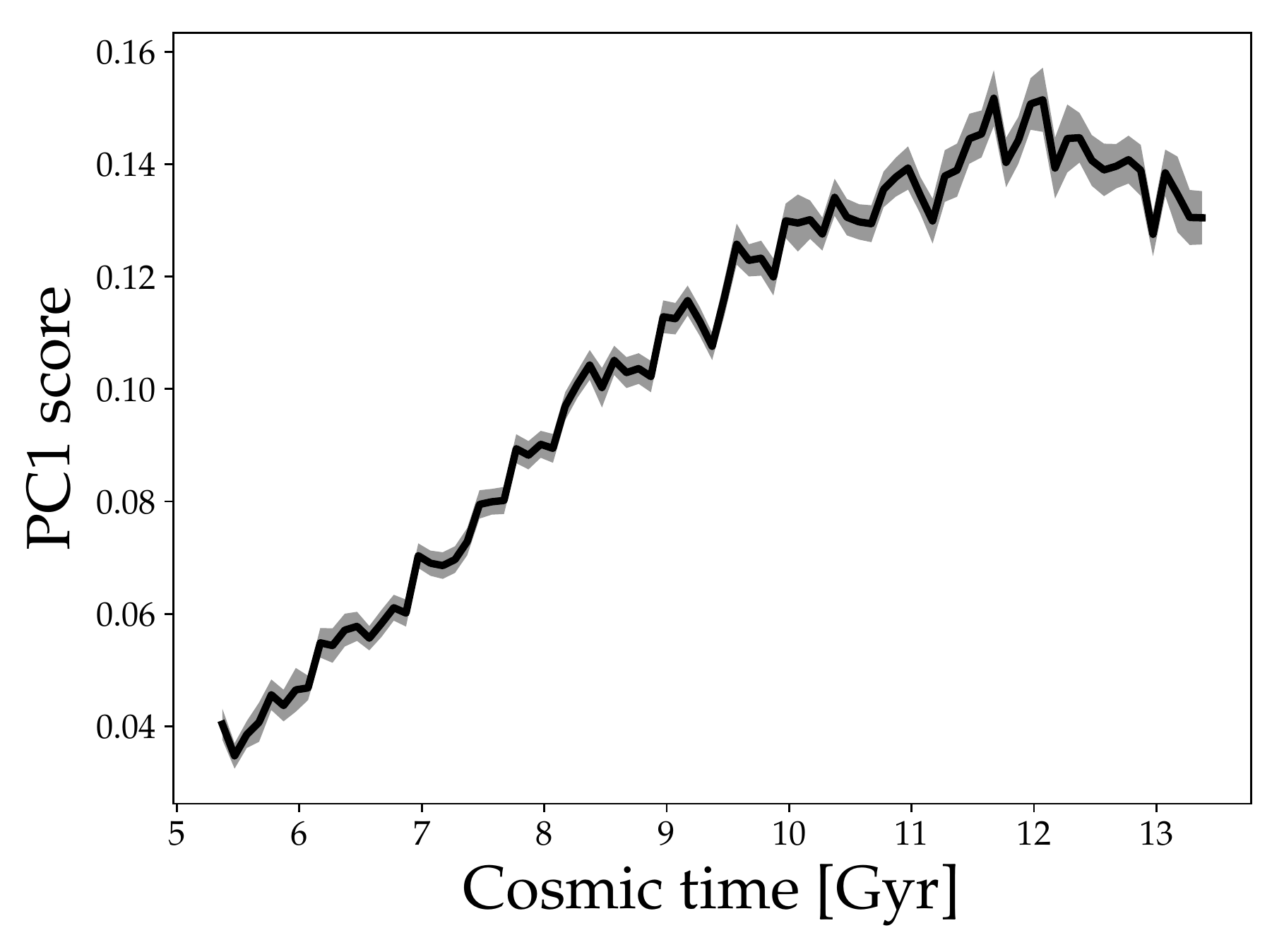} &
\hspace{-0.55cm}\includegraphics[width=6cm]{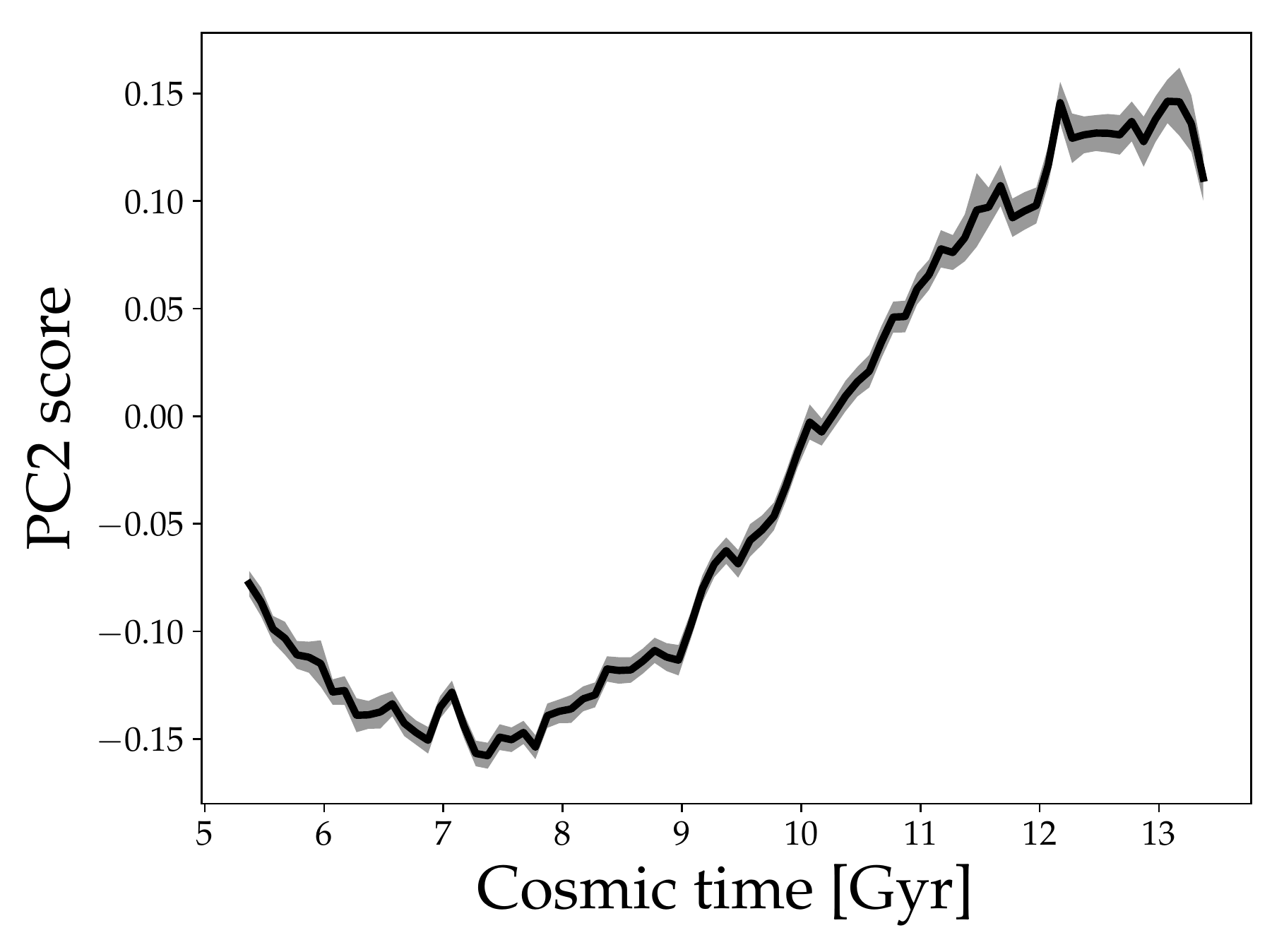} &
\hspace{-0.55cm}\includegraphics[width=6cm]{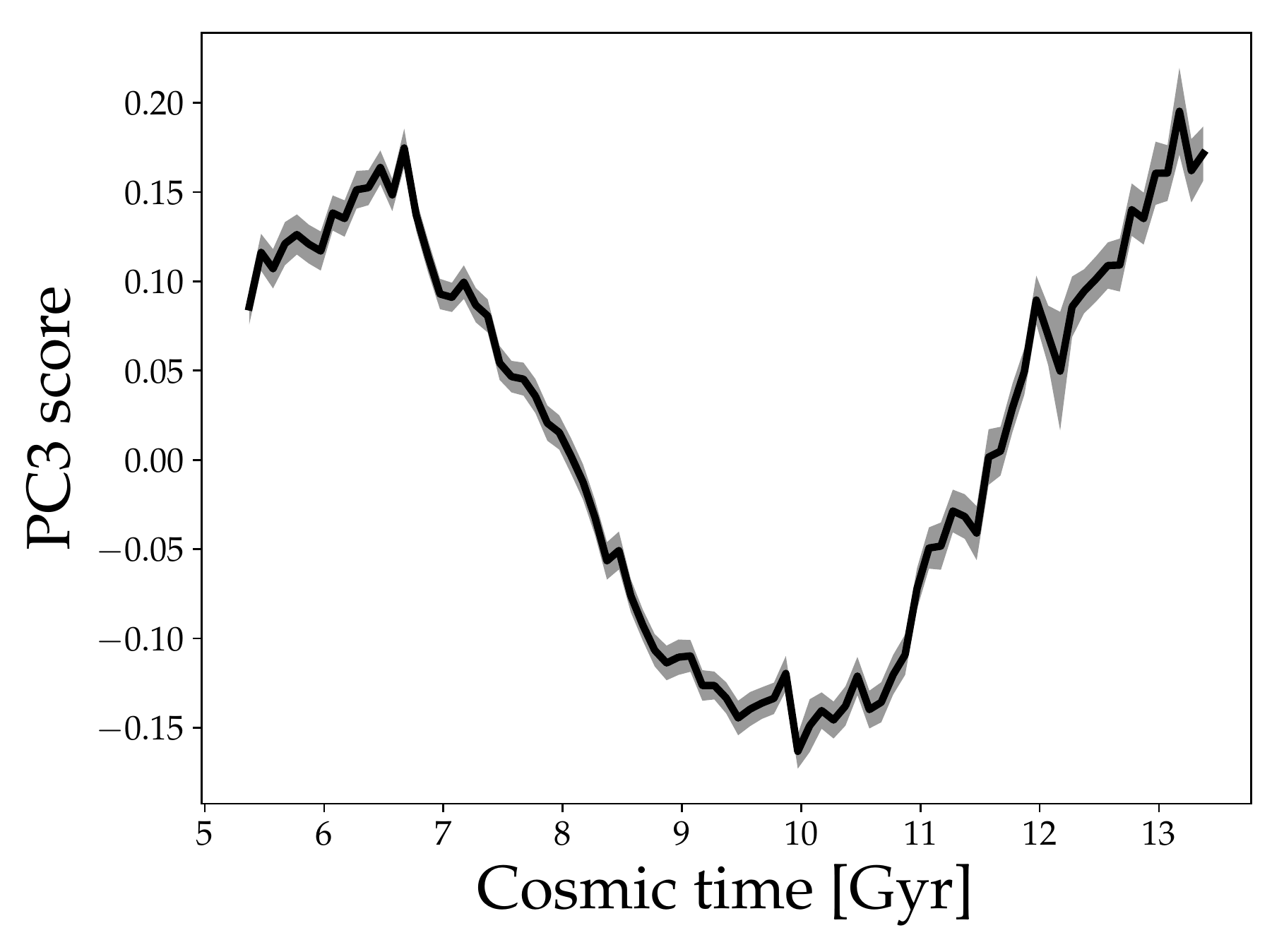} \\
\includegraphics[width=6cm]{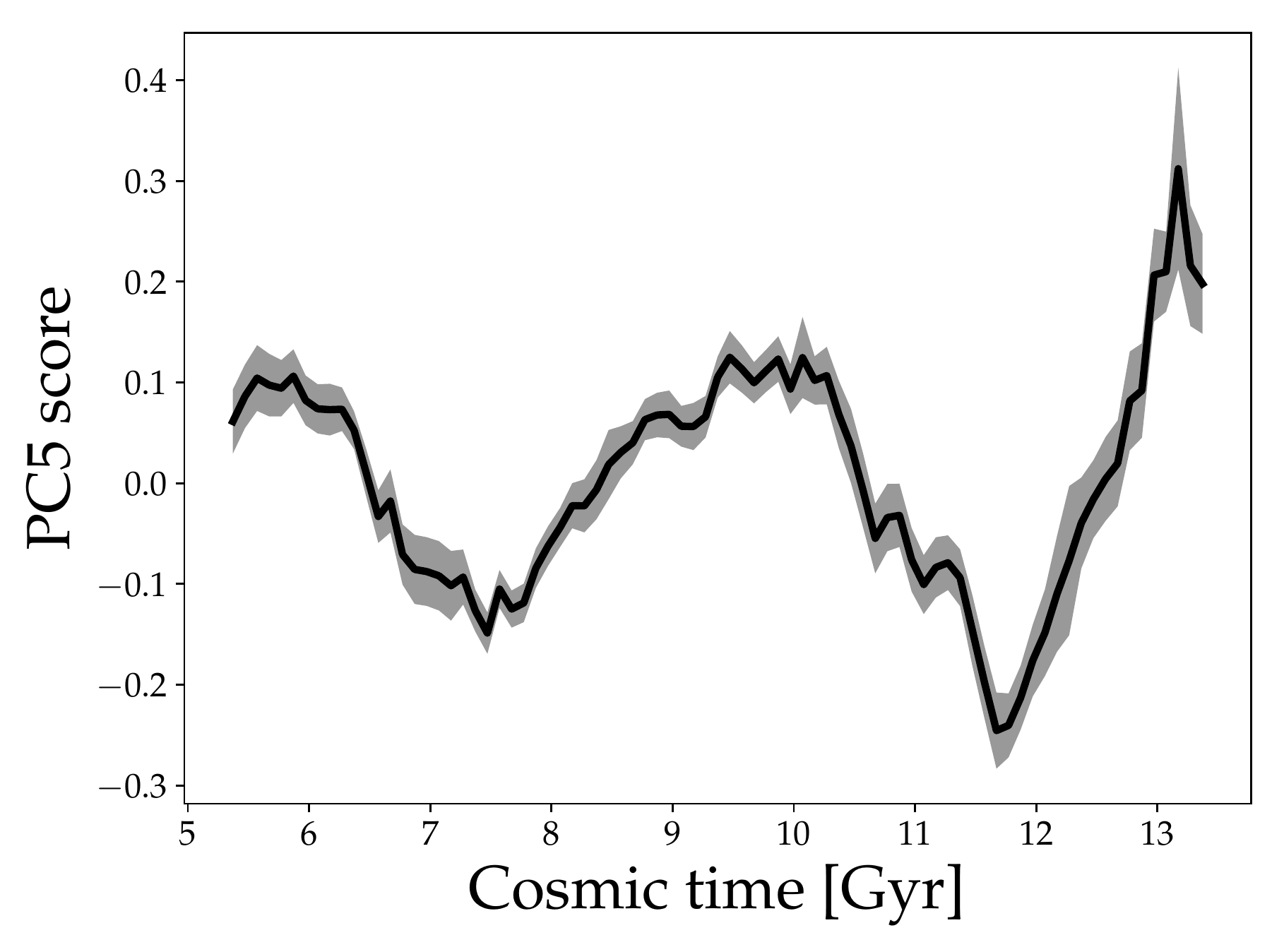} &
\hspace{-0.55cm}\includegraphics[width=6cm]{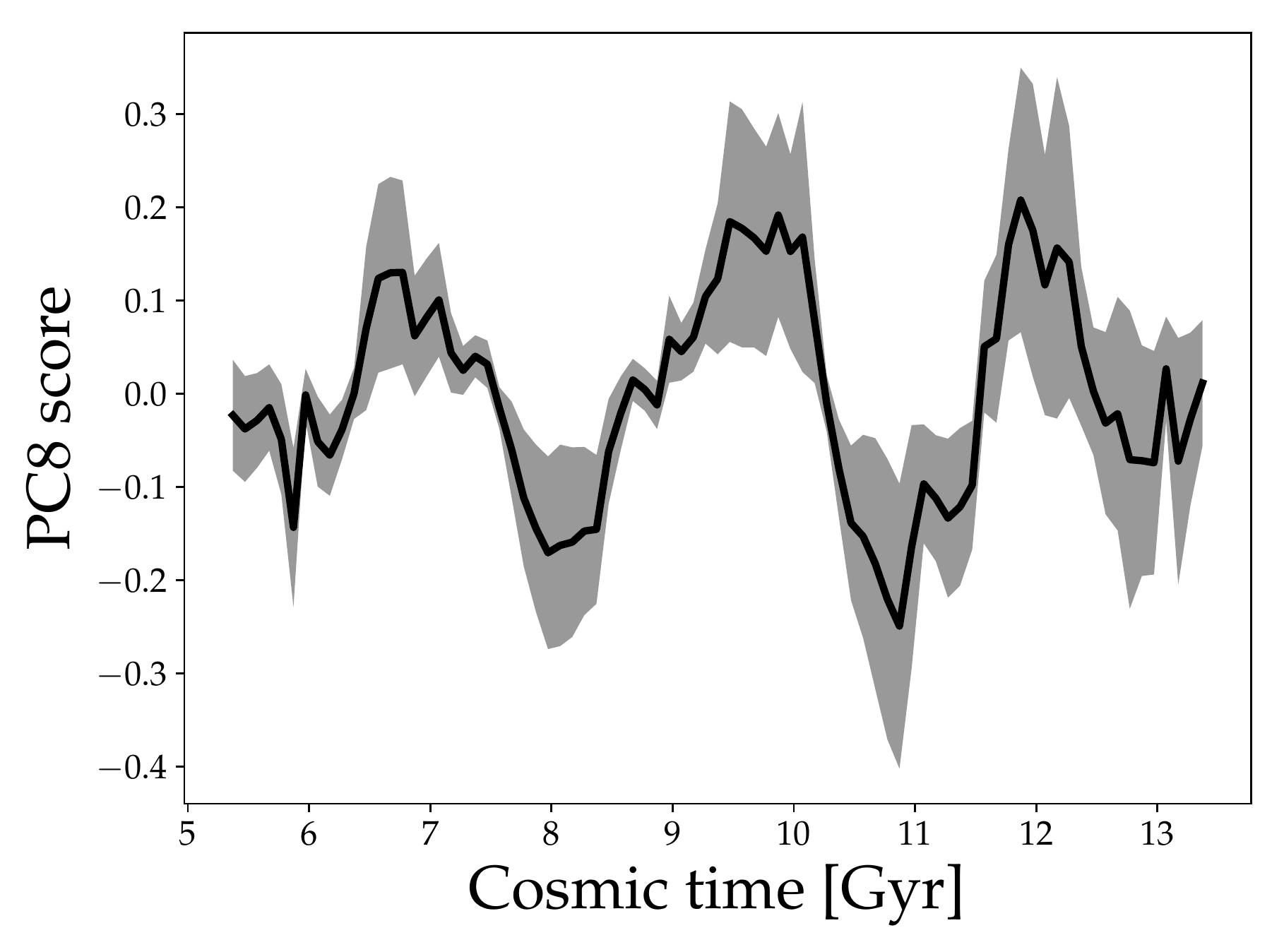} &
\hspace{-0.55cm}\includegraphics[width=6cm]{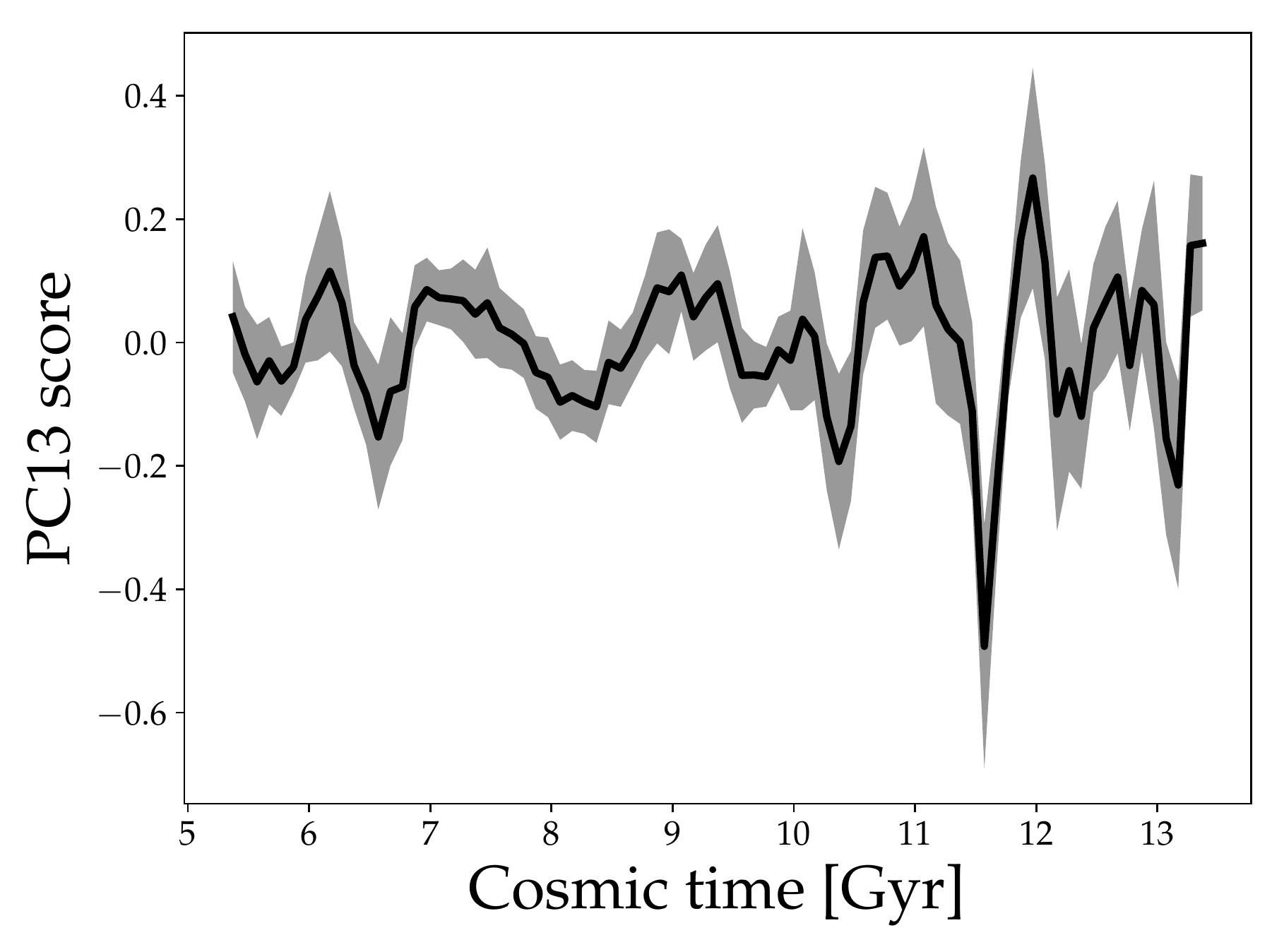} \\
\includegraphics[width=6cm]{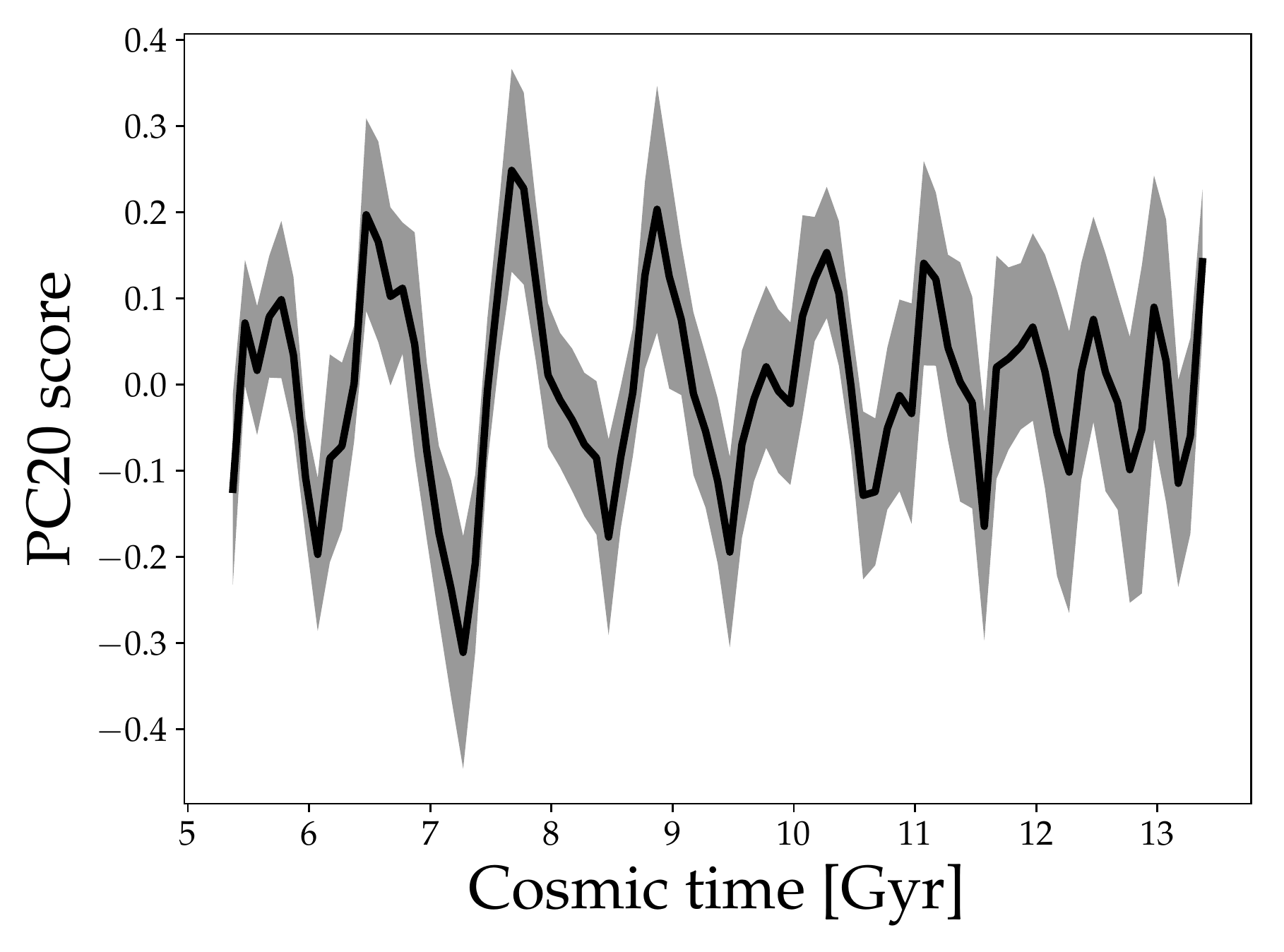} &
\hspace{-0.55cm}\includegraphics[width=6cm]{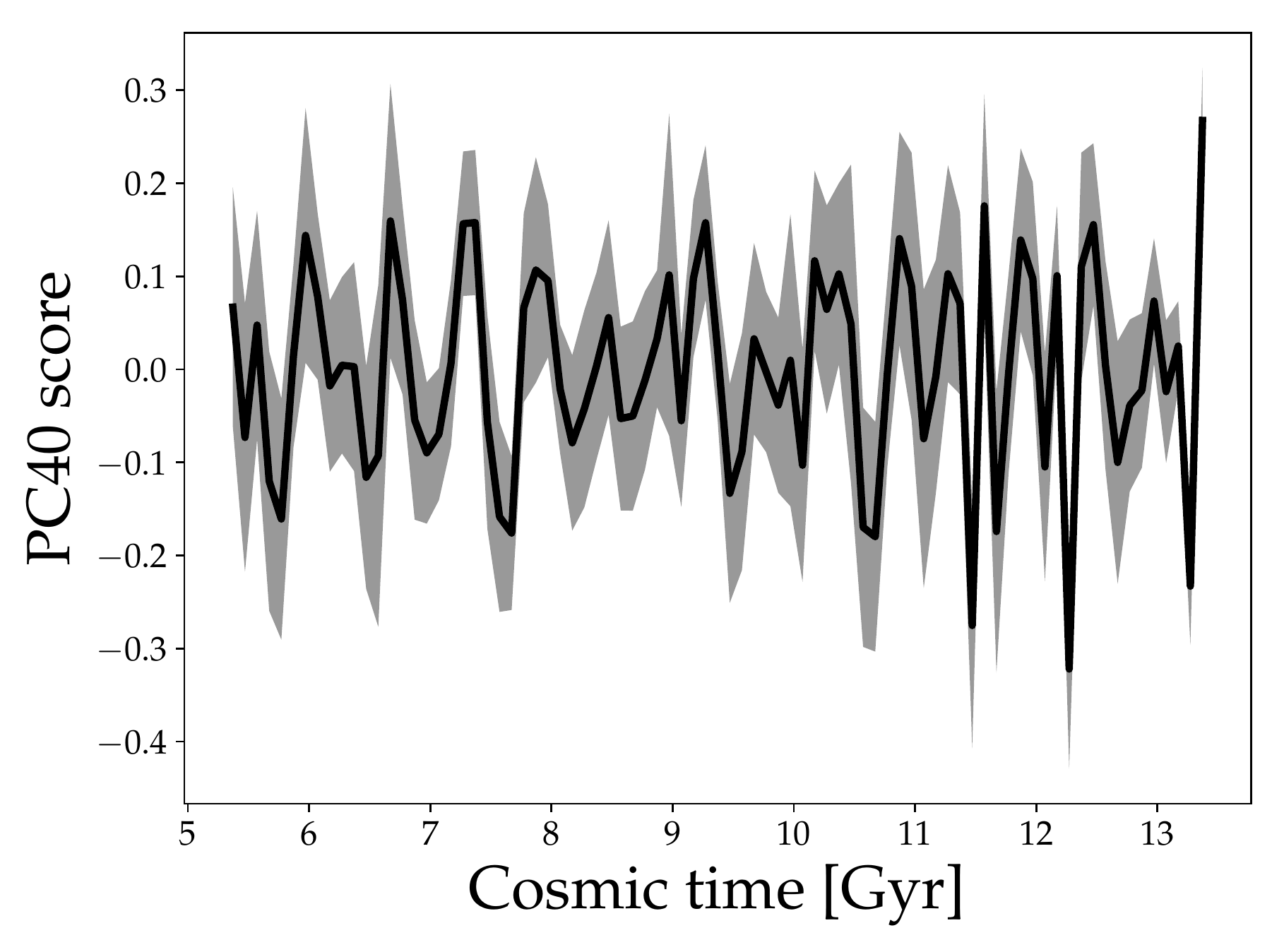} &
\hspace{-0.55cm}\includegraphics[width=6cm]{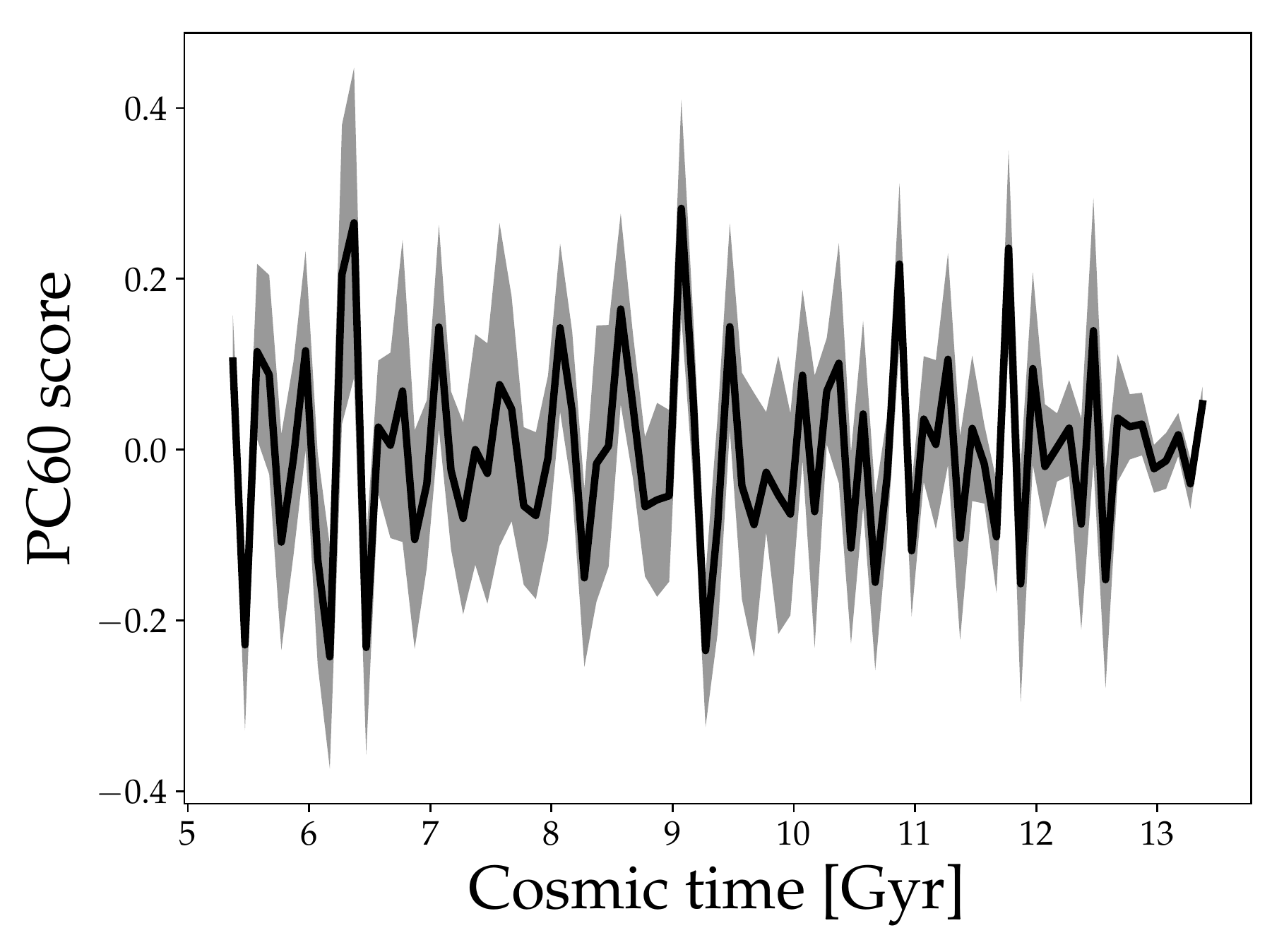} \\
\end{tabular}
\caption{\small{The principal components of the tracks of star-forming galaxies with M$_{\rm star}=10^{9.5\pm0.1}$ at $z=0$ through $\Delta$log$_{10}$ SFR - time space. The grey shaded area shows the standard deviation in the scores of each PC in 10,000 realisations of individual SFHs. We show the first three principal components (responsible for 47, 9 and 4.5 \% of the variance, respectively), and, for illustration, arbitrarily selected principal components with higher PC number (responsible for 1.9, 1.3, 1.0, 0.7, 0.4 and 0.2 \% of the variance, respectively). This figure illustrates that the PCs can be interpreted as the eigenvectors of the fluctuations through $\Delta$log$_{10}$ SFR - time space, where the fluctuation period depends on the PC number, $p$, approximately as $T\approx16 p^{-1}$ Gyr. }} \label{fig:PC_projections}
\end{figure*}

We note that the SFRs in Fig. $\ref{fig:path_individual}$ are measured as SFR$_{i} = \frac{{\rm M}_{i+1}-{\rm M}_{i}}{\Delta \rm t}$, where M$_{i}$ is the sum of the birth (i.e. zero age main sequence) stellar mass of star particles in a galaxy at simulation output time $i$ and $\Delta$t is the output spacing. This is different from the rest of the paper (except for the rest of this section and \S $\ref{sec:relative_timescales}$), where we measure SFR directly from the star-forming gas. The instantaneous SFRs were however not saved in the higher time-resolution output of the simulation. This means that mass growth due to (minor) mergers is included in the SFR as well\footnote{We also note that due to this limitation, it is not possible to isolate the contribution to SFR fluctuations on short time-scales induced by minor mergers.}, but as can be seen from Fig. $\ref{fig:path_individual}$, the highest SFRs are typically $<2-3$ M$_{\odot}$ yr, corresponding to a typical maximum mass growth of $\approx 2-3\times10^8$ M$_{\odot}$ per timestep, i.e. maximum merger mass fractions of $\lesssim 1/25$. The presence of oscillations with frequencies higher than the snapshot spacing of 100 Myr cannot be tested. An additional source of scatter in these SFRs derived from the difference in stellar masses are Poisson fluctuations due to discreteness in star formation implementation (see \citealt {SchayeVecchia2008} for details). We calculate the contribution of Poisson noise to the scatter as follows: for a given SFR, the expected number of star particles formed per timestep is $N= {\rm SFR} \frac{\Delta t}{m}$, where $\Delta t$=100 Myr and $m=1.81\times10^6$ M$_{\odot}$, the birth mass of a star particle. Therefore, assuming Poisson noise, the scatter in the SFR measured over such a time step is $\sigma({\rm SFR}) = \sqrt{N} \frac{m}{\Delta t} = \sqrt{{\rm SFR}\,m/\Delta t}$. As we study logarithmic SFRs, we convert this to $\sigma$(log$_{10}$SFR)$  = \frac{1}{\rm ln10} \sqrt{\frac{m}{{\rm SFR} \Delta t}}$. For the typical SFRs in Fig. $\ref{fig:path_individual}$ the Poissonian noise is $0.05-0.10$ dex, meaning that Poisson fluctuations do not dominate the measured variation in log$_{10}$SFR of $\approx0.2-0.3$ dex.

\subsection{The relative importance of different time-scales} \label{sec:relative_timescales}
As discussed above, we find that the SFRs of galaxies fluctuate on both short and long time-scales. Here, we examine quantitatively which of these time-scales drive the majority of scatter in the SFR-M$_{\rm star}$ relation. As the topic of interest are fluctuations with respect to the median SFR-M$_{\rm star}$ relation at a given epoch, we analyse the tracks of individual galaxies through $\Delta$log$_{10}$ SFR - time space (i.e. star formation histories as in Fig. $\ref{fig:path_individual}$, but normalised to the median SFR of galaxies with the same stellar mass at each cosmic epoch). 

We use a principal component analysis (PCA; \citealt{Pearson1901}) of the tracks of star-forming galaxies in bins of stellar mass at $z=0$. PCA compresses these individual tracks in orthogonal eigenvectors (principal components; PCs) that are ordered by their eigenvalues (see also \citealt{Sparre2015} for a PCA of SFHs of simulated galaxies in the Illustris simulation). As we will see, it turns out that each PC can be associated with a fluctuation time-scale following $T\propto p^{-1}$, where $p=1,2,...$ is the PC number.\footnote{We note that our numbering of PCs differs from \cite{Sparre2015} as we normalise SFR fluctuations with respect to the evolution of the general population before performing the PCA.} PCA of the evolutionary tracks of individual galaxies can be seen as a non-parametric version of a Fourier analysis. The eigenvalue of each PC can be converted into the fraction of the total variance that is accounted for in $\Delta$log$_{10}$ SFR - time space. As shown in e.g. \cite{Gao2003}, the shape of the eigenvalue spectrum of a PCA of a set of evolutionary tracks contains information on the nature of these tracks. In particular, the eigenvalue spectrum depends on the degree of long-term memory in evolutionary tracks (for example quantified by the Hurst exponent in a fractional Brownian motion process; see also \citealt{Kelson2014}). A steep spectrum corresponds to tracks that are dominated by correlated long time-scale fluctuations, while a flat spectrum is found for tracks that are dominated by uncorrelated, stochastic fluctuations. 

In practice, we select the subset of star-forming galaxies at $z=0$ in stellar mass ranges of 0.1 dex and compute their evolutionary tracks through $\Delta$log$_{10}$ SFR - time space as described in \S $\ref{sec:variety}$. In order to assess the importance of Poisson noise on the SFRs, we run a PCA on 10,000 realisations of the tracks of individual galaxies, where in each realisation we perturb the SFR of each galaxy at each time-step with its associated Poisson uncertainty. Then, we compute the median and standard deviation of the PCs and their eigenvalues. As PCAs of different realisations may result in flipping the signs of the eigenvectors, we force the sign of projection of the $\Delta$log$_{10}$ SFR$_{z=0}$ axis to be the same in each of the realisations. 

We show the first three and six other example PCs in Fig. $\ref{fig:PC_projections}$ for star-forming galaxies in the mass range M$_{\rm star}=10^{9.5\pm0.1}$ M$_{\odot}$. The first PCs correspond to fluctuations with period $T\approx16 p^{-1}$ Gyr. This trend continues for the other PCs, even up to high PC numbers. These results show that, to first order, the evolutionary tracks of star-forming galaxies with M$_{\rm star}=10^{9.5\pm0.1}$ at $z=0$ through $\Delta$log$_{10}$ SFR - time space are characterised by a slow evolution in the zero point (PC1, which accounts for $\approx50$ \% of the variance). This implies that the evolution of individual galaxies through SFR - M$_{\rm star}$ space is not exactly parallel to the evolution of the median of the galaxy population (in which case no zero point evolution should be seen). The grey shaded regions in Fig. $\ref{fig:PC_projections}$ show that the uncertainty associated to the Poisson errors in SFRs is relatively small, meaning that the results from the PCA are robust to these uncertainties.

\begin{figure*}
\begin{tabular}{cc}
\includegraphics[width=8.6cm]{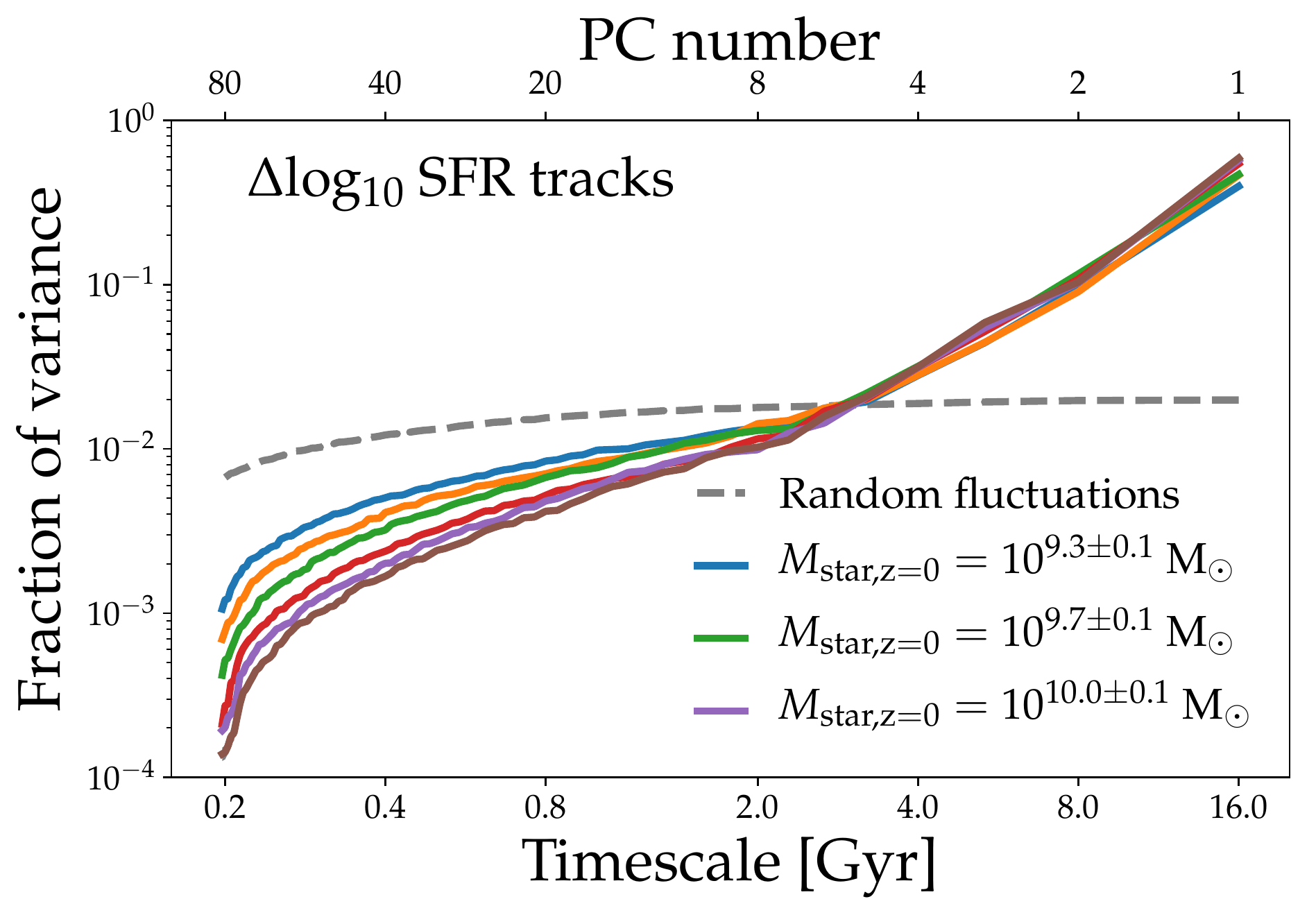}  &
\includegraphics[width=8.6cm]{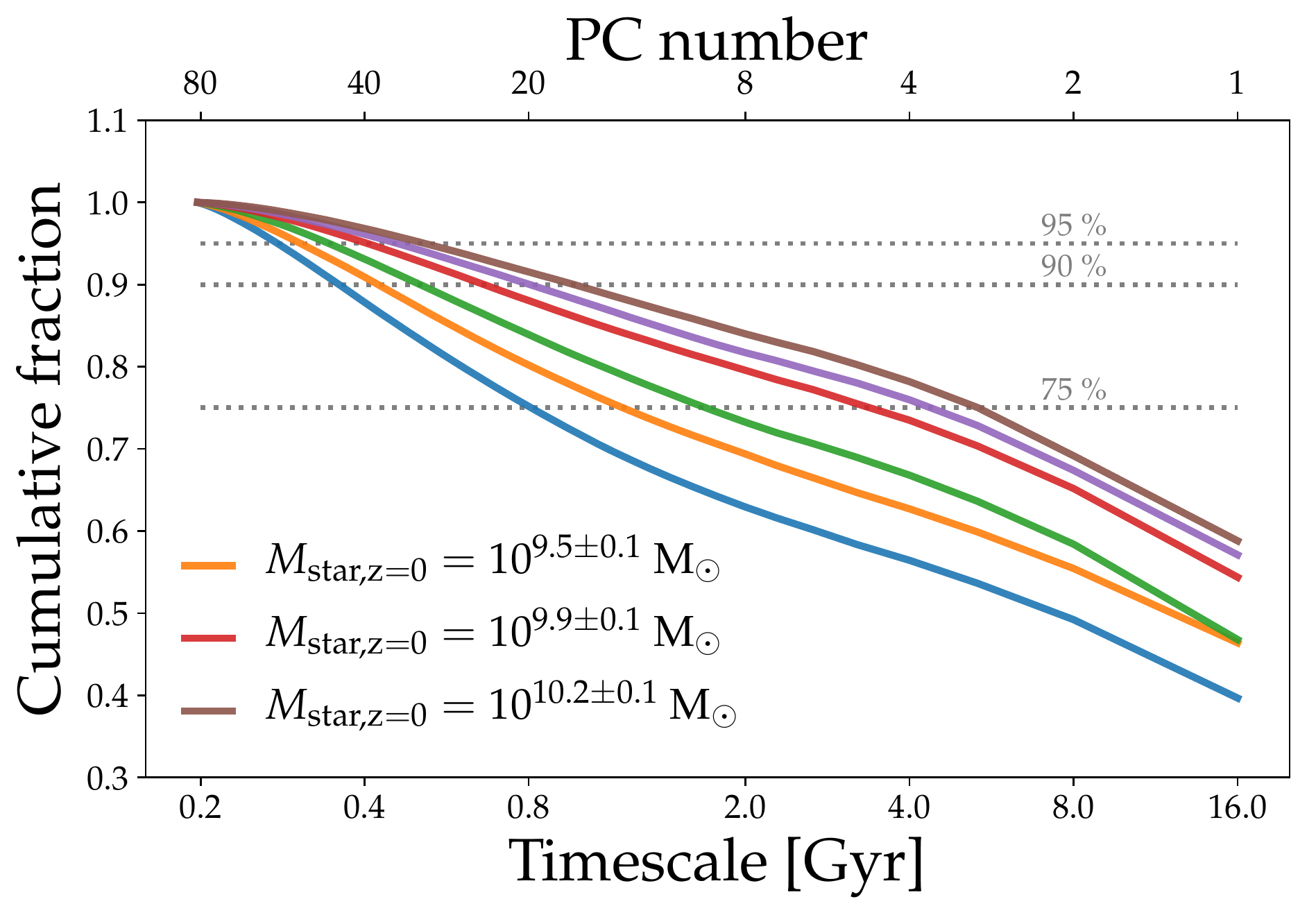} 
\end{tabular}
\caption{\small{{\it Left}: The eigenvalue spectrum of the principal component analysis of the tracks of individual in three mass ranges through $\Delta$log$_{10}$ SFR - time space (coloured lines; note the legend is shared with the right panel). The y-axis is normalised to show the fraction of the variance accounted for by each PC. The x-axis is converted to the fluctuation period associated to each PC. Tracks are characterised by correlated long time-scale fluctuations, while stochastic fluctuations contribute variations on short time-scales. The grey dashed line illustrates the eigenvalue spectrum of simulated tracks with pure random (uncorrelated between different time-steps) fluctuations mapped on the same time-grid. {\it Right}: Cumulative fraction of variance as a function of PC number (associated time-scale) in the same mass ranges. Fluctuations on longer time-scales are relatively more important in higher-mass galaxies.  }} \label{fig:PC_spectrum}
\end{figure*}

The eigenvalue spectrum of the PCs is shown in the left panel of Fig. $\ref{fig:PC_spectrum}$ for three representative mass ranges. The eigenvalues are normalised to show the fraction of the variance in the tracks that is accounted for by fluctuations on different time-scales. The eigenvalue spectrum shows that fluctuations on long time-scales are most important, but that there is a continuous contribution from shorter time-scales, down to the shortest time-scales resolved in the simulation. For comparison, we also show the flat eigenvalue spectrum of a PCA of simulated tracks that are characterised by white noise (i.e. what would be the case for random fluctuations). These simulated tracks are discretised with the same time-stepping. The fact that the eigenvalue spectrum of simulated white noise tracks bends down at time-scales below 0.8 Gyr illustrates that the discreteness of the simulation output slightly influences the PCA results.  For more massive galaxies, the eigenvalue spectrum flattens at slightly shorter time-scales (meaning stochastic/random fluctuations are less important). The slope of the eigenvalue spectrum at long time-scales is relatively insensitive to mass.

For star-forming galaxies with M$_{\rm star}=10^{9.5\pm0.1}$ M$_{\odot}$ at $z=0$, the cumulative fraction of the variance in the tracks through $\Delta$log$_{10}$ SFR - time space that is explained by time-scales $>2$ Gyr is 70\%, while 90 (95) \% of the variation can be accounted for by including PCs that oscillate on periods of $\approx$0.4 (0.3) Gyr.  The right panel of Fig. $\ref{fig:PC_spectrum}$ illustrates how these results depend on mass. Long time-scale fluctuations contribute to a larger fraction of the variance in the tracks of more massive galaxies. For example, the minimum fluctuation time-scale that needs to be included to account for 75 \% of the scatter increases from $\approx0.8$ Gyr at M$_{\rm star, z=0} = 10^{9.3\pm0.1}$ M$_{\odot}$ to $\approx7$ Gyr at M$_{\rm star, z=0} = 10^{10.2\pm0.1}$ M$_{\odot}$.

\begin{figure}
\includegraphics[width=8.6cm]{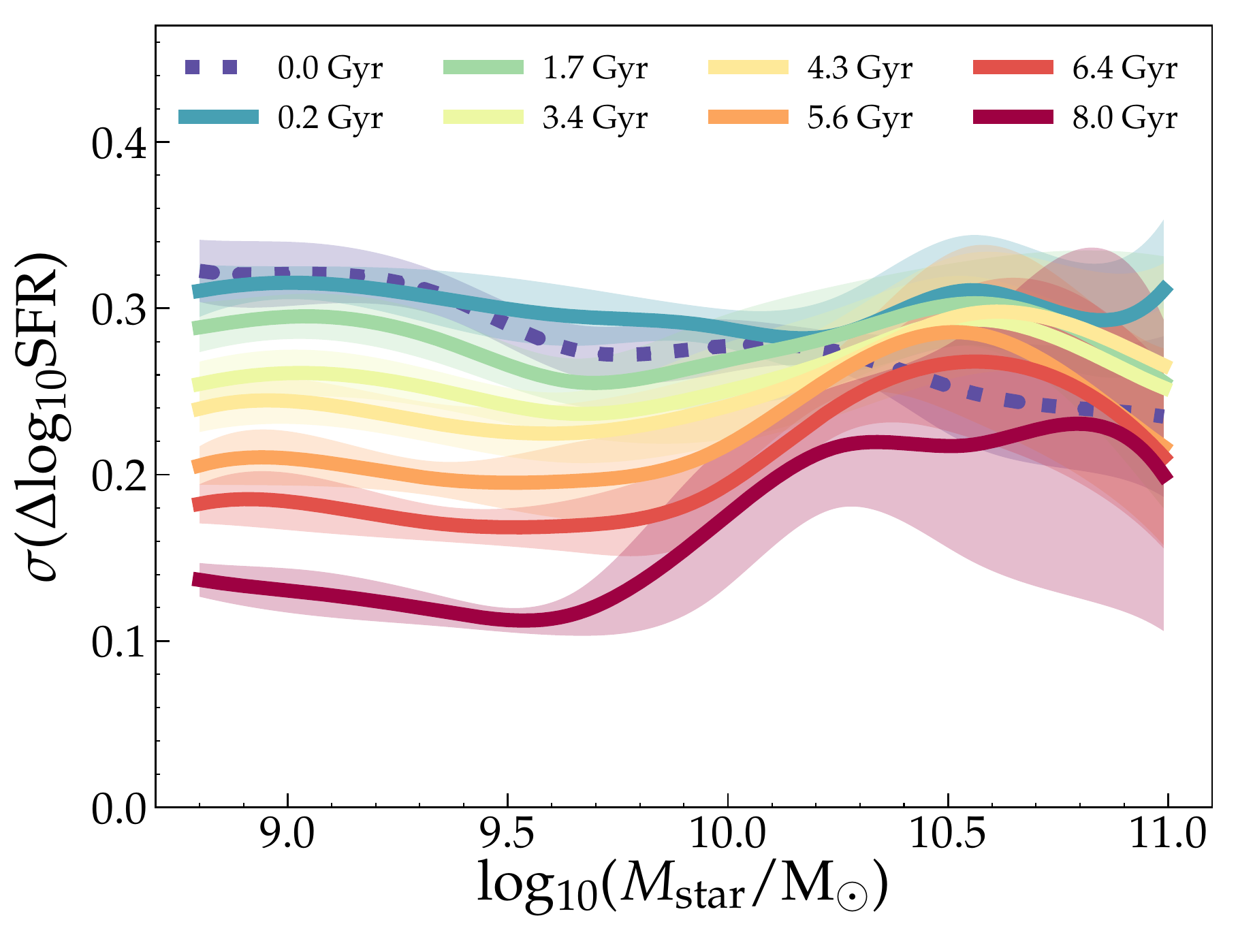} 
\caption{\small{The dependence of the scatter in SFR(M$_{\rm star}$) at $z=0$ on the time-scale over which the SFR of each galaxy is averaged. The instantaneous SFR is shown as a dashed line, while averaged SFRs (computed using the merger-tree as in \S $\ref{sec:variety}$) are shown as solid lines. Time-averaged SFRs are corrected for Poisson noise associated with the SF recipe. We removed galaxies that underwent a major merger (1:4 ratio) during the time over which the SFR is averaged. We find that short time-scale fluctuations drive the scatter in the SFR-M$_{\rm star}$ relation, but long 8 Gyr fluctuations still contribute $\approx0.15$ dex of scatter at M$_{\rm star}\lesssim10^{10}$ M$_{\odot}$.  }}
\label{fig:scatter_on_time-scales}
\end{figure}

Another way to address the relative importance of different time-scales in causing scatter in the SFR-M$_{\rm star}$ relation at a certain point in cosmic time is to measure the scatter after averaging the SFR over a range of time-scales. This is done in Fig. $\ref{fig:scatter_on_time-scales}$. Here, time-averaged SFRs for individual galaxies are computed as in \S $\ref{sec:variety}$ using the difference in initial stellar mass of the main progenitor between $z=0$ and earlier times. We corrected the scatter for Poisson noise associated to the SF implementation. For each time-interval, we removed galaxies that underwent a major merger (1:4 ratio) in any 100 Myr interval during the time over which the SFR is averaged. This is necessary as the SFR is inferred from the change in stellar mass. The fraction of galaxies that underwent a major merger increases with stellar mass and with the time-scale  over which SFR is averaged. For galaxies M$_{\rm star}<10^{10}$ M$_{\odot}$ a fraction of 10 (20) \% is removed  on 2 (8) Gyr time-scales, while 25 (60) \% of the galaxies is removed on these respective time-scales at stellar masses M$_{\rm star}>10^{10}$ M$_{\odot}$. We find that the scatter is largest for the instantaneous SFR and short time-scales $<1$ Gyr ($\approx0.3$ dex) and decreases to a non-negligible scatter of $\approx0.15$ dex for long, 8 Gyr, time-scales, particularly at M$_{\rm star}\lesssim10^{10}$ M$_{\odot}$.

\section{The origin of long time-scale correlations} \label{sec:cosmological_origin}
What is driving the coherence in the long time-scale SFHs of galaxies and its dependence on their present-day sSFR? Star formation is fuelled by the inflow of (cold) gas. Naively, it is therefore expected that the star formation history of a galaxy is related to its halo mass accretion history \citep[e.g.][]{Moster2013}. In this section, we therefore explore how the halo and stellar mass assembly histories are related to the star formation histories of galaxies, and the positions of galaxies on the SFR - M$_{\rm star}$ plane. Specifically, we use the merger tree to map out the dark matter mass history of halos (in the matched DMO simulation, which is not affected by baryonic processes). We measure $z_{1/2}$, the redshift at which half of the halo mass (M$_{200}$) at $z=0.1$\footnote{For the remainder of the paper, we study central galaxies at $z=0.1$ in the EAGLE simulation.} was first assembled in the main progenitor as a way to quantify formation time.

\begin{figure}
\includegraphics[width=8.6cm]{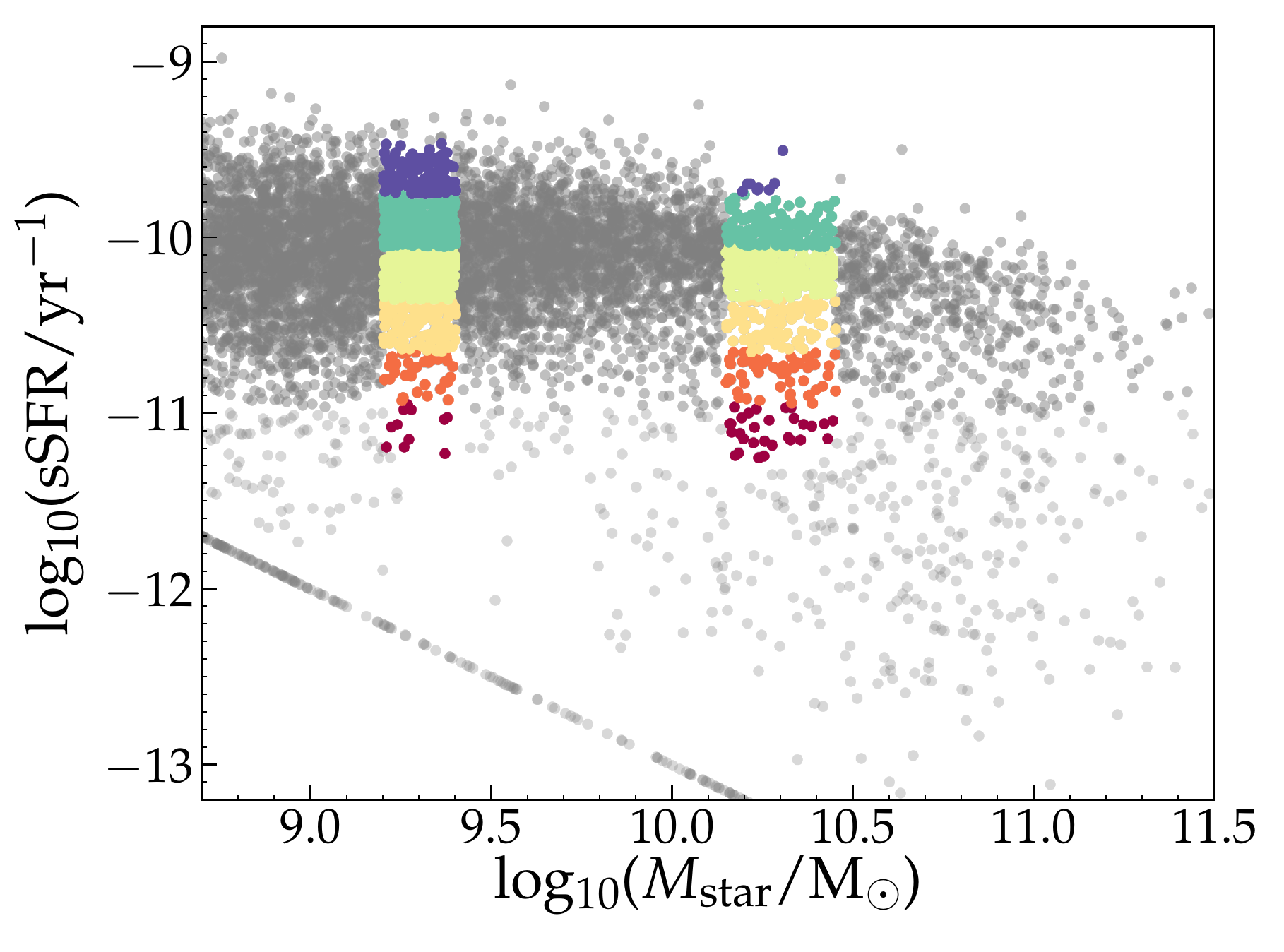} 
\caption{\small{Relation between sSFR and stellar mass for central galaxies at $z=0.1$. Galaxies that contain no star-forming gas particles are placed at a SFR of 0.001 M$_{\odot}$ yr$^{-1}$ for visualisation purposes. We use color-coding to highlight the galaxies that have been binned in Fig. $\ref{fig:MstarSFR_EVO}$, where we show their median SFR, M$_{\rm star}$ and M$_{200}$ histories. }}
\label{fig:SFRMstar}
\end{figure}

\begin{figure}
\includegraphics[width=8.6cm]{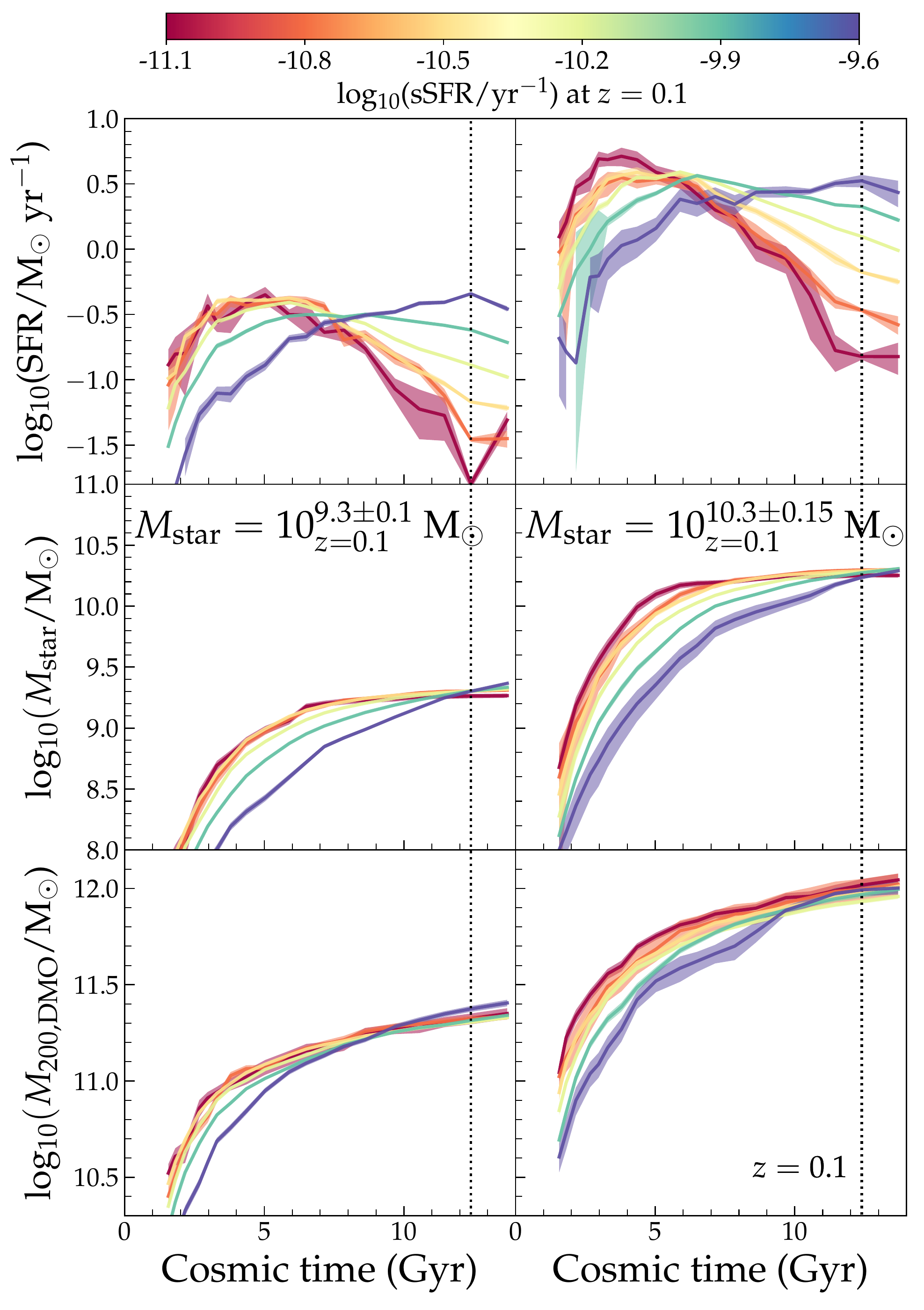}
\caption{\small{Evolution of the median SFR ({\it top row}), stellar mass ({\it middle row}) and DMO halo mass ({\it bottom row}) in bins of $z=0.1$ stellar mass ({\it different columns}), sub-divided in bins of $z=0.1$ sSFR (colour-coding). The shaded regions indicate the formal errors ($\sigma_{\rm bin}$/$\sqrt{N_{\rm bin}}$). At fixed stellar mass, galaxies with higher SFR at $z=0.1$ have a delayed SFH compared to galaxies with low SFR at $z=0.1$. Galaxies with a higher SFR at $z=0.1$ typically have had a lower stellar mass throughout cosmic history, compared to other galaxies with the same final stellar mass. In the first $\approx 8$ Gyr, their halo masses have also been lower than the typical halo mass of galaxies with similar $z=0.1$ stellar mass.  }}
\label{fig:MstarSFR_EVO} 
\end{figure}

\subsection{The joint evolution of SFR, M$_{\rm star}$ and M$_{200, \rm DMO}$}
We show the joint evolution of the median stellar mass, SFR and halo mass (in the matched DMO simulation) in bins of stellar mass and SFR (illustrated in Fig. $\ref{fig:SFRMstar}$) in Fig. $\ref{fig:MstarSFR_EVO}$. Here, each column shows a bin in stellar mass and in each panel different coloured lines correspond to different bins in specific SFR. We note that the stellar mass bins are chosen to be representative and that their exact values do not influence the results. We also note that for visualisation purposes we have binned galaxies by their mass and SFR at $z=0.1$, while the plots show the history down to $z=0$. 

\begin{figure}
\includegraphics[width=8.7cm]{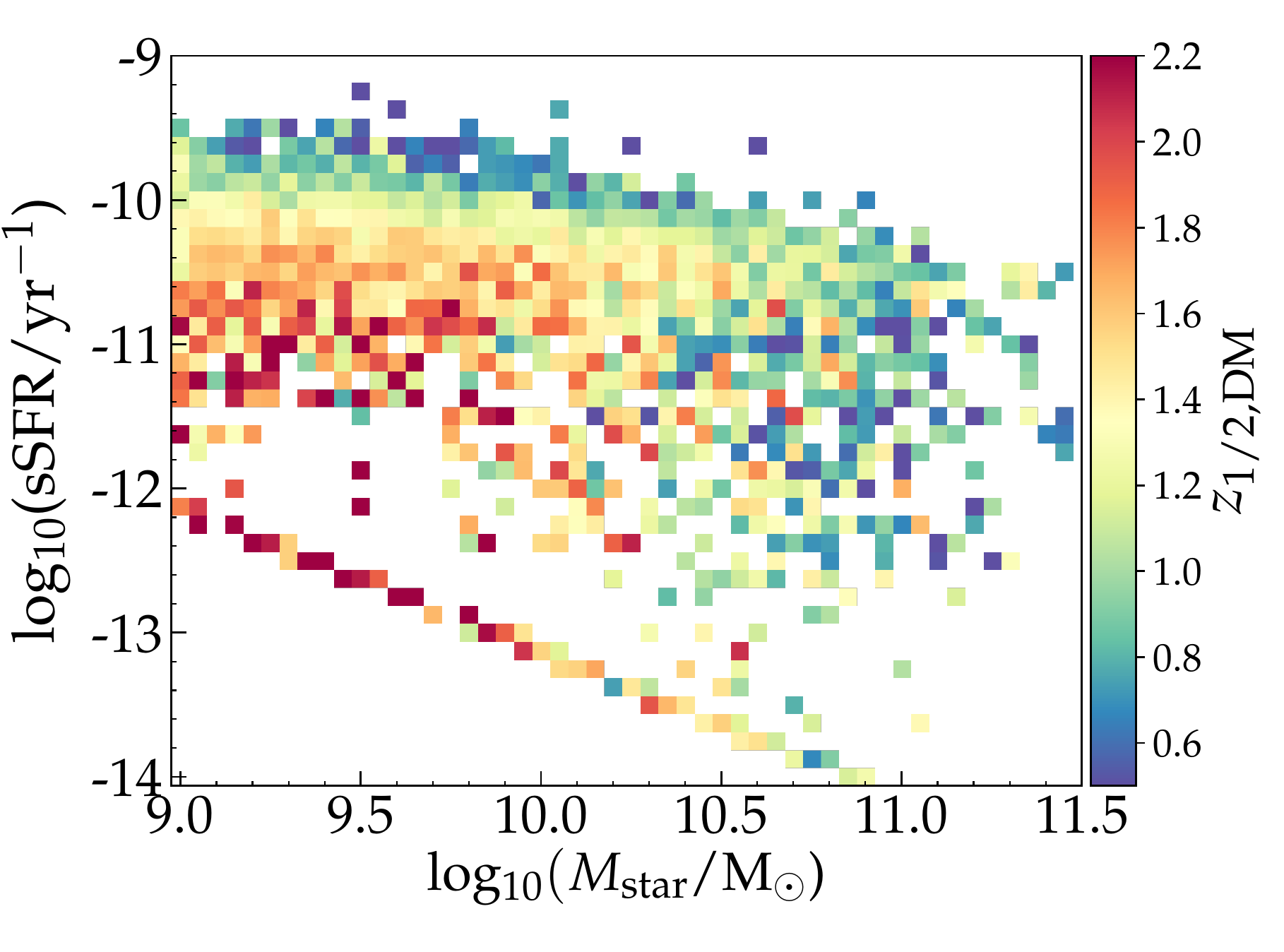}
\caption{\small{The relation between sSFR and stellar mass for central galaxies at $z=0.1$, color-coded by the median halo formation time in the matched DMO simulation, in bins of sSFR and M$_{\rm star}$. Galaxies that contain no star-forming gas particles are placed at a SFR of $10^{-3}$ M$_{\odot}$ yr$^{-1}$ for visualisation purposes. The scatter in sSFR is related to halo formation time for stellar masses of $\lesssim10^{10}$ M$_{\odot}$. }}
\label{fig:SFRMstar_z05}
\end{figure}

In general, for both masses shown here, the following characteristics can be seen for galaxies with high/low sSFR at $z=0.1$: i) they have a SFH that peaks later/earlier and is more extended/compressed, ii) they have had a lower/higher stellar mass during most of the history of the Universe, compared to other galaxies that end up with the same stellar mass, iii) they had a relatively low/high halo mass during the first $\approx8$ Gyr of the Universe, but not necessarily at later times. These results clearly indicate that galaxies that occupy similar regions in the SFR-M$_{\rm star}$ plane have similar SFHs, confirming the results from \S $\ref{burstiness}$. 

Observationally, \cite{Pacifi2016} find similar results by measuring the SFHs of a large sample of galaxies in the local Universe (see also \citealt{Pacifi2013}). They find that low-mass galaxies formed their stars over a longer time-scale than massive galaxies (`downsizing'; e.g. \citealt{Gallazzi2005,Thomas2005}) and that at fixed stellar mass, galaxies with relatively low SFRs have a more compressed SFH than galaxies with relatively high SFRs (see also \citealt{Dressler2016}). More recently, \cite{Chauke2018} confirm this result with a detailed analysis of high-resolution spectra at $z=1$. They find that at fixed stellar mass, star-forming galaxies have a SFH that is more extended and shifted towards later times compared to passive galaxies, and that the ongoing SFR is correlated with the SFH on $\sim3$ Gyr time-scales, a significant fraction of the Hubble time at $z=1$.

One subtle difference between the different mass bins in Fig. $\ref{fig:MstarSFR_EVO}$ is that galaxies in the higher stellar mass bin with a high sSFR at $z=0.1$ have had a relatively low halo mass throughout cosmic history, instead of only during the first $\approx 8$ Gyr, and vice versa for galaxies with low sSFRs. The explanation is that at these mass scales additional physical processes start to play a more prominent role, such as stellar mass growth due to merging and AGN feedback. These processes may result in a different relation between stellar and halo mass growth. We explore this further in \S $\ref{sec:BH}$.

\subsection{The relation with halo accretion history} \label{sec:scatter_z05} 
Next, we investigate in more detail how halo assembly influences the scatter in the SFR-M$_{\rm star}$ relation. In Fig. $\ref{fig:SFRMstar_z05}$ we show that the sSFR\footnote{We show sSFR instead of SFR as it visualises our results better, but we note that the same results are obtained for the residuals of the SFR-M$_{\rm star}$ relation.} of galaxies at fixed M$_{\rm star}$ is related to $z_{1/2}$, the redshift at which the dark matter halo mass of the main progenitor had half its $z=0.1$ mass (left panel). In this figure, we have created bins in 2D space of width 0.07 dex in M$_{\rm star}$ and 0.125 dex in sSFR and computed the median halo formation time of all galaxies in each bin. Galaxies with zero SFR are shown with a sSFR that would correspond to a SFR$=10^{-3}$ M$_{\odot}$ yr$^{-1}$. For stellar masses $\lesssim10^{10}$M$_{\odot}$, it is clear that haloes that form later tend to host galaxies with a higher sSFR. As discussed above, additional physical processes play a role at high masses (such as AGN feedback, see \S $\ref{sec:BH}$). We quantify the strength of the correlation between the residuals of the SFR-M$_{\rm star}$ relation and halo formation time using the Spearman rank correlation coefficient (R$_S$, where R$_S = (-)1$ for a perfect (anti)-correlation and R$_S = 0$ for no correlation). We find that R$_S \approx -0.45$ for stellar masses of $10^9-10^{10}$M$_{\odot}$, and R$_S \approx -0.10$ at M$_{\rm star}>10^{10.5}$ M$_{\odot}$.\footnote{We test other ways to quantify the dark matter halo accretion history in Appendix $\ref{test_massgrowth}$, but find none that correlate more strongly with the scatter in SFR-M$_{\rm star}$ than $z_{1/2}$ does. We have also tested whether short time-scale fluctuations in the growth rate of M$_{200}$ are related to the short time-scale fluctuations in the SFR that were discussed in \S $\ref{burstiness}$ and illustrated in Fig. $\ref{fig:path_individual}$. Typical fluctuations in the halo growth rate on $\sim 100$ Myr time-scales have a spread of $\approx0.4$ dex, larger than fluctuations in SFRs. For the majority of galaxies, we do not find a strong correlation between the two growth rates, neither measured simultaneously nor measured with a time-delay between halo growth and stellar mass growth of $0-1$ Gyr in steps of 100 Myr. }
 
Our results show that the scatter in the SFR-M$_{\rm star}$ relation is correlated with dark matter halo formation time. As halo formation time is related to halo clustering \citep[e.g.][]{Gao2005,Gao2007}, the scatter in the SFR-M$_{\rm star}$ relation is (partly) connected to assembly bias (see also \citealt{Tinker2017b,Diemer2017}) and the formation of large-scale structure. These results are consistent with the results from the semi-analytical model of \cite{Dutton2010}, who found that the scatter in the SFR-M$_{\rm star}$ correlates with halo concentration (which correlates strongly with $z_{1/2}$, see e.g. \citealt{Matthee2017}). Our results are also consistent with observations from \cite{Coil2017}, who found that at fixed stellar mass, galaxies with a higher sSFR are less clustered (indicating that they form later), although that could also be due to a lower halo mass.

Connecting these results with the results from Fig. $\ref{fig:MstarSFR_EVO}$, we find that the SFHs of haloes that form earlier also tend to peak earlier, while haloes that form later have a more extended SFH with a later peak. Quantitatively, we find that the median SFR of a central galaxy in a halo with M$_{200} = 10^{11.8}$ M$_{\odot}$ within the earliest halo formation time quartile ($z_{1/2} \approx 2.5$) has increased by a factor $\approx 10$ in the last 10 Gyr, while the median SFR of a galaxy within the latest halo formation time quartile ($z_{1/2} \approx 0.2$) has only increased by a factor $\approx1.5$ over the same period. Hence, part of the diversity in SFHs of galaxies at fixed halo mass is driven by differences in halo formation time.\footnote{We note that the halo formation time is strongly correlated with the projection of a galaxy on the first PC vector of the PCA described in \S $\ref{sec:relative_timescales}$.}  This drives a coherence in the long time-scale fluctuations in galaxies' SFRs that correlates with their current positions in the SFR-M$_{\rm star}$ diagram.

\subsection{How much scatter can be explained?} \label{howmuchscatter}
\begin{figure}
\includegraphics[width=8.6cm]{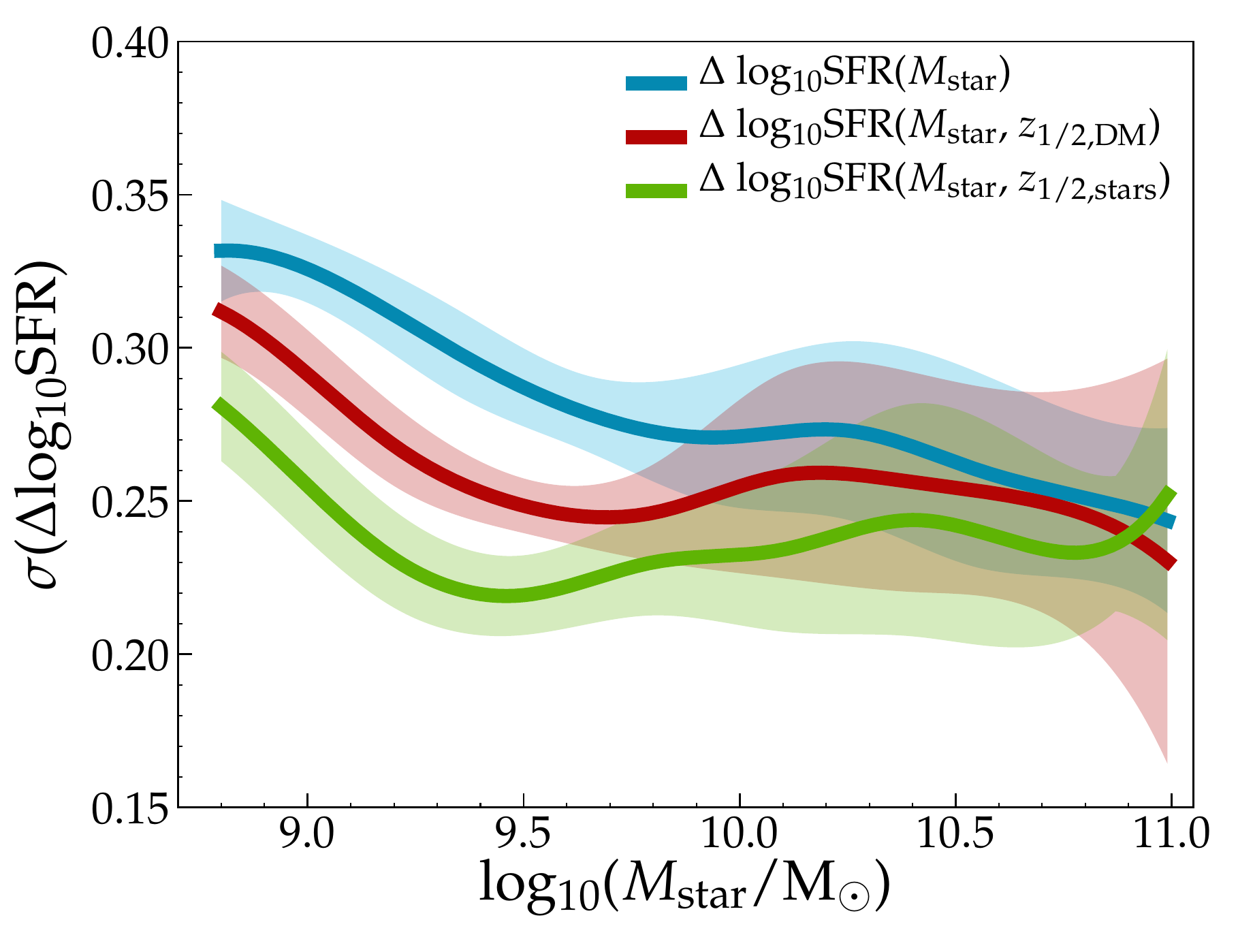} 
\caption{\small{Scatter in SFR at fixed stellar mass for central star-forming galaxies at $z=0.1$. The blue line shows the full scatter, while the red and green lines show the scatter after taking halo formation time (in the matched DMO simulation) and stellar mass assembly time into account, respectively. Shaded regions show the uncertainty associated to cosmic variance.}}
\label{fig:scatter_SFR_formationtime}
\end{figure}

What fraction of the scatter in the SFR-M$_{\rm star}$ relation can be attributed to variations in halo formation time? We measure this following the method described in detail in \cite{Matthee2017}: in each stellar mass bin, we fit a linear relation between the residuals in the SFR-M$_{\rm star}$ relation and halo formation time (in the matched DMO simulation). Hence, we compute the residual SFR (which is similar to residual sSFR) corrected for formation time:
\begin{equation}
\rm log_{10} \,sSFR_{\rm corrected} = log_{10} \,sSFR + \alpha + \beta\, z_{1/2}.
\end{equation}
The normalisation $\alpha$ and slope $\beta$ are weakly mass dependent, with (for example) $\alpha=9.8$ at $\beta=0.2$ at M$_{\rm star}\approx10^{9.5}$ M$_{\odot}$. We use this relation to compute $\Delta$log$_{10}$ SFR (M$_{\rm star}, z_{1/2}$), the residuals in the SFR-M$_{\rm star}$ relation after taking halo formation time into account. We then compute the standard deviation of these residuals in bins of stellar mass and show the results in Fig. $\ref{fig:scatter_SFR_formationtime}$. Note that this method is similar to measuring the scatter perpendicular to a three-dimensional plane of sSFR, M$_{\rm star}$ and $z_{1/2}$.

It can be seen that for stellar masses of $10^9-10^{10}$ M$_{\odot}$ the scatter is significantly reduced, by $\approx 0.05$ dex, after taking halo formation time into account. Assuming that the total scatter can be written as $\sigma_{\rm tot}^2 = \sigma_{z_{1/2}}^2 + \sigma_{\rm other}^2$, we find that the scatter due to fluctuations in $z_{1/2}$ alone is $\approx0.15$ dex at M$_{\rm star} < 10^{10}$ M$_{\odot}$. No significant reduction in the scatter is found for higher stellar masses. This result means that scatter in halo formation time is the cause of only part of the scatter in the SFR-M$_{\rm star}$ relation and its magnitude is similar to the scatter that we measured when SFRs were averaged on long time-scales (Fig. $\ref{fig:scatter_on_time-scales}$). We have varied our definition of formation time to mass fractions in the range 0.2-0.7 instead of 0.5, but found no significant difference. We have also tested whether the scatter is related to other DMO halo properties such as halo sphericity, triaxiality, spin or the number of neighbours within 2 and 10 Mpc (see \citealt{Matthee2017} for their exact definitions), but find that they do not reduce the scatter by more than 0.01 dex. Similarly, neither of these properties is related strongly to the scatter in the stellar mass - halo mass relation in EAGLE \citep{Matthee2017}.

As shown in Fig. $\ref{fig:scatter_SFR_formationtime}$, the stellar half-mass assembly time is (over the full mass-range) more closely related to the scatter in the SFR-M$_{\rm star}$ relation than halo formation time is. This is not surprising, as $z_{1/2, \rm stars}$ is a measure of the star formation history and we already showed that the long-term SFH is correlated with the scatter in SFR-M$_{\rm star}$\footnote{At fixed stellar mass, PC1 in \S $\ref{sec:relative_timescales}$ correlates very strongly with $z_{1/2, \rm stars}$ (with an absolute Spearman rank R$_s = 0.85$), while the scatter in this relation correlates strongly with PC2. This means that the stellar half-mass assembly time is strongly related to fluctuations around the main sequence on time-scales similar to the age of the Universe.}. Assuming that the total scatter in the SFR-M$_{\rm star}$ relation is due to $z_{1/2, \rm stars}$  and additional `stochastic' scatter combined in quadrature (see \S $\ref{burstiness}$), we find that $\approx 0.20$ dex of scatter can be attributed to fluctuations in $z_{1/2, \rm stars}$ at fixed stellar mass at M$_{\rm star} < 10^{10}$ M$_{\odot}$, and $\approx0.10$ dex at M$_{\rm star} \approx 10^{10-11}$ M$_{\odot}$. Alternatively, as mentioned in \S $\ref{sec:relative_timescales}$, about 70 \% of the variance in tracks through $\Delta$ SFR(M$_{\rm star}$) - time space can be explained by variations on time-scales $>2$ Gyr. This means that $\approx 0.21$ dex of scatter could be attributed to fluctuations on the longest time-scales, while the remaining fluctuations on shorter time-scales account for the additional $\approx 0.2$ dex of scatter, whose nature is more stochastic (Fig. $\ref{fig:PC_spectrum}$).

\begin{figure}
\includegraphics[width=8.6cm]{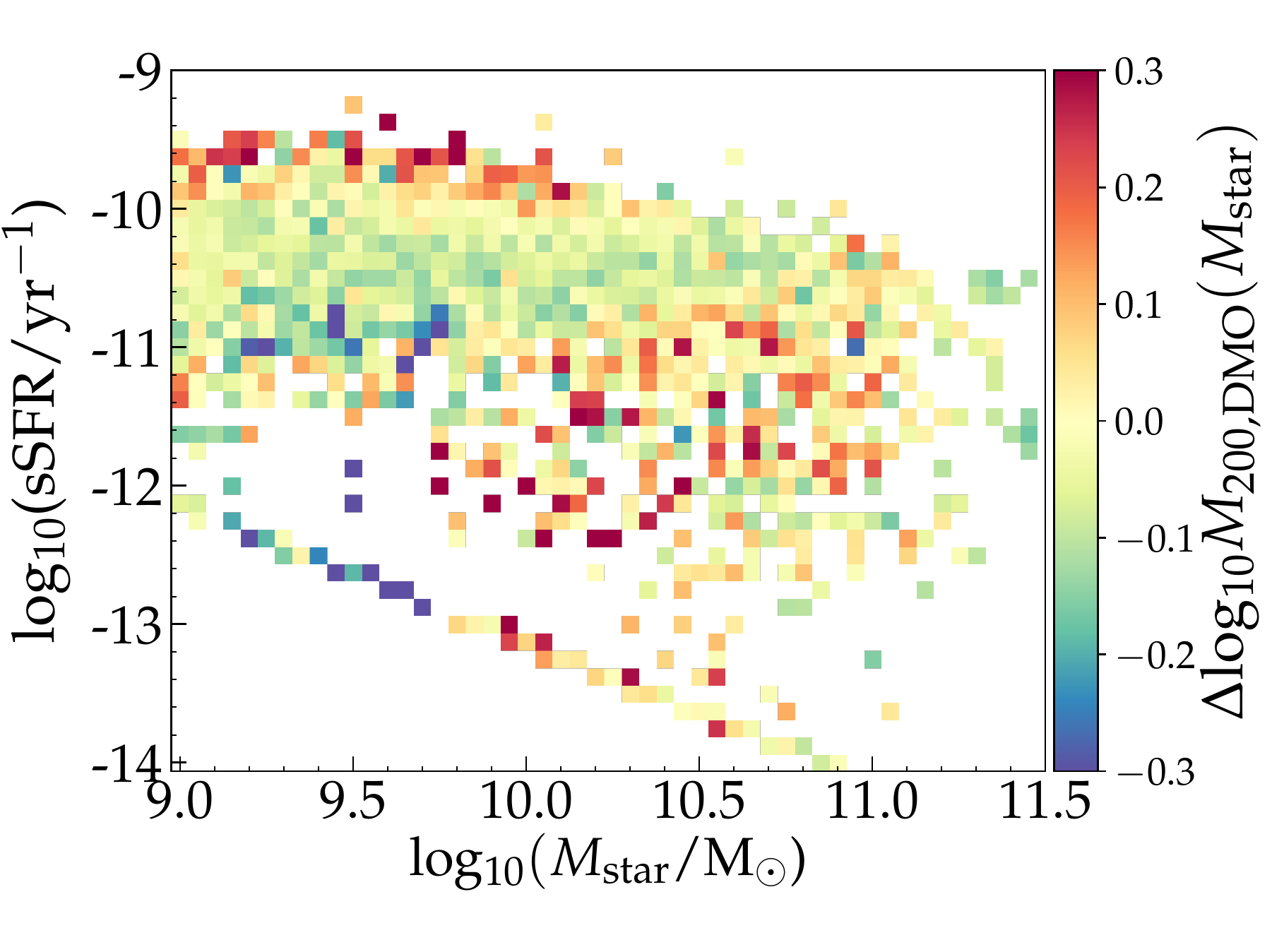}
\caption{\small{The relation between sSFR and stellar mass for central galaxies at $z=0.1$, color-coded by the median deviation from the relation between halo and stellar mass. We find that galaxies with higher/lower SFRs at fixed stellar mass tend to have relatively high/low halo masses, for galaxies with stellar masses $\approx10^9-10^{10}$ M$_{\odot}$. At M$_{\rm star}> 10^{10}$ M$_{\odot}$ the trend reverses because galaxies with relatively massive haloes at fixed stellar mass tend to have higher black hole masses, reducing their SFRs (see \S $\ref{sec:BH}$).}}
\label{fig:SFRMstar_SMHM}
\end{figure}

\subsection{Connection to the M$_{\rm star}$-M$_{\rm halo}$ relation}
In this subsection, we explore how our results are related to the origin of scatter in the M$_{\rm star}$-M$_{\rm halo}$ relation, which is a measure of the efficiency of galaxy formation. In \cite{Matthee2017} we showed that at a fixed halo mass M$_{200}<10^{12.6}$ M$_{\odot}$ in the matched DMO simulation, a galaxy with a relatively high stellar mass tends to reside in a halo that was assembled relatively early (such that the halo is relatively concentrated). This means that the haloes of these galaxies have a relatively low recent growth rate (and vice versa for haloes with relatively low stellar mass). Therefore, it is likely that the scatter in the stellar mass - halo mass relation is related to the scatter in the SFR-M$_{\rm star}$ relation. We show this in Fig. $\ref{fig:SFRMstar_SMHM}$. For central galaxies with M$_{\rm star}<10^{10}$ M$_{\odot}$, we find that at fixed stellar mass galaxies with higher SFRs tend to have higher halo masses. This is consistent with the result of \cite{Matthee2017}, as these galaxies also have relatively late halo formation times, such that they have a relatively low stellar mass compared to their halo. This means that observational samples with a selection based on SFR (and hence likely with relatively high sSFR) may yield samples with biased stellar mass to halo mass ratios. At M$_{\rm star}>10^{10}$ M$_{\odot}$, galaxies with a relatively high halo mass tend to have a relatively low SFR, inverting the trend between SFR and the relative halo mass to stellar mass. As we discuss in more detail in the next section, the stellar mass at which the transition occurs corresponds to the stellar mass where feedback associated with the growth of supermassive black holes starts to influence the star formation activity in galaxies.

\section{Relation with the growth of black hole mass} \label{sec:BH}
In this section, we investigate how the mass of super-massive black holes influences the scatter in the SFR-M$_{\rm star}$ relation. While correlations between the scatter and black hole mass do not imply causation, such correlations may still reveal clues with regards to the physical processes driving the scatter at high galaxy masses.

At the highest masses, the majority of galaxies have ceased their star formation and have much lower sSFRs than a typical main sequence galaxy. This means that the `pseudo-equilibrium growth' of galaxies, where their SFR typically only fluctuates within a factor of two, ends at some mass-scale. \cite{Bower2016} show that, in the EAGLE simulation, this is due to the increasing importance of AGN feedback, as the hot corona around galaxies in haloes with mass M$_{200} \gtrsim 10^{12}$ M$_{\odot}$ traps winds driven by star formation, leading to runaway black hole growth until the AGN becomes sufficiently luminous to drive a large-scale outflow (see \citealt{Trayford2016} for how this affects simulated galaxy colours). 

Around the transition regime, some haloes grow their black holes earlier than others, influencing their galaxies' SFHs and hence their location on the SFR-M$_{\rm star}$ plane \citep[e.g.][]{Sijacki2015,Terrazas2016,McAlpine2017}. The black hole (BH) mass is a measure for the accumulated amount of AGN feedback that was injected into a galaxy (assuming some of the released energy coupled efficiently to the gas). To first order, BH mass scales with halo mass and stellar mass, at least for halo masses $\gtrsim 10^{12}$ M$_{\odot}$. However, to second order more subtle effects can take place, for example in how the BH mass scales with stellar mass at fixed halo mass.  

\begin{figure}
\includegraphics[width=8.7cm]{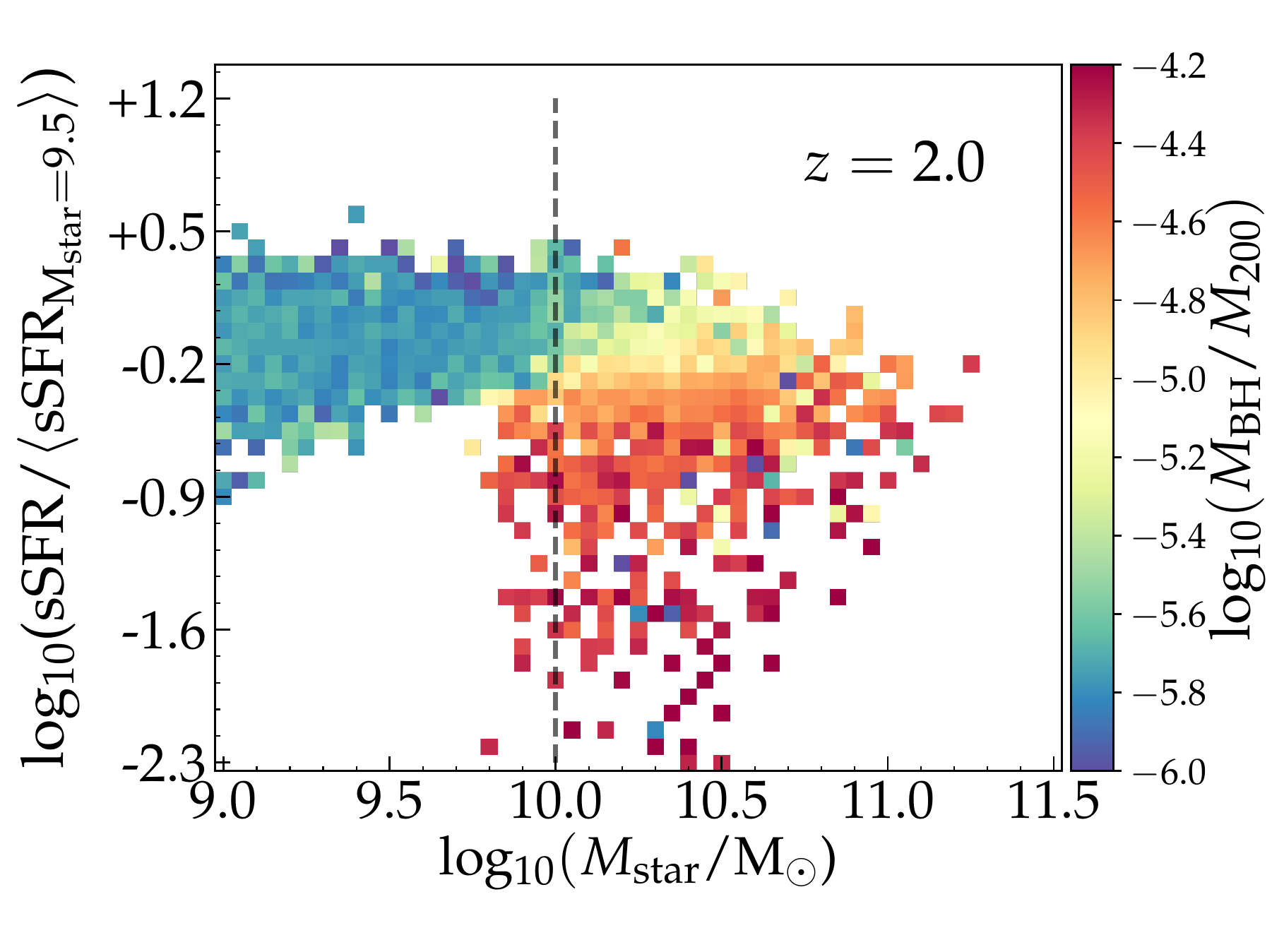}\vspace*{-2mm} \\ 
\includegraphics[width=8.7cm]{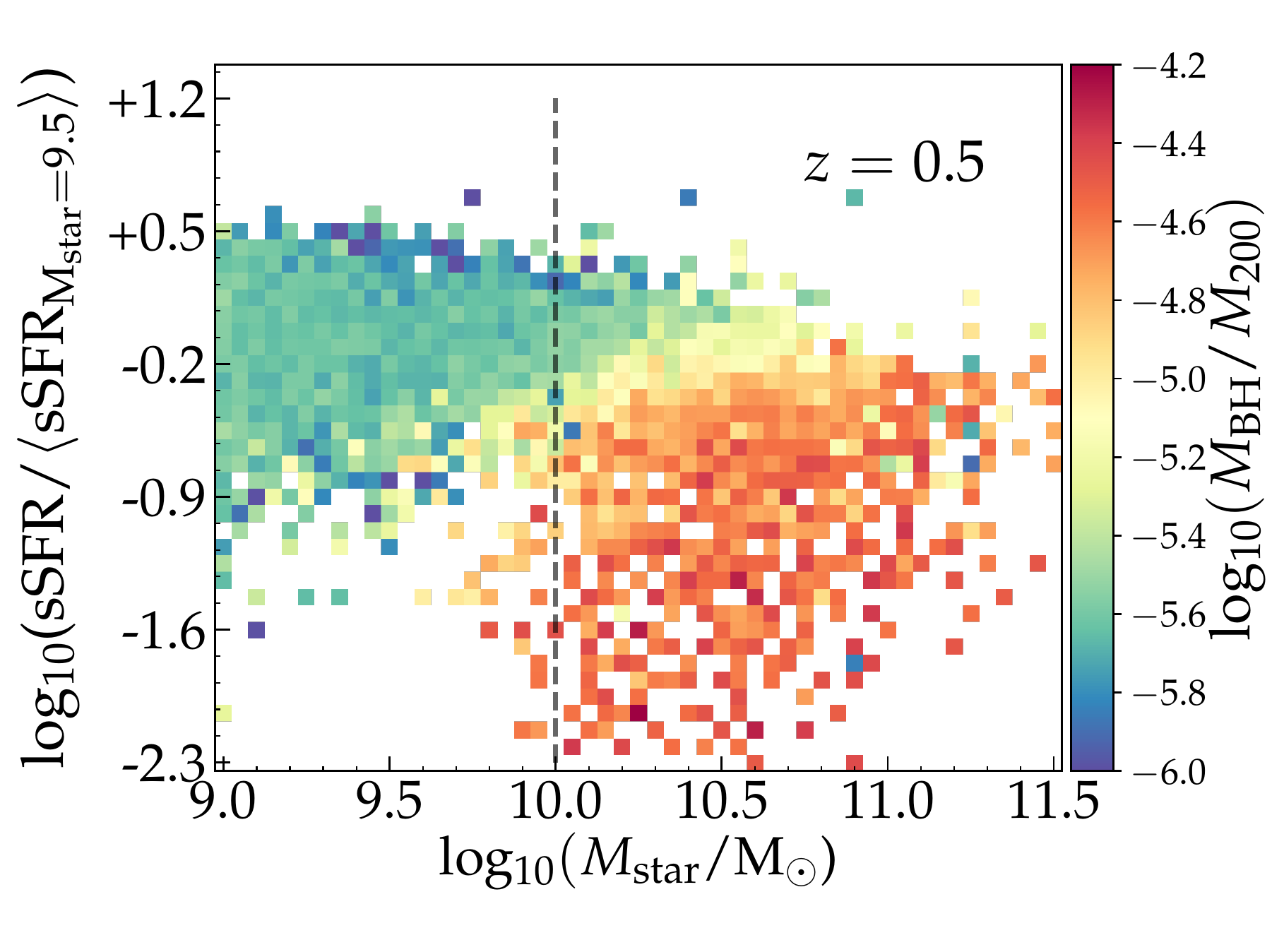} \vspace*{-5mm}\\
\includegraphics[width=8.7cm]{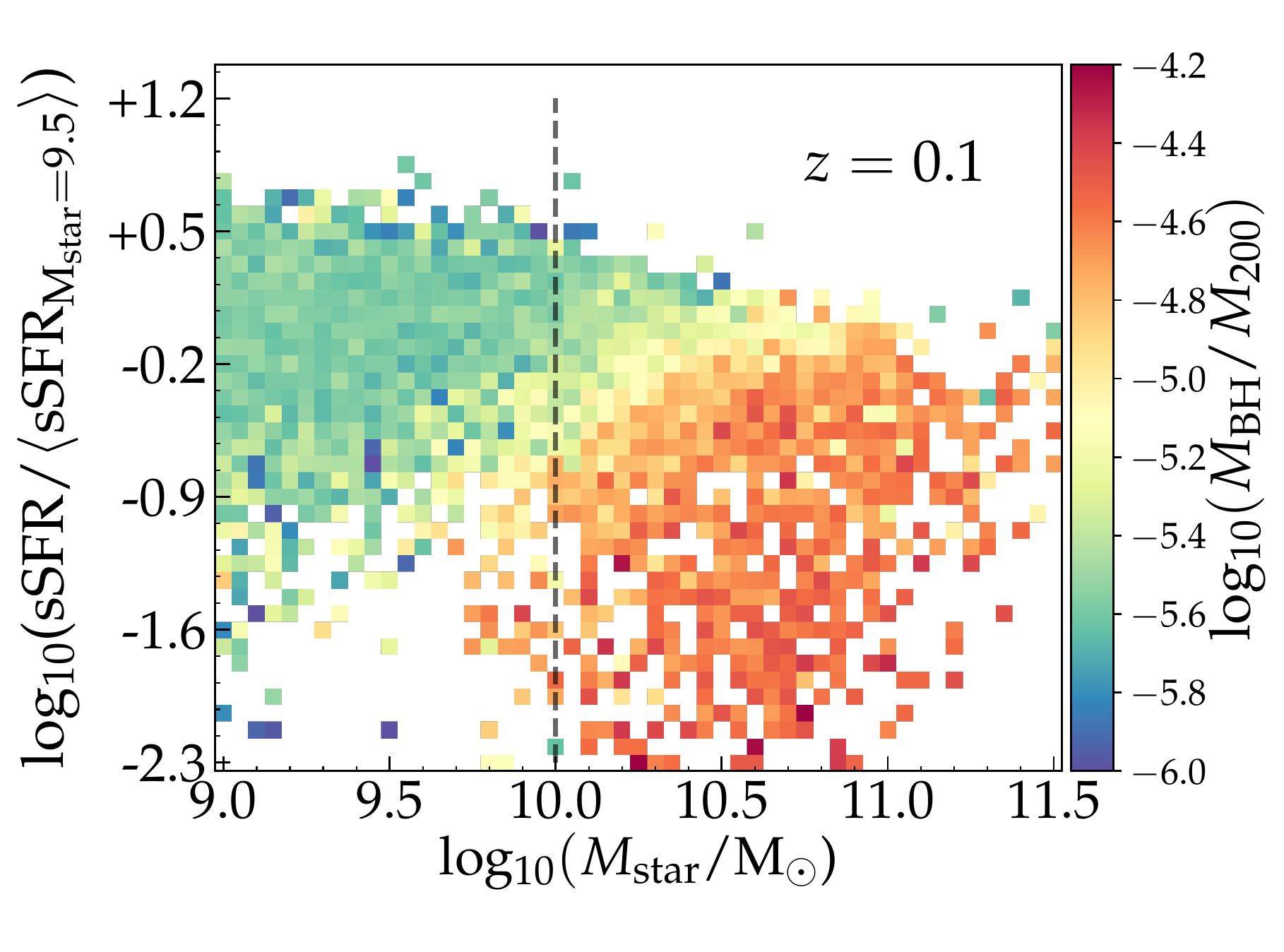}\vspace*{-5mm}\\
\caption{\small{Relation between sSFR and stellar mass for central galaxies at $z=2.0, 0.5, 0.1$ ({\it top, middle} and {\it bottom} rows, respectively). We normalise the sSFR to the median sSFR of galaxies with M$_{\rm star}=10^{9.50\pm0.05}$ M$_{\odot}$ and colour code the bins by the M$_{\rm BH}$/M$_{200}$ ratio, which highlights which halos have formed a BH efficiently. The vertically dashed line indicates a stellar mass of 10$^{10}$ M$_{\odot}$. }}
\label{fig:SFRMstar_BH_EVO}
\end{figure}

\begin{figure}
\includegraphics[width=8.7cm]{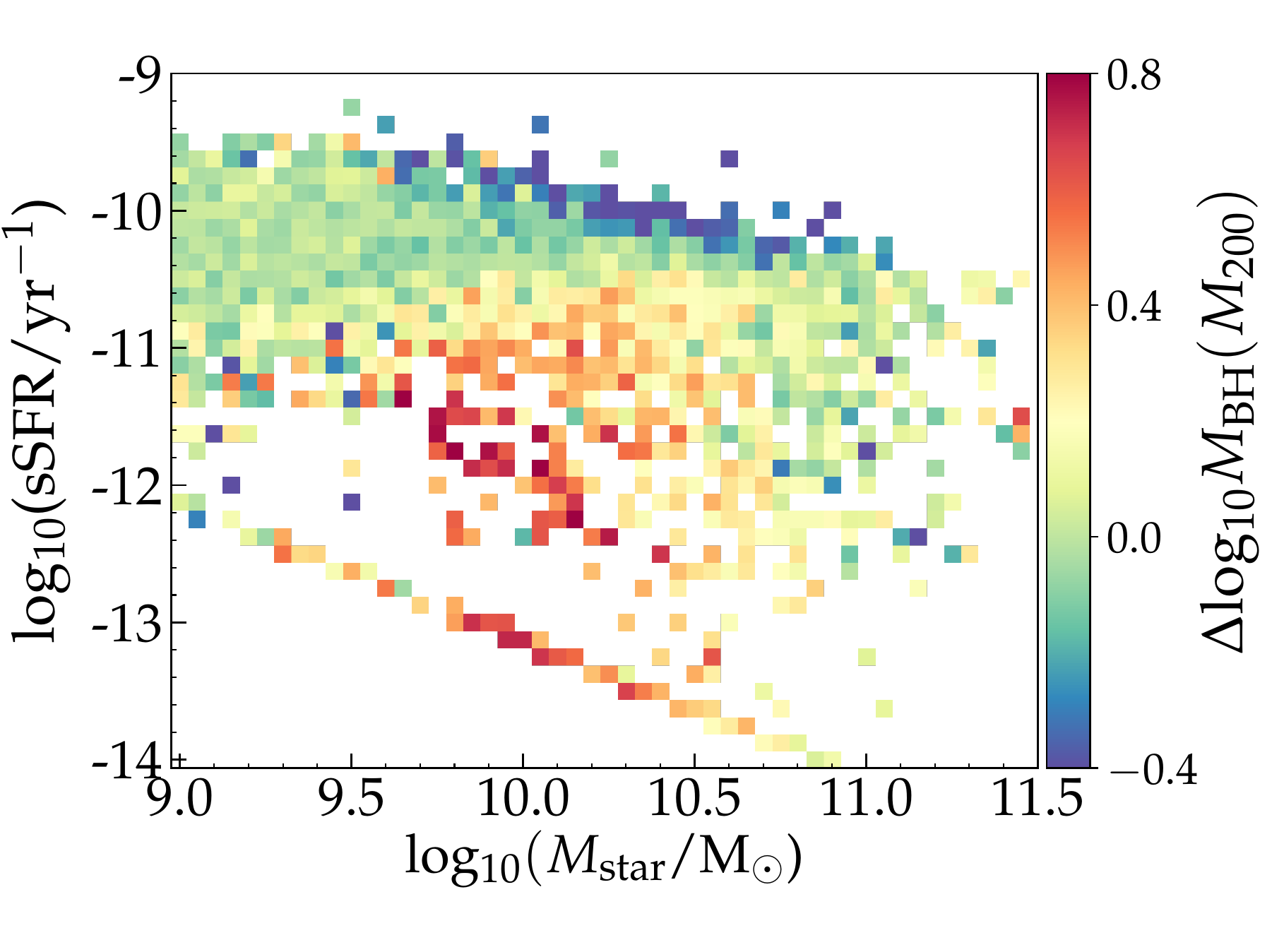}
\caption{\small{Relation between sSFR and stellar mass for central galaxies at $z=0.1$. Here, we colour code the grid of bins with the median residual of the BH - halo mass relation ($\Delta$log$_{10}$ M$_{\rm BH}(M_{200})$), which highlights the {\it relative} efficiency of BH formation. Part of the scatter at stellar mass $\gtrsim10^{10}$ M$_{\odot}$ can be attributed to the mass of the BH relative to that of the halo.}}
\label{fig:SFRMstar_BH}
\end{figure}

In Fig. $\ref{fig:SFRMstar_BH_EVO}$ we show that the SFRs of central galaxies with M$_{\rm star}\ll10^{10}$ M$_{\odot}$ are uncorrelated with M$_{\rm BH}$. However, galaxies with M$_{\rm star}>10^{10}$ M$_{\odot}$ residing in haloes that formed a BH efficiently (resulting in a high M$_{\rm BH}$/M$_{200}$ ratio) have relatively low SFRs at fixed stellar mass at $z=0.1, 0.5$ and $z=2.0$, and have thus stopped forming stars at the MS (or are in the process of ceasing their star formation). A relation between the relative BH growth and the SFH is also seen observationally by \cite{MartinNavarro2018}, who found that galaxies with over-massive black holes (at fixed velocity dispersion) formed their stellar mass earlier.

The stellar mass scale at which BH mass starts to drive scatter in the SFR-M$_{\rm star}$ relation is slightly smaller at $z=2$. This likely reflects the fact that the stellar mass - halo mass relation in EAGLE is lower at $z=2$ (see \citealt{Matthee2017}), meaning that galaxies with M$_{\rm star} \approx 10^{10}$ M$_{\odot}$ reside in higher-mass haloes at $z=2$, compared to galaxies with this mass at $z=0$. This effect explains why the scatter in the SFR-M$_{\rm star}$ relation increases with mass around M$_{\rm star}\approx10^{9.7}$ M$_{\odot}$ at high redshift (see Fig. $\ref{fig:scatter_SFR_evolution}$), and why the transitioning stellar mass at which this happens decreases slightly with increasing redshift. 

Fig. $\ref{fig:SFRMstar_BH_EVO}$ also shows that a sample of galaxies selected above a simple sSFR threshold includes galaxies in which BH growth is already affecting galaxy properties, particularly at high masses. The selection of ``main sequence galaxies'' in a simulation, i.e. galaxies where AGN feedback is not yet important, could therefore impose an additional selection threshold of log$_{10}$(M$_{\rm BH}$/M$_{200}$) $<-5.0$.

\begin{figure}
\includegraphics[width=8.6cm]{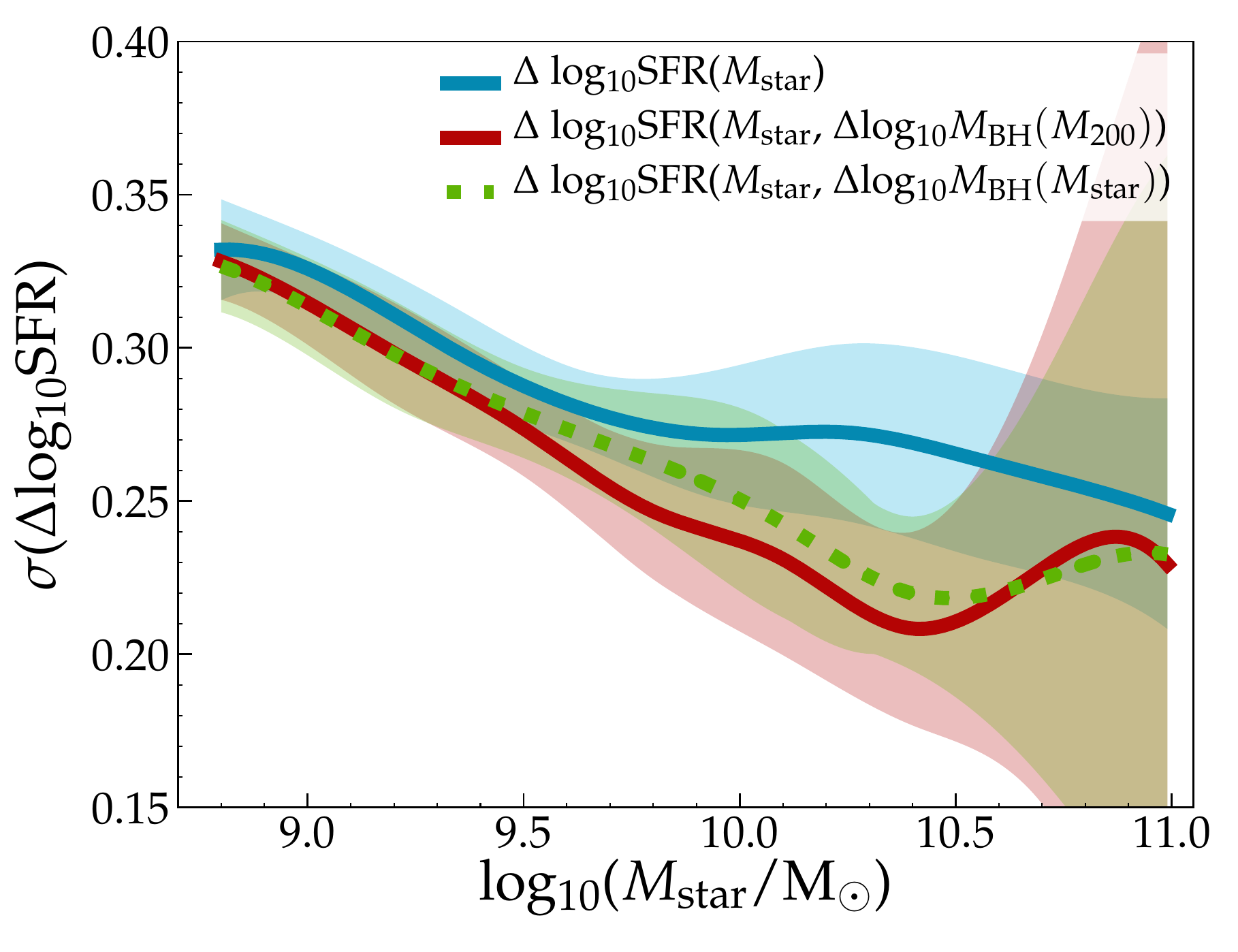} 
\caption{\small{Scatter in SFR at fixed stellar mass for $z=0.1$ central star-forming galaxies only (sSFR$>10^{-11}$ yr$^{-1}$), and after correcting for the effect of the relative BH mass (using the residuals of the M$_{\rm BH}$-M$_{200}$ and M$_{\rm BH}$-M$_{\rm star}$ relations). Although the uncertainties due to cosmic variance are high at large masses, we find clear indications that a significant part of the scatter in the SFR-M$_{\rm star}$ relation at M$_{\rm star}>10^{10}$ M$_{\odot}$ is due to the differences in the relative efficiency of BH growth. Note that the galaxies for which the residuals of the M$_{\rm BH}$-M$_{200}$ relation are the highest (around M$_{\rm star}\approx10^{10}$ M$_{\odot}$) are not included in this analysis as these are not classified as star-forming galaxies due to their reduced sSFR. }}
\label{fig:scatter_SFR_MBH}
\end{figure}

\begin{figure*}
\begin{tabular}{cc}
\includegraphics[width=8.65cm]{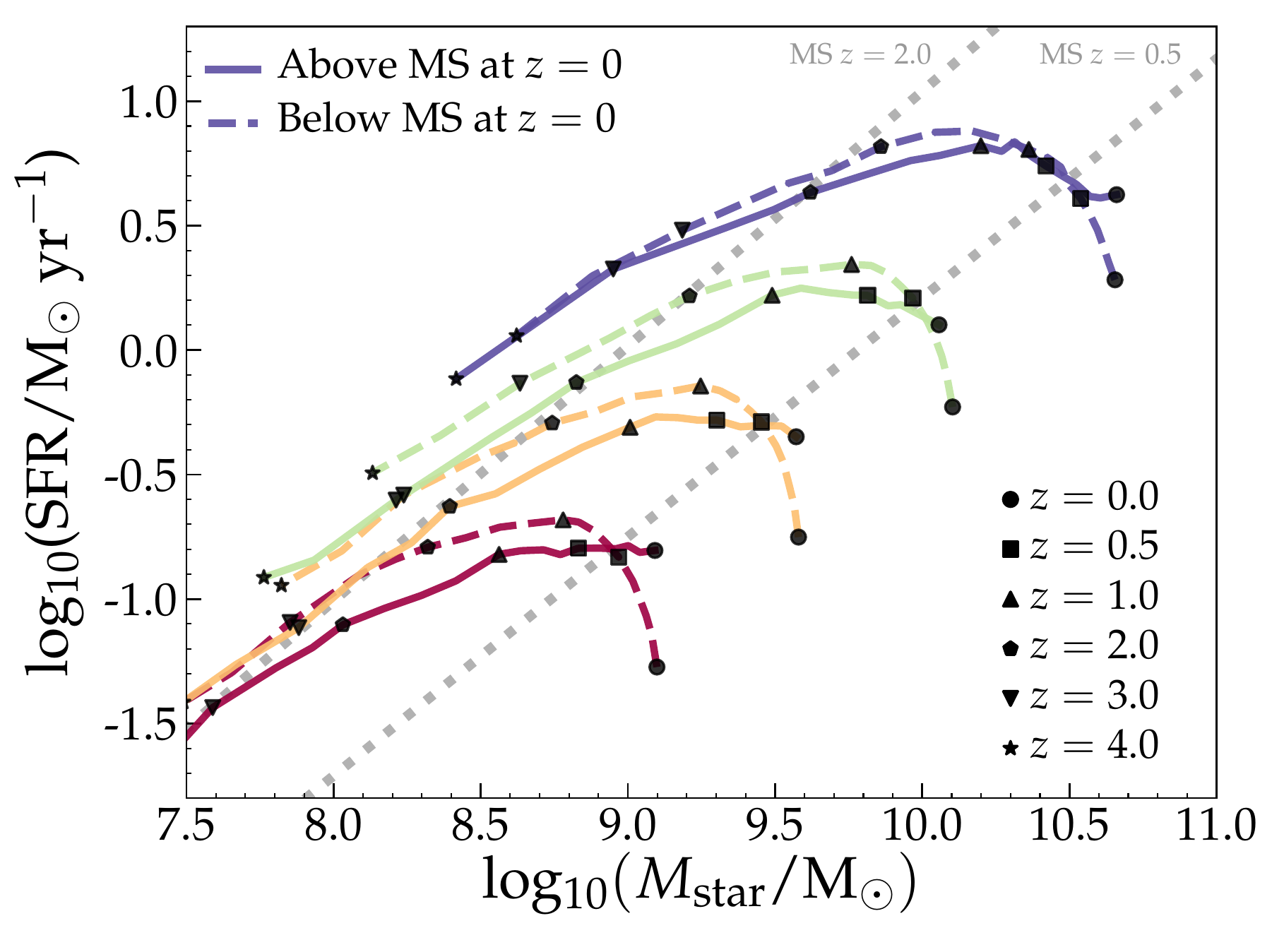} &
\includegraphics[width=8.65cm]{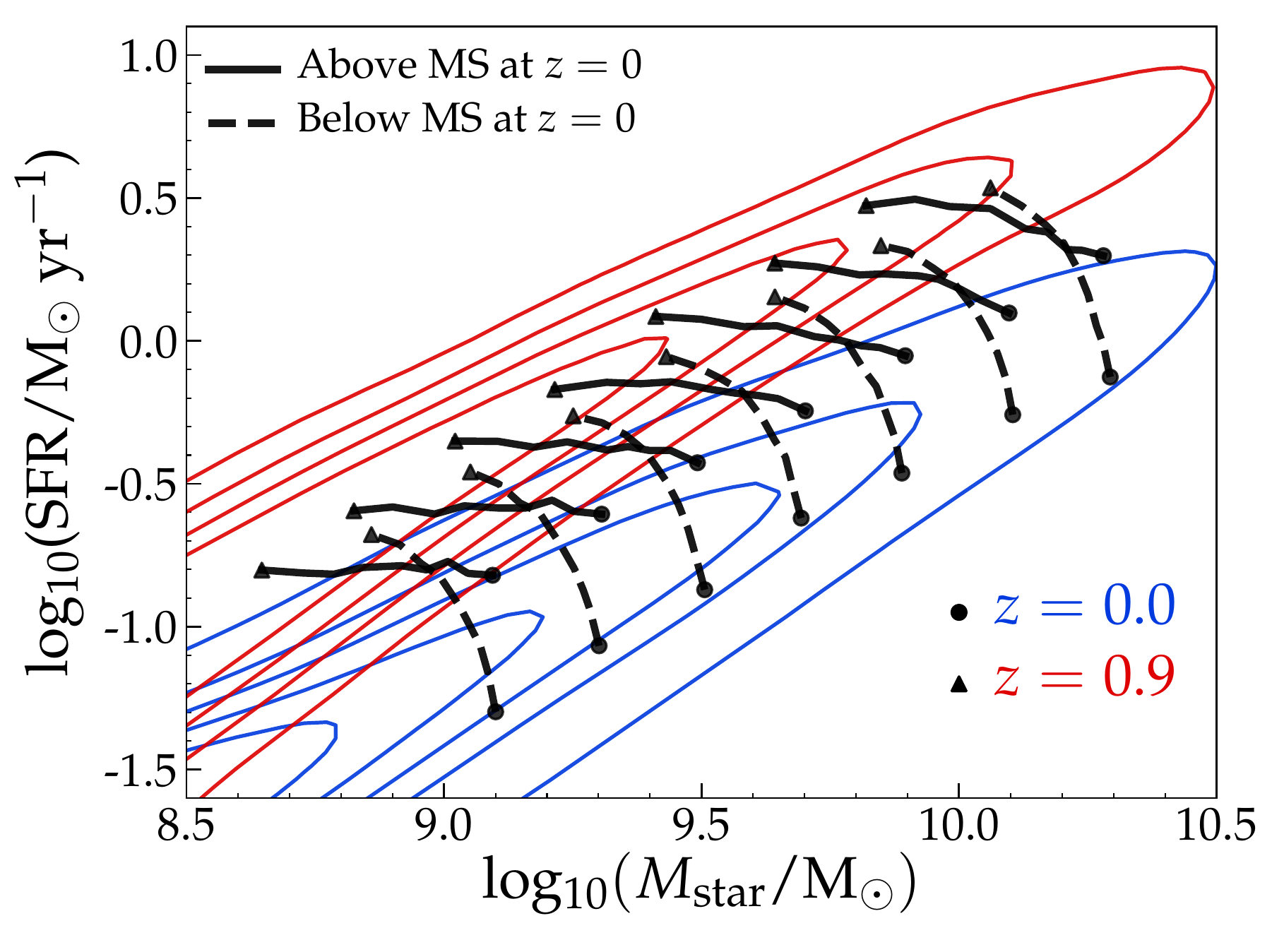} 
\end{tabular}
\caption{\small{{\it Left:} The median tracks of central star-forming galaxies in stellar mass bins (different colours), split by their sSFR at $z=0$ (different line-styles) in the SFR-M$_{\rm star}$ plane. The positions of galaxies at specific redshifts are indicated by different symbols. The grey dashed lines in the background show the median relations for star-forming galaxies at $z=0.5$ and $z=2.0$. There are clear differences between the median paths of galaxies at fixed mass, depending on their present-day SFR. {\it Right:} The median evolutionary tracks of galaxies in mass and SFR bins between $z=0.0$ and $z=0.9$ (roughly half of the history of the Universe). The contours show the full star-forming galaxy population at both redshifts. At all masses, the median SFR of the galaxies that are above/below the main sequence at $z=0$ was also above/below the main sequence at $z=0.9$. }}
\label{fig:path_plots}
\end{figure*} 

While the relation between the scatter in the SFR-M$_{\rm star}$ relation and BH growth shown in Fig. $\ref{fig:SFRMstar_BH_EVO}$ illustrates mostly ``first order effects'' (i.e. at fixed stellar mass galaxies with lower SFRs tend to reside in higher-mass halos which tend to have higher BH to halo mass ratios), a ``second order'' effect of how BH growth affects galaxies SFRs is clearly illustrated in Fig. $\ref{fig:SFRMstar_BH}$. Here, the colour coding shows the residuals of the M$_{\rm BH}$-M$_{200}$ relation, and thus highlights haloes that have grown their central super-massive black hole  {\it relatively} efficiently. This figure shows that particularly at the transition (stellar) mass scale (M$_{\rm star} \sim 10^{10}$ M$_{\odot}$, $z=0.1$), haloes that have a higher BH mass relative to their halo mass tend to have a lower SFR at fixed stellar mass. For the highest and lowest stellar masses there is no clear relation with the relative black hole mass as galaxies typically have been quenched already (highest masses) or AGN feedback is unimportant (lowest masses). 

As shown in Fig. $\ref{fig:scatter_SFR_MBH}$, accounting for black hole mass reduces the scatter in the SFR of star-forming galaxies by $\approx 0.05$ dex for masses M$_{\rm star} \approx 1-3\times10^{10}$ M$_{\odot}$ (although the differences are within the uncertainties associated with cosmic variance). Assuming that the total scatter is the quadratic sum of the scatter due to relative BH formation efficiency and other sources of scatter, this means that variations in BH formation efficiency lead to $\approx0.15$ dex of scatter in the SFR-M$_{\rm star}$ relation at these mass scales. The results are similar when we compare the relative BH mass to stellar mass instead of to halo mass.

Which property determines whether halos grow a BH relatively efficiently? We find that the residuals of the M$_{\rm BH}$-M$_{200, \rm DMO}$ relation (particularly for halo masses M$_{200, \rm DMO}>10^{12}$ M$_{\odot}$) are correlated with halo formation time (and thus with concentration, see \citealt{BoothSchaye2010,BoothSchaye2011}). As a result, at fixed halo mass, haloes that form earlier typically end up receiving a larger amount of AGN feedback (once AGN feedback becomes efficient), resulting in a lower sSFR and eventually lower stellar mass. Simultaneously, at fixed M$_{200}$, but at M$_{200, \rm DMO}<10^{12}$ M$_{\odot}$, an earlier halo formation time tends to yield a larger M$_{\rm star}$ (see \citealt{Matthee2017}). As a result, galaxies with high SFR at fixed M$_{\rm star}$ at stellar masses above $10^{10}$ M$_{\odot}$ tend to have a relatively low halo mass, while galaxies at lower stellar mass with high SFR typically have a high halo mass (see Fig. $\ref{fig:SFRMstar_SMHM}$). The question that remains is: at fixed halo mass and fixed halo formation time, what determines whether a halo forms stars more efficiently? This is a topic for future investigations.

\section{Discussion} \label{sec:discussion}
\subsection{What drives the scatter in the main sequence?}
In the introduction, we posed the question whether the galaxy main sequence is an ``attractor-solution'', with the scatter originating from rapid fluctuations around a median relation, or whether it represents a ``population-average'', with the scatter reflecting the diversity in star formation histories. As presented in \S $\ref{burstiness}$  and quantified in \S $\ref{howmuchscatter}$, we find that it most likely is a combination of both. Around $\approx 0.2$ dex of scatter in the SFR-M$_{\rm star}$ relation is due to star formation burstiness on short $\lesssim1$ Gyr time-scales, originating from the self-regulating interplay between gas cooling, star formation and feedback from star formation, and $\approx0.2$ dex of scatter is due to long $\sim 10$ Gyr time-scale differences related to the formation time of dark matter haloes (which originate from large-scale dark matter density fluctuations in the early universe). At masses M$_{\rm star} > 10^{10}$ M$_{\odot}$ the residuals in the SFR-M$_{\rm star}$ relation are anti-correlated with the relative efficiency of past BH growth (which is a proxy for the accumulated AGN feedback energy). Since the BH mass is correlated with halo mass and formation time, see \S $\ref{sec:BH}$, this is an alternative manifestation of how halo assembly bias (i.e. variations in halo formation time at fixed mass) impacts the scatter in the SFR-M$_{\rm star}$ relation.

We illustrate the implications of these results for the {\it median} tracks of central star-forming galaxies in the SFR-M$_{\rm star}$ plane in Fig. $\ref{fig:path_plots}$. In the left panel, we show the tracks of galaxies that are binned by their $z=0$ mass (line-colour) and sSFR (line-style). Their positions at specific redshifts are indicated by different symbols. The figure shows that there are clear differences between the median paths of galaxies at fixed mass, depending on their present-day SFR. For example, galaxies with M$_{\rm star}=10^{10}$ M$_{\odot}$ that are above the main sequence at $z=0$ had a median stellar mass of $\approx10^{9.4}$ M$_{\odot}$ at $z=1$, while galaxies with the same mass at $z=0$, but with a relatively low sSFR, already had a median stellar mass of $\approx10^{9.7}$ M$_{\odot}$ at $z=1$. As a result, the median tracks that galaxies follow are not parallel to the main sequence (which typically has a slope $\approx1$, as illustrated with grey dashed lines at $z=0.5$ and $z=2.0$) at specific redshifts. 

The right panel of Fig. $\ref{fig:path_plots}$ shows another illustration of the long-term coherence of the median SFHs of galaxies depending on their position along the main sequence (see also Fig. $\ref{fig:memory}$). We show the positions of galaxies on the SFR-M$_{\rm star}$ relation at $z=0.0$ and $z=0.9$ (roughly half of the age of the Universe) and compare them to the general population of star-forming galaxies at these redshifts. At all masses, galaxies (at fixed mass at $z=0$) that are above the main sequence at $z=0$ have had a median SFR above the MS for roughly half of the age of the Universe, and vice versa for galaxies below the main sequence at $z=0$. We emphasise that Fig. $\ref{fig:path_plots}$ shows median tracks. As discussed in \S $\ref{burstiness}$, the SFRs of individual galaxies fluctuate around these tracks on shorter time-scales.

\begin{figure}
\includegraphics[width=8.6cm]{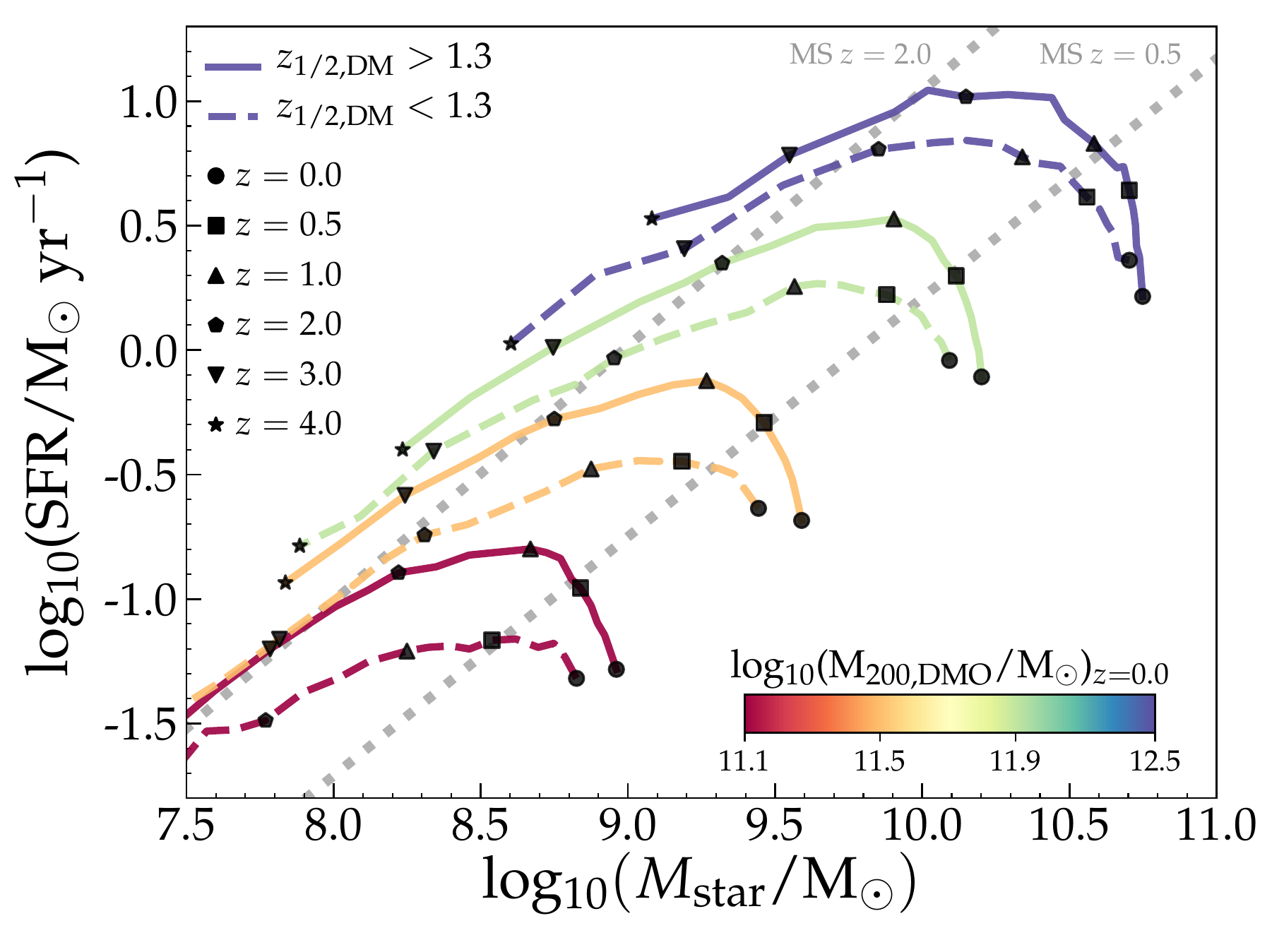} 
\caption{\small{As Fig. $\ref{fig:path_plots}$, the paths of central star-forming galaxies at $z=0$ through the SFR-M$_{\rm star}$ plane in bins of $z=0$ DMO halo mass (colour scale) and halo formation time (different line-styles). At fixed final halo mass, haloes that assembled more than half of their mass at $z>1.3$ (the median formation redshift of haloes included in this analysis) end up with a higher stellar mass (except for the highest halo masses). These early forming haloes had a higher SFR for most of the history of the Universe, but not at $z=0$, where they have a lower SFR (except for the lowest halo masses). }}
\label{fig:path_Mhalo}
\end{figure}

Fig. $\ref{fig:path_Mhalo}$ illustrates that the origin of these long-term ``fast-track'' and ``slow-track'' SFHs lies in the different halo formation times and is thus a manifestation of assembly bias. At fixed halo mass, haloes that formed earlier end up with a higher stellar mass and underwent a period with a higher SFR in the past. While short time-scale fluctuations in the SFHs of individual galaxies that are not in phase cancel each other out when computing the median (as discussed in \S $\ref{burstiness}$), we note that the typical fluctuations in log$_{10}$(SFR) are $\approx 0.2-0.3$ dex on time-scales of $\lesssim1$ Gyr (see Fig. $\ref{fig:path_individual}$). Therefore, individual galaxies likely cross the main sequence multiple times, but they fluctuate around different median relations that are related to their halo formation times (resulting in differences in the median paths). Hence, at a specific snapshot in the history of the Universe, the scatter in the main sequence reflects a combination of the average paths of galaxies in the SFR-M$_{\rm star}$ plane related to their halo mass and formation time (see Fig. $\ref{fig:path_Mhalo}$), and short time-scale fluctuations around these paths that are related to specific gas cooling and feedback events, bound by self-regulation.

\subsection{Observational tests} \label{sec:tests}
While the EAGLE simulation reproduces several key global galaxy properties, such as the stellar mass function, galaxy sizes, build-up of stellar mass, passive fraction and the black hole mass - stellar mass relation \citep[e.g.][]{Schaye2014,Furlong2015,Furlong2017}, more detailed tests of the model can be performed using second order relations. Here we provide several second-order observational tests that can be performed to test the model:
\begin{enumerate}
\item As shown in Fig. $\ref{fig:scatter_SFR_evolution}$, EAGLE predicts that there is significantly less scatter in the SFR-M$_{\rm star}$ relation at $z\gtrsim1.5$ than at $z=0$ for galaxies with stellar masses M$_{\rm star}<10^{9.5}$ M$_{\odot}$ (0.2 dex, versus 0.35 dex). At $z\approx1.5-4$, the scatter is currently mostly constrained at higher masses (where observations find little evolution, similar to the results in the simulation). Deep surveys of mass-selected galaxies (we illustrate in Appendix $\ref{appendix_SDSS}$ that a SFR-selected sample under-estimates the scatter significantly) that is complete down to these low masses with accurate measurements of their SFRs can test whether there is indeed less scatter at these redshifts.
\item As shown in Fig. $\ref{fig:MstarSFR_EVO}$, the SFHs of galaxies at fixed stellar mass depend strongly on their present-day sSFR. The median SFHs of galaxies with the lowest $z=0$ sSFRs have been declining for more than $\approx 8$ Gyr, while the SFHs of galaxies with high sSFR have a much more extended SFH. These qualitative trends can be tested with observationally inferred SFHs, for example based on detailed modelling of high-resolution spectra \citep[e.g.][]{Chauke2018}. Detailed comparisons with observations would also require the mimicking of observational techniques on simulated galaxies.
\item An interesting consequence of the correlation between present-day sSFR and galaxies' SFHs is that the stellar and gas-phase $\alpha$-enhancements of star-forming galaxies in the local Universe depend on sSFR \citep{MattheeSchaye2018}. Although the gas-phase iron abundance and $\alpha$-enhancements of young stellar populations are challenging to measure, a correlation between sSFR and $\alpha$-enhancement at fixed stellar mass would provide an independent constraint on differences in SFHs.
\item Most challenging to measure will be the relation between the scatter in the SFR-M$_{\rm star}$ relation and dark matter halo mass and black hole mass (such as the relative BH growth efficiency; Fig. $\ref{fig:scatter_SFR_MBH}$). While (statistical) measurements of halo masses will be possible with lensing surveys such as {\it Euclid}, direct measurements of the masses of the supermassive black holes will remain challenging, particularly in non-active galaxies.
\end{enumerate}

\section{Summary} \label{sec:summary}
We have used the cosmological hydrodynamical EAGLE simulation to study the magnitude, mass dependence and origin of scatter in the SFR-M$_{\rm star}$ relation at $z=0$, and its evolution. In order to allow for a proper comparison to observations, we have also measured the magnitude and mass dependence of the scatter in a sample of galaxies in the local Universe from the Sloan Digital Sky Survey. 

At $z=0$, we find that the scatter in the SFR of star-forming galaxies at fixed M$_{\rm star}$ decreases slightly with stellar mass (Fig. $\ref{fig:scatter_SFR_SDSS}$) from 0.35 dex at M$_{\rm star} \approx 10^9$ M$_{\odot}$ to 0.30 dex at M$_{\rm star} \gtrsim 3\times10^{10}$ M$_{\odot}$. Accounting for measurement errors in the data results in disagreement with the observed slope of the scatter as a function of stellar mass at the low stellar mass end (M$_{\rm star}\ll10^{10}$ M$_{\odot}$). This highlights the importance of understanding the mass- and SFR- dependency of measurement uncertainties. Consistent with observational constraints, there is little evolution in the scatter in SFR at fixed stellar mass for galaxies with masses M$_{\rm star} \gtrsim 10^{10}$ M$_{\odot}$ between $z=0$ and $z=2$. At lower masses, however, the scatter is predicted to decrease from $\approx 0.35$ dex at $z=0$ to $\approx 0.20$ dex at $z>2.5$, see Fig. $\ref{fig:scatter_SFR_evolution}$. Excluding star-forming satellite galaxies only removes $\approx 0.04$ dex of scatter (Fig. $\ref{fig:scatter_SFR_satellites}$). This indicates that satellite-specific processes are either weak, or strong and rapid (such that the satellites quickly drop out of the sample of star-forming galaxies).

We find that for galaxies that are `above' the main sequence at $z=0.1$ the median SFR tends to have been `above' the main sequence for $\sim 10$ Gyr, see Figs. $\ref{fig:memory}$ and $\ref{fig:path_plots}$. On top of these long time-scale differences (of $\approx 0.20$ dex) tracked by the median SFHs, we find that the SFRs of individual galaxies typically fluctuate by $\approx0.2$ dex on time-scales of $\lesssim 2$ Gyr (Fig. $\ref{fig:path_individual}$ and \S $\ref{sec:relative_timescales}$). As these short time-scale fluctuations in different galaxies are typically not in phase, median stacking analyses cannot identify them as sources of scatter. 

A principal component analysis of tracks of individual SFGs through $\Delta$SFR(M$_{\rm star}$) - time space (i.e. the evolution relative to the evolution of the main sequence) reveals that the majority of the variance is due to fluctuations with long ($>2$ Gyr) periods. Fluctuations on shorter periods account for progressively smaller fractions of the variance. (Fig. $\ref{fig:PC_spectrum}$). We find that the importance of fluctuations on long time-scales increases with stellar mass. For example, 75 \% of the variance can be explained by including fluctuations with time-scales $\gtrsim2$ Gyr for galaxies with M$_{\rm star, z=0} = 10^{9.5\pm0.1}$ M$_{\odot}$, while fluctuations with time-scales down to $\approx0.8$ Gyr are required to account for a similar amount of variance in the tracks of galaxies with M$_{\rm star, z=0} = 10^{9.3\pm0.1}$ M$_{\odot}$. 

As a consequence, we find that the scatter in the SFR-M$_{\rm star}$ relation depends on the time-scale over which the SFR is averaged, particularly for M$_{\rm star}\lesssim10^{10}$ M$_{\odot}$. The scatter is $\approx0.3$ dex when SFR is averaged over 1 Gyr and it decreases to a non-negligibile $\approx0.15$ dex when SFR is averaged over 8 Gyr, highlighting the significant contribution of long-term SFHs to the scatter in the SFR-M$_{\rm star}$ plane (Fig. $\ref{fig:scatter_on_time-scales}$).
  
The origin of the long time-scale fluctuations lies in variations in dark matter halo formation times. For M$_{\rm star} \lesssim 10^{10}$ M$_{\odot}$, scatter in the formation times of dark matter halos contributes 0.15 dex of scatter in the main sequence (at $z=0.1$; central galaxies only), but its effect is smaller at higher stellar masses (Figs. $\ref{fig:SFRMstar_z05}$ and $\ref{fig:scatter_SFR_formationtime}$). As halo formation time is related to large-scale structure, this contribution to scatter in the SFR-M$_{\rm star}$ relation is a manifestation of assembly bias. Hence, while individual galaxies cross the main sequence multiple times during the history of the Universe, they fluctuate around different median relations that are related to their halo mass and their halo formation times (i.e. a fast-track associated to haloes that form relatively early and a slow-track for haloes that form relatively late, see Fig. $\ref{fig:path_Mhalo}$).

At high masses the BH mass relative to the halo mass/stellar mass is strongly correlated with the scatter in the SFR-M$_{\rm star}$ relation (Figs. $\ref{fig:SFRMstar_BH_EVO}$ and $\ref{fig:scatter_SFR_MBH}$). Galaxies with a relatively high black hole mass for their halo/stellar mass tend to have a low SFR at fixed stellar mass (even when the galaxies are still classified as star-forming, see Fig. $\ref{fig:SFRMstar_BH}$). As the relative BH formation efficiency is also higher in halos that assemble earlier, galaxies residing in haloes with a `fast-track' evolution are affected earlier by AGN feedback.
 
Our results imply that the scatter in the SFR-M$_{\rm star}$ relation reflects the diversity in SFHs. Most of this scatter is driven by long time-scale ($\sim10$Gyr) differences related to halo mass and halo formation time (i.e. assembly bias), but scatter also originates from short ($\lesssim1$Gyr) time-scale fluctuations that do not simply trace back to changes in halo accretion, but are likely controlled by self-regulation of star formation through feedback.

\section*{Acknowledgments}
JM acknowledges the support of a Huygens PhD fellowship from Leiden University. We thank Camila Correa for help analysing snipshot merger-trees. We thank the anonymous referee for constructive comments. We also thank Jarle Brinchmann, Rob Crain, Antonios Katsianis, Paola Popesso and David Sobral for discussions and suggestions. We also thank the participants of the Lorentz Center workshop `A Decade of the Star-Forming Main Sequence' held on September 4-8 2017, for discussions and ideas. We have benefited from the public available programming language \texttt{Python}, including the \texttt{numpy, matplotlib} and \texttt{scipy} \citep{Hunter2007} packages and the \texttt{Topcat} analysis tool \citep{Topcat}.




\bibliographystyle{mnras}

\bibliography{bibliography_pceagle.bib}




\label{lastpage}
\appendix
\section{On selection biases and the measured scatter} \label{appendix_SDSS}
As discussed in the main text (\S $\ref{sec:eagle_sdss}$), several observational biases complicate the measurement of scatter in the SFR-M$_{\rm star}$ relation for the observational sample from SDSS. Here, we motivate the redshift- and mass-limits of the sample used and explain why it is necessary to perform these cuts. 

\begin{figure*}
\begin{tabular}{cc}
\includegraphics[width=8.5cm]{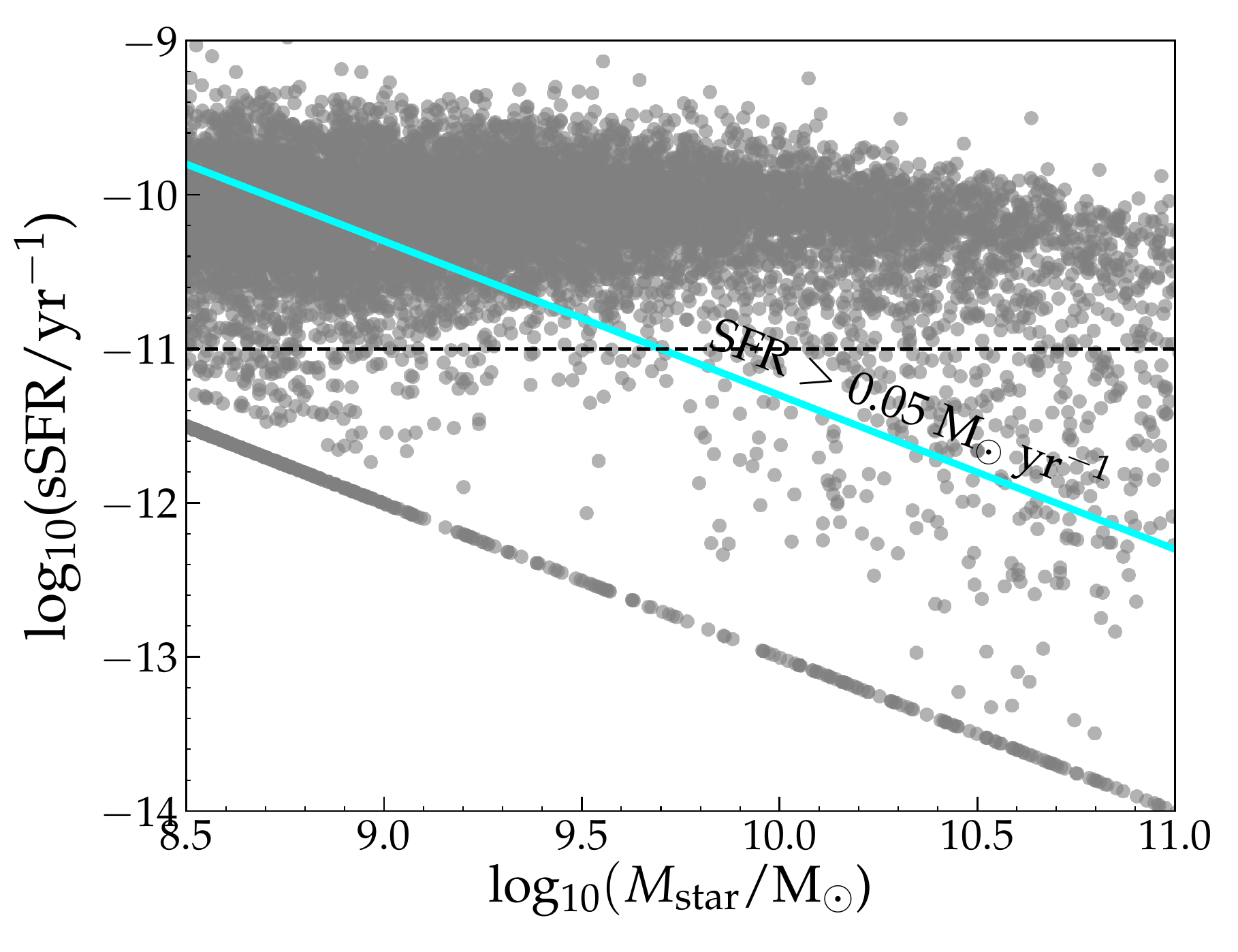}&
\includegraphics[width=8.5cm]{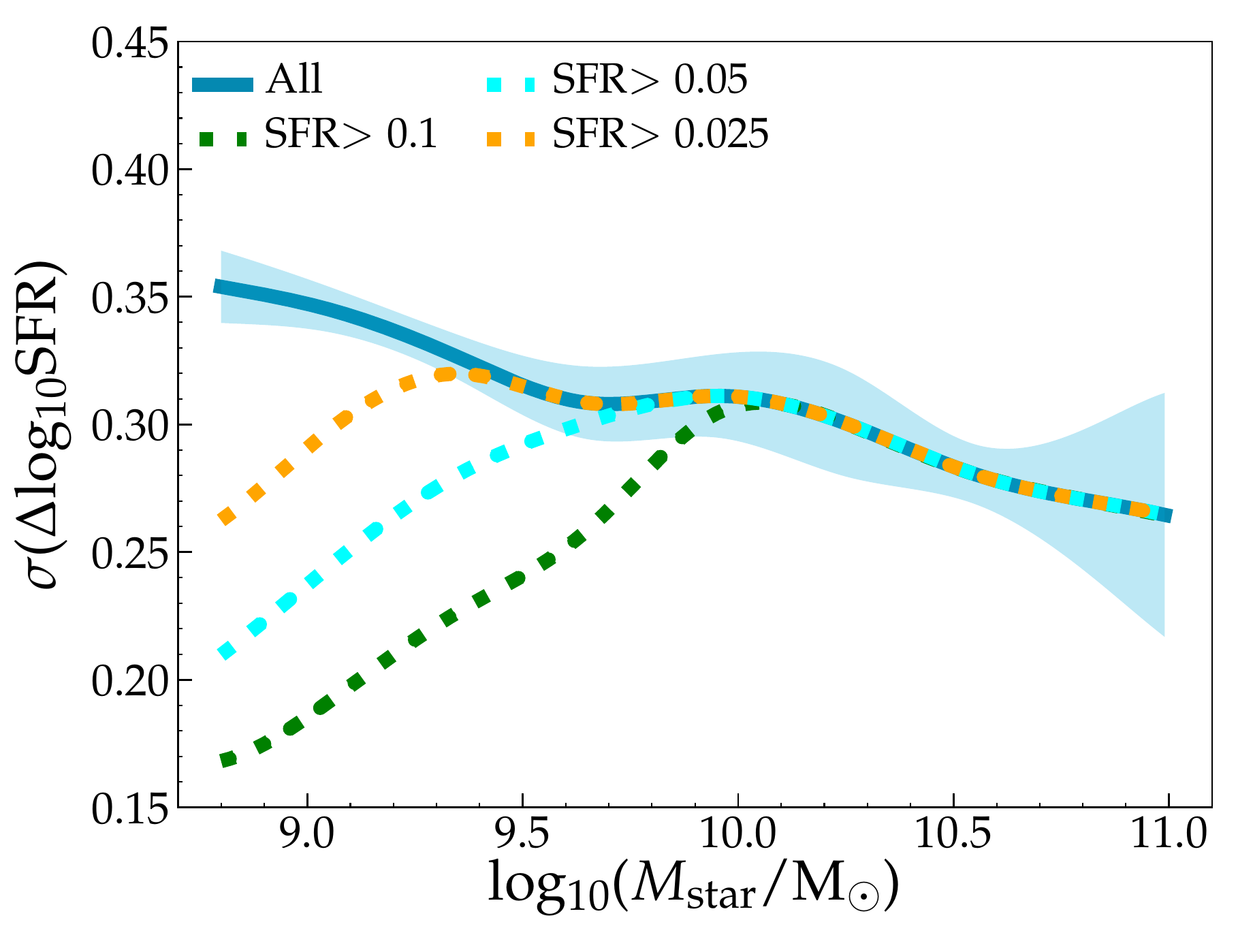}\\
\end{tabular}
\caption{\small{The effect of observational selection biases on the scatter in the SFR-M$_{\rm star}$ relation, mimicked using galaxies in EAGLE. The {\it left panel} illustrates a pure stellar mass selection (all grey points) and a SFR selection (cyan line). The {\it right panel} shows the scatter in the SFR-M$_{\rm star}$ relation that is measured from galaxy samples with these selection biases (note that we show various thresholds for SFR-selected samples).  }} \label{fig:incompleteness}
\end{figure*}

First, we use galaxies in the EAGLE simulation to show how sample incompleteness may influence the measured scatter in Fig. $\ref{fig:incompleteness}$. In the left panel we illustrate two different selections of galaxies which depend on either SFR or stellar mass and in the right panel we show the corresponding scatter in the SFR-M$_{\rm star}$ relation measured with such selections. It is clear that any selection that is somewhat dependent on SFR results in an under-estimate of the scatter at low stellar masses, with the mass-scale at which this bias becomes important depending on the SFR threshold. In order to prevent such a bias from becoming important, the required SFR threshold can be computed as: 
log$_{10}$sSFR $>$ med(log$_{10}$sSFR) - 4$\sigma$(sSFR),
where med(log$_{10}$sSFR) is the median sSFR (assuming a main sequence with a slope of 1) and $\sigma$(sSFR) the scatter. From this, it follows that the scatter can be measured reliably for a SFR threshold of $>0.05$ at M$_{\rm star}\approx10^{9.7}$ M$_{\odot}$ and above.

\begin{figure*}
\begin{tabular}{cc}
\includegraphics[width=8.5cm]{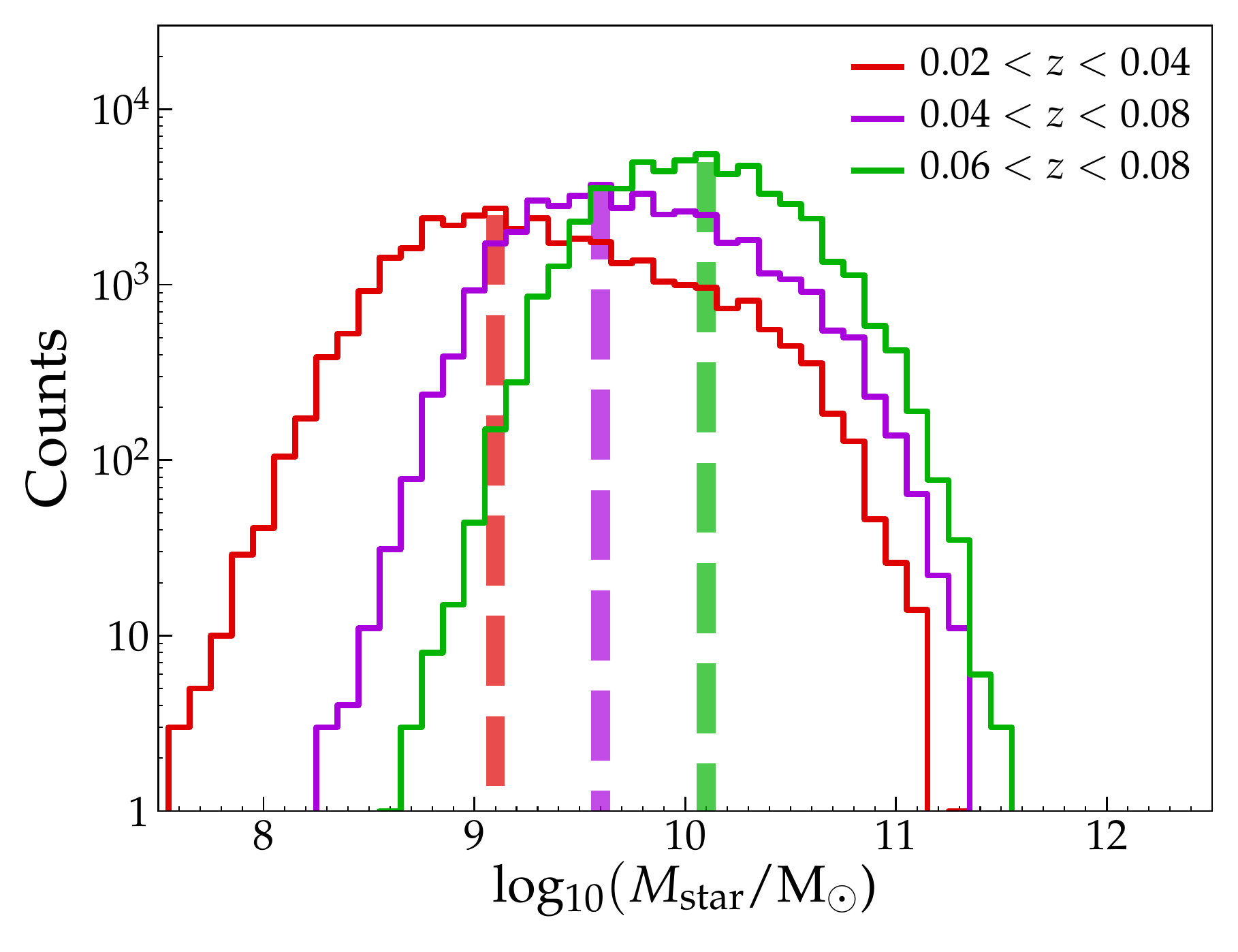}&
\includegraphics[width=8.5cm]{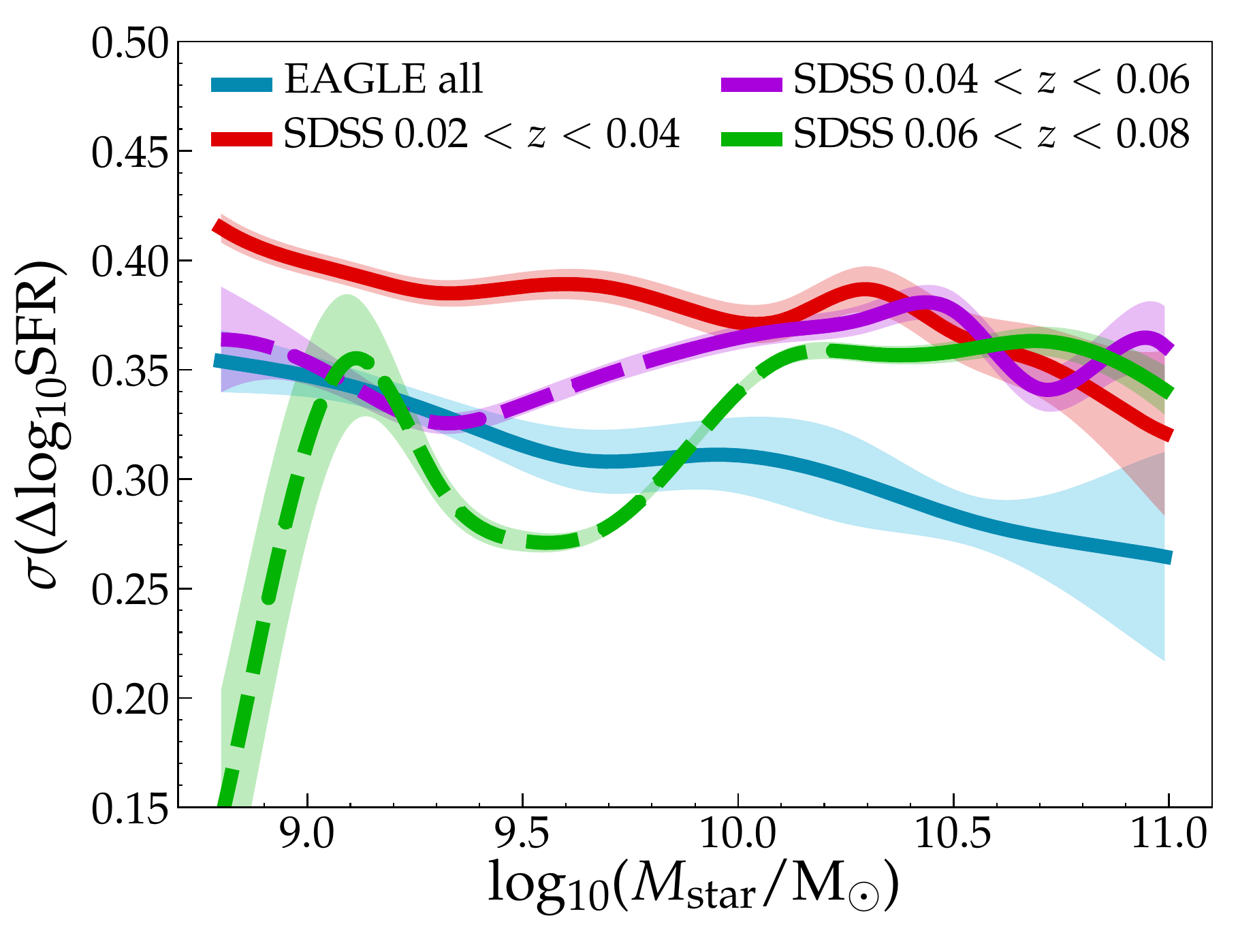}
\end{tabular}
\caption{\small{Galaxy counts in the SDSS sample as a function of stellar mass in three bins in redshift ({\it left panel}) and the measured scatter in the SFR-M$_{\rm star}$ relation in each redshift bin, where the line-style changes to dashed at masses below the completeness threshold  ({\it right panel}). The left panel illustrates the stellar mass above which the bins are close to complete (which we define as the mass at which the histogram peaks, illustrated with a vertical dashed line). The right panel shows that incomplete galaxy samples result in an under-estimate of the scatter. }} 
\label{fig:histogram_SDSS}
\end{figure*}

Second, as observations are limited by observed flux, the mass above which at which a sample is complete increases with redshift. In Fig. $\ref{fig:histogram_SDSS}$, we show histograms of stellar masses for the SDSS galaxies of \cite{Chang2015} in three narrow redshift slices (left panel) and the corresponding measured scatter in the SFR-M$_{\rm star}$ relation in each redshift slice (right panel). We identify the lowest stellar mass at which each redshift slice is close to complete by the peak in the histogram, and show that indeed, due to the selection offset, the measured scatter is biased low at incomplete masses, similar to the mimicked observations in the right panel of Fig. $\ref{fig:incompleteness}$. In the paper we therefore restrict the sample to $0.02<z<0.04$, which is complete at stellar masses $\gtrsim10^9$ M$_{\odot}$.

\section{Testing different measures of halo mass growth}\label{test_massgrowth}
\begin{figure}
\includegraphics[width=8.6cm]{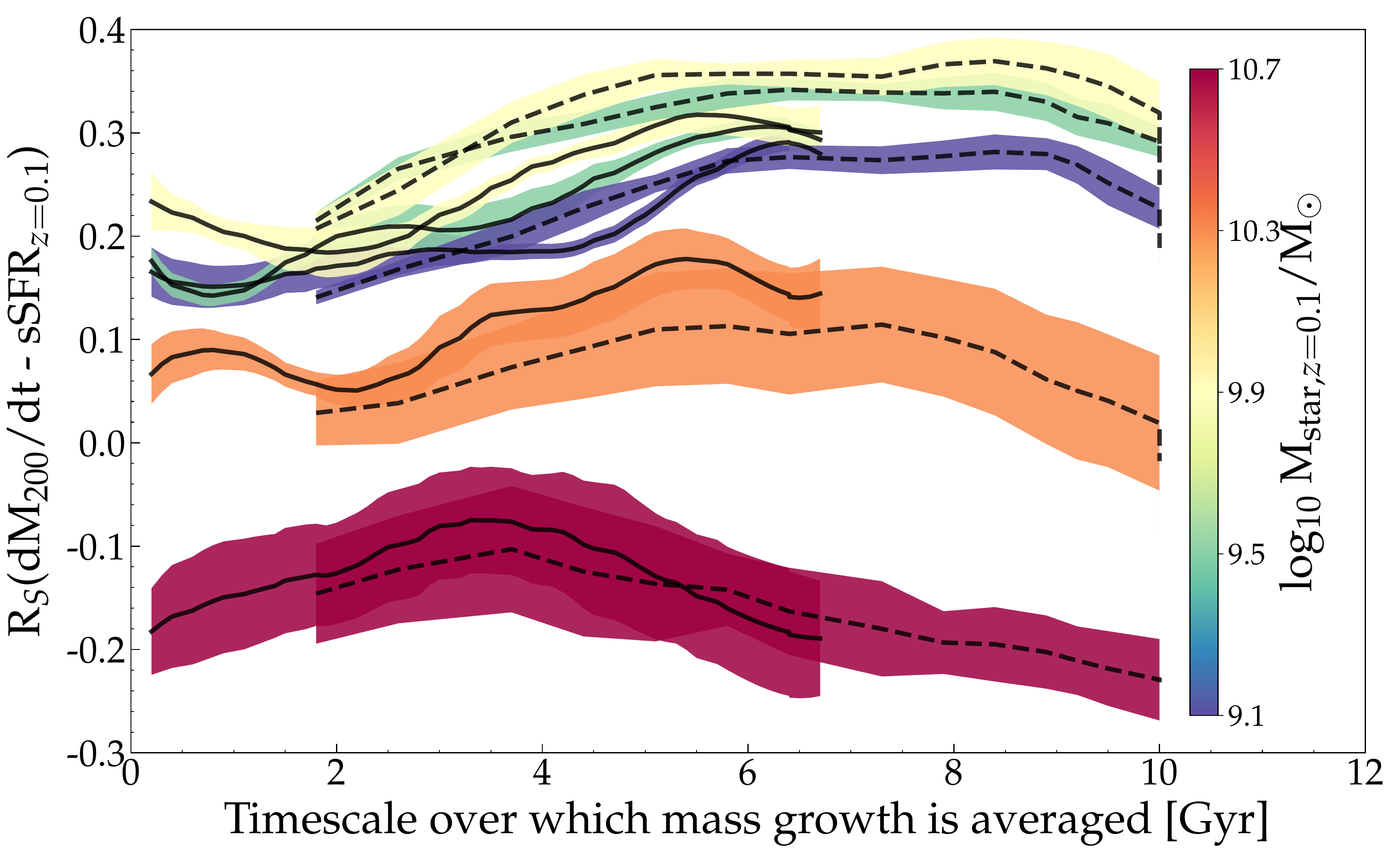}
\caption{\small{The strength of the correlation between the sSFR and the halo mass growth, where the latter is averaged over a range of time-scales, quantified by the Spearman correlation rank, R$_S$. Results are shown in bins of stellar mass at $z=0.1$. Solid lines show the results for the high time-resolution merger tree, while dashed lines show the results for the low time-resolution merger tree. The shaded regions indicate errors that are estimated through jackknife resamples of eight sub-volumes in the simulation volume. At fixed stellar mass, the specific SFR is related most strongly to halo mass growth if it is averaged over long, $\sim 7$ Gyr time-scales. }}
\label{fig:SFRMstar_M200snip}
\end{figure}

In \S $\ref{sec:scatter_z05}$, we quantified the halo accretion history using $z_{1/2}$, the redshift at which half of the (dark matter) halo mass at $z=0.1$ had been assembled in the main progenitor. In addition to only looking at $z_{1/2}$, which may be an overly simplistic parametrisation of the halo assembly history, we measure the halo mass growth averaged over a range of time-scales. We combine the low time-resolution (with resolution of 0.5-1 Gyr) merger tree up to $z=4$ with a high time-resolution (with a resolution of $\approx100$ Myr) merger tree up to $z=2$ and measure the halo mass growth averaged over $0.15-10$ Gyr. In Fig. $\ref{fig:SFRMstar_M200snip}$ we examine the strength of the relation between the sSFR at $z=0.1$ and the mass growth as a function of the time-scale over which the mass growth is averaged, in bins of stellar mass. 

The most important result from Fig. $\ref{fig:SFRMstar_M200snip}$ is that none of the time-scales yield a correlation with an absolute Spearman rank |R$_S|>0.4$ (which is the strength of the correlation between the scatter in the SFR-M$_{\rm star}$ relation and $z_{1/2}$). This means that there is no simple measurement of accretion history that correlates better with sSFR at fixed stellar mass than $z_{1/2}$ (R$_S \approx -0.45$). For masses between $10^9$ and $10^{10}$ M$_{\odot}$ we find that the strongest correlation is found when mass growth is averaged over long time-scales of $\approx 7$ Gyr. Perhaps surprisingly, on average, the current sSFR does not depend strongly on the most recent halo mass growth. This suggests that there is a gas reservoir in and/or around galaxies that dampens short time-scale fluctuations in the growth of M$_{200}$ and weakens the direct influence of halo accretion on SFRs.

\end{document}